\documentclass[acmsmall]{acmart}

\AtBeginDocument{%
  }

\setcopyright{cc}
\setcctype{by}
\acmJournal{PACMHCI}
\acmYear{2025} \acmVolume{9} \acmNumber{7} \acmArticle{CSCW321} \acmMonth{11} \acmPrice{}\acmDOI{10.1145/3757502}

\usepackage{enumitem}
\usepackage{framed}
\usepackage{multirow}
\usepackage{wrapfig}
\usepackage{subcaption}
\usepackage{hyperref}
\usepackage{xcolor}

\acmSubmissionID{3891}

\newcommand{\revision}{\textcolor{black}}

\newcommand{\yuansong}{\textcolor{black}}

\definecolor{customcolor}{RGB}{123, 191, 210}
\definecolor{DGcolor}{RGB}{78, 181, 255}

\newcommand{\mybox}[1]{%
  \fcolorbox{customcolor}{customcolor!10}{\strut #1}%
}
\newcommand{\DG}[1]{%
  \colorbox{DGcolor!10}{\strut #1}%
}

\newcommand{\darkcircled}[1]{\textcircled{\raisebox{-0.9pt}{\scriptsize \textbf{#1}}}}
\newcommand{\transparentcircled}[1]{\textcircled{\raisebox{-0.9pt}{\scriptsize #1}}}

\begin{document}

\title{ReviseMate: Exploring Contextual Support for Digesting \yuansong{STEM} Paper Reviews}

\author{Yuansong Xu}
\orcid{0009-0005-1630-6279}
\affiliation{%
  \institution{School of Information Science and Technology, ShanghaiTech University}
  \city{Shanghai}
  \country{China}
}
\email{xuys2023@shanghaitech.edu.cn}

\author{Shuhao Zhang}
\orcid{0009-0008-1933-1869}
\affiliation{%
  \institution{School of Information Science and Technology, ShanghaiTech University}
  \city{Shanghai}
  \country{China}
}
\email{zhangshh12024@shanghaitech.edu.cn}

\author{Yijie Fan}
\orcid{0009-0001-9025-3719}
\affiliation{%
  \institution{College of Computing and Data Science, Nanyang Technological University}
  \city{Singapore}
  \country{Singapore}
}
\email{yfan013@e.ntu.edu.sg}
\authornote{This work was done when Yijie Fan was an undergraduate student at ShanghaiTech University.}

\author{Shaohan Shi}
\orcid{0009-0004-3384-8304}
\affiliation{%
  \institution{School of Information Science and Technology, ShanghaiTech University}
  \city{Shanghai}
  \country{China}
}
\email{shishh2023@shanghaitech.edu.cn}

\author{Zhenhui Peng}
\orcid{0000-0002-5700-3136}
\affiliation{%
  \institution{School of Artificial Intelligence, Sun Yat-sen University}
  \city{Zhuhai}
  \country{China}
}
\email{pengzhh29@mail.sysu.edu.cn}

\author{Quan Li}
\orcid{0000-0003-2249-0728}
\affiliation{
  \institution{School of Information Science and Technology, ShanghaiTech University}
  \city{Shanghai}
  \country{China}
}
\email{liquan@shanghaitech.edu.cn}
\authornote{Quan Li is the corresponding author.}

\renewcommand{\shortauthors}{Yuansong Xu et al.}

\begin{abstract}
Effectively assimilating and integrating reviewer feedback is crucial for researchers seeking to refine their papers and handle potential rebuttal phases in academic venues. However, traditional review digestion processes present challenges such as time consumption, reading fatigue, and the requisite for comprehensive analytical skills. Prior research on review analysis often provides theoretical guidance with limited targeted support. Additionally, general text comprehension tools overlook the intricate nature of comprehensively understanding reviews and lack contextual assistance. To bridge this gap, we formulated research questions to explore the authors' concerns and methods for enhancing comprehension during the review digestion phase. Through interviews and the creation of storyboards, we developed \textit{ReviseMate}, an interactive system designed to address the identified challenges. A controlled user study (N=31) demonstrated the superiority of \textit{ReviseMate} over baseline methods, with positive feedback regarding user interaction. Subsequent field deployment (N=6) further validated the effectiveness of \textit{ReviseMate} in real-world review digestion scenarios. These findings underscore the potential of interactive tools to significantly enhance the assimilation and integration of reviewer feedback during the manuscript review process.
\end{abstract}

\ccsdesc[500]{Human-centered computing~Empirical studies in HCI}

\keywords{Review Comprehension, STEM Papers, Human-centered Design}

\received{July 2024}
\received[revised]{December 2024}
\received[accepted]{March 2025}

\begin{teaserfigure}
 \centering 
 \includegraphics[width=\columnwidth]{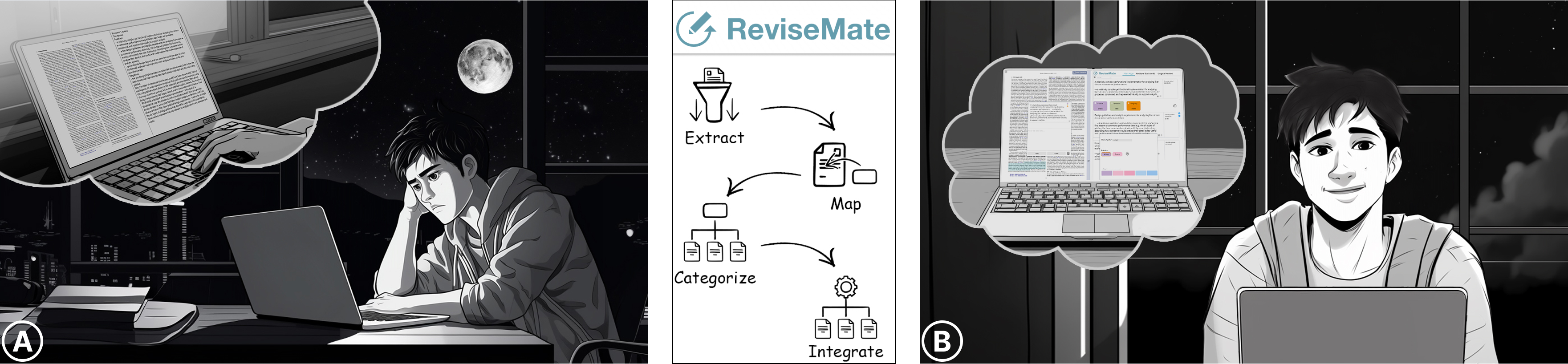}
 \caption{An example scenario illustrating how authors can effectively digest reviews of their papers using \textit{ReviseMate}: (A) The author encounters challenges when attempting to comprehend the feedback provided by reviewers for his paper. (B) In search of assistance, the author turns to \textit{ReviseMate}. By leveraging the system's capabilities in the analysis process, including comment extraction, aligning comments with relevant paragraphs in the original paper, comment categorization, and the integration of analyzed information, the author successfully engages in a contextual and tailored review text digestion process. \yuansong{This figure was created with the collaboration of an AI-powered image generation tool \textit{Midjourney}.}}
 \label{fig:title}
\end{teaserfigure}

\maketitle

\section{Introduction}

\par \yuansong{In the Science, Technology, Engineering, and Mathematics (STEM) field,} publishing academic papers is crucial for driving technological innovation and knowledge dissemination~\cite{picarra2015open}. Researchers receive and analyze reviewer feedback, making necessary revisions to their manuscripts. Effectively understanding and addressing reviewer comments enhances authors' efficiency, improves communication with reviewers, reduces misunderstandings, and leads to high-quality revisions. This process ultimately contributes to the publication of high-quality papers, promoting knowledge sharing within the academic communities.


\par Despite the crucial role of assimilating and integrating reviewer feedback in the review process, authors encounter distinct challenges in their conventional review digestion processes. These include both perceived difficulties and challenges of which they may be unaware. \textbf{First, revising a paper based on reviewer comments is inherently a time-consuming task with an unpredictable duration. During this period, \yuansong{the organization and analysis in the STEM field are generally expected to be thematically organized and follow a structured format~\cite{williams2004reply,hidouri2024key},} while the lack of systematic organization in traditional information processing can lead to cognitive limitations and biases ($\boldsymbol{\Gamma1}$).} Constrained by the limited capacity of human working memory~\cite{Ross2010ThePO}, the review analysis process is prolonged and continuous. Authors may encounter interruptions and pauses during the extended revision period, and their memory retention is limited by the forgetting curve~\cite{Ebbinghaus18852013MemoryAC,Tversky1974JudgmentUU}. These factors necessitate repeatedly reading and analyzing feedback. Additionally, as time elapses, authors' perspectives on revisions prompted by the same reviewer comments may evolve. Without efficient and systematic organization and presentation, these evolving ideas are difficult to document in a timely manner, potentially leading to the oversight of details in the reviewer comments. \textbf{Second, during the continuous process of digesting and extracting key information from review comments, authors are inevitably influenced by psychological factors, such as the emergence of phenomena like reading fatigue ($\boldsymbol{\Gamma2}$).} Prolonged exposure to information can lead to decreased motivation, particularly when the feedback is predominantly negative~\cite{Baumeister2001BadIS}. Consequently, authors may exhibit selective attention~\cite{Lane1982TheDO}, focusing more on familiar or less challenging comments while giving less attention to others. Additionally, there is a risk of confirmation bias~\cite{Nickerson1998ConfirmationBA}, where authors favor feedback that aligns with their existing beliefs and ignores contradictory information. The reading fatigue caused by these factors could lead to inaccurate estimations of the revision workload and consequently affect the final revision plan. \yuansong{Although this is essential, providing assistance to alleviate these factors and improve efficiency during the process of digesting and extracting key information would be more beneficial.} \textbf{Third, the analysis of review comments, akin to academic reading, requires strong comprehension and analytical skills for a comprehensive and effective evaluation. Insufficient experience and support may impact the thoroughness and efficiency of authors' analysis ($\boldsymbol{\Gamma3}$).} Engaging with review comments involves understanding complex information, identifying key points, overcoming personal biases, and responding effectively. This demands the development and application of skills like information synthesis and critical thinking, typically requiring long-term experience and training~\cite{Yuan2023CriTrainerAA}. Moreover, accessible and targeted support tools for the review comment analysis process are currently limited. \yuansong{Therefore, there is a gap in assisting individuals without experience or long-term development of skills in comprehensive review digestion.}

\par Several guidelines and training modules have been proposed by researchers to aid authors in comprehending and responding to reviewers' feedback. For instance, Day and Gastel~\cite{Day1980HowTW} provided guidelines on handling criticism and responding to reviewers, while Noble~\cite{Noble2017TenSR} proposed rules for effectively engaging with reviewer comments. However, existing solutions for analyzing review texts primarily offer general principles and broad instructions, \textbf{but they do not adequately address authors' practical needs and concerns} (research gap \textbf{RG1}). Furthermore, the nature of review texts necessitates prolonged review periods and sustained attention in single sessions, requiring not only a thorough understanding of the context but also personalized and consistent support throughout the continuous digestion process. While general tools assist with text comprehension and facilitate knowledge acquisition~\cite{Kim2016SimpleScienceLS, Head2020AugmentingSP, Chang2023CiteSeeAC, Tayeh2014OpenCL, park2022exploring}, they \textbf{often overlook the specific characteristics of review texts, failing to accommodate the nuanced contextual understanding required for comprehensive retrospective analysis} (\textbf{RG2}).
Moreover, there is \textbf{a lack of comprehensive insight into effectively integrating the identified support using appropriate interaction and representation techniques}. This is especially crucial when incorporating both AI methods and interactive features, considering users' acceptance and collaboration with these integrated features (\textbf{RG3}). Overall, there is a gap in research concerning the \underline{design}, \underline{efficacy}, and \underline{user experience} associated with the assimilation and integration of reviewer feedback during the process of review comprehension, analysis, and preparation for manuscript revision.

\par To comprehensively address the research gap and understand authors' needs and barriers during the review analysis process, we formulated the following research questions (\textbf{RQ1-RQ4}). To recognize authors' concerns during the review digestion phase, we proposed and investigated two research questions: \textbf{RQ1}: ``\textit{What are the factors that authors are most concerned about when digesting reviews?}'' and \textbf{RQ2}: ``\textit{What methods are acceptable for authors to enhance their comprehension while digesting the reviews?}'' To answer \textbf{RQ1}, we conducted a preliminary interview with 12 researchers from a local university to understand the difficulties during review comments digestion. We identified their demand for a targeted and practical system to assist in the digestion of \yuansong{STEM} paper reviews (\textbf{RG1}). Following the procedure of ``initial information construction, then integrated analysis''~\cite{Kintsch1991TheCM}, we split the review text digestion process into the \textbf{Preprocessing} and \textbf{Analysis} phases, corresponding to the information preprocessing of raw review text and in-depth analysis preparing for further revision and responses. To answer \textbf{RQ2}, we developed 10 storyboards based on interview findings to explore the design space for an interactive tool supporting review digestion. (\textbf{RG1}). Finally, We introduced \textit{ReviseMate}, an interactive system designed to aid users in review digestion process (\textbf{RG2}). Based on the literature on human-AI collaboration and decision-making~\cite{wang2019human, gu2021lessons,cai2019hello,mackeprang2019discovering}, we integrate human-AI collaborative features into \textit{ReviseMate}. This approach aligns with the insight that ``AI can organize, filter, and analyze raw data to enable humans to examine it more efficiently and gain deeper insights''~\cite{chen2023marvista}, consistent with previous studies on human-AI collaboration~\cite{chen2023marvista,tschandl2020human,wang2019human}. Specifically, \textit{ReviseMate} facilitates various steps including assisted comment extraction ($\boldsymbol{\Gamma2}$), user-engaged customizing categorization ($\boldsymbol{\Gamma3}$), establishing connections between comments and the original text ($\boldsymbol{\Gamma2-\Gamma3}$), recording timely insights upon analysis ($\boldsymbol{\Gamma1}$), and structuring the analyzed information for revision planning ($\boldsymbol{\Gamma2}$).


\par To assess our prototype, we further formulated and explored the following research questions: \textbf{RQ3}: ``\textit{How are the usability and effectiveness of the system in review text analysis?}'' and \textbf{RQ4}:``\textit{How do authors interact and accept the system during their review analysis?}'' Through a controlled user study and field deployment study, we thoroughly analyzed the results to answer the research questions (\textbf{RG3}). In a controlled user study (N=31), participants analyzed paper reviews and made a revision outline under two conditions: using \textit{ReviseMate} and using \yuansong{a baseline system presenting raw review text and original paper with limited interactions}. Through the analysis of the results, we found that \textit{ReviseMate} was significantly more effective than the baseline from various perspectives. We also gathered positive insights on users' interaction and acceptance of the system. The results of the field deployment study provided qualitative insights into how \textit{ReviseMate} supports users in real-world review digestion of their submitted papers. We discussed the analysis space for \textit{ReviseMate}, its generalizability, and the usability. In summary, our contributions can be outlined as follows:
\begin{itemize}
    \item We identified paper authors' challenges in comprehending reviewer feedback through a formative study and establishing design goals.
    \item We developed an interactive system to help authors understand reviews, focusing on time commitment and feedback comprehension challenges.
    \item A controlled user study (N=31) compared our system with traditional review text, confirming its significant usefulness.
    \item A field deployment study (N=6) focused on user experience completeness, gathering feedback through interviews, and providing valuable insights into \textit{ReviseMate}'s usability, effectiveness, interaction, and acceptance.
\end{itemize}

\section{Related Work}
\par \yuansong{The related work section is organized into three subsections: \autoref{sec:1} introduces the background and relevant theories of academic writing and review, \autoref{sec:2} presents tools that facilitate text comprehension, and \autoref{sec:3} summarizes relevant technical approaches involved in the review comment analysis process.}

\subsection{Context and Comprehension on Peer Review}
\label{sec:1}
\par \yuansong{Peer review is the process where experts evaluate academic papers to assess their quality for publication, with some variation across different domains. In STEM fields, reviewers focus on validating data reproducibility and methodological rigor against technical and theoretical standards. Some conferences also use rolling reviews. \revision{For instance, they adopt a monthly deadline approach to process submissions in batches and even utilize an open submission platform\footnote{https://aclrollingreview.org/} that enables continuous public iteration of papers.} In humanities and social sciences, the focus is on societal relevance and theoretical contributions, often requiring multiple revision rounds and extensive communication~\cite{nature_peer_review,british_academy_peer_review}. Peer review has been widely studied across fields, with research highlighting gender bias in STEM, social sciences, and humanities~\cite{squazzoni2021peer,Biolkov2023InvestigationOP}. In STEM, studies also examine factors specific to computer science, such as fairness in randomly assigned reviewers~\cite{stelmakh2019peerreview4all} and bias from reviewers knowing authors' resubmission history~\cite{Stelmakh2020PriorAP}. Recent studies explored the use of AI tools like \textit{ChatGPT} in peer review and the concerns associated with their application~\cite{zou2024chatgpt}. Research also examines authors' experiences, including their viewpoints~\cite{Jansen2016WhatDA}, satisfaction~\cite{Weber2002AuthorPO}, and challenges with review comprehension~\cite{RodrguezBravo2017PeerRT}.}

\par \revision{To facilitate the comprehension of reviews, several theories and models grounded in general text comprehension have been proposed as the foundation for this process.} For example, Bartlett et al. introduced the \textit{Schema Theory}, which posits that comprehension ensues through the activation and application of pre-existing knowledge structures~\cite{Wolters1933RememberingAS,Anderson1988ASV}. Stanovich emphasized the interactive nature of reading, suggesting strategies such as questioning and summarizing to compensate for comprehension difficulties~\cite{stanovich1980toward}. Kintsch et al.~\cite{Kintsch1991TheCM} framed text comprehension around construction and integration stages, highlighting the continual updating of mental representations by authors throughout this process. \revision{However, the gaps between general text comprehension and review analysis make them not directly applicable to review digestion.} \yuansong{Review digestion goes beyond traditional academic reading by requiring contextual analysis to link comments with the manuscript and create actionable revision plans~\cite{happell2011responding}.} \revision{Research also proposed theoretical guidelines to facilitate review analysis.} Guidelines from Day and Gastel provide insights on deconstructing criticisms and formulating responses to reviewers~\cite{Day1980HowTW}, while Nobel outlines rules for authors in responding effectively~\cite{Noble2017TenSR}. Although these sources offer valuable general insights, they primarily focus on overarching principles and may not always meet the nuanced demands of authors during the analysis process.


\par In our study, we focus on identifying the specific requirements and challenges authors face when comprehending and analyzing reviews. \revision{We adopt Kintsch's Construction-Integration (CI) model as our framework for interpreting research questions, considering the specific characteristics of the review text. This model emphasizes the construction and integration of information, aligning with review analysis where various reviewers' comments are extracted and integrated with the original content to formulate responses.} We propose solutions derived from a thorough analysis of actual review content, supported by relevant theoretical frameworks. Our research aims to contribute practical assistance to the existing body of work, benefiting authors in the review process.

\subsection{Tools for Augmenting Text Comprehension}
\label{sec:2}
\par Previous literature has proposed various methods and tools to enhance comprehension in different contexts, including academic content~\cite{Cachola2020TLDRES, Kim2016SimpleScienceLS, Head2020AugmentingSP, park2022exploring}, online reviews~\cite{Zhu2022BiasAwareDF, Yatani2011ReviewSA, Wang2014ReCloudSW}, and interconnected texts~\cite{Chang2023CiteSeeAC, Tayeh2014OpenCL, Tayeh2015ADE}. 

\par In academic materials, the focus has been on aiding readers in acquiring knowledge more efficiently through summarization and simplification methods. For example, Cachola et al. introduced ``Too Long; Didn't Read'' (TLDR) summaries for scientific content~\cite{Cachola2020TLDRES}, while SimpleScience aimed to make scientific terms more understandable for a wider audience~\cite{Kim2016SimpleScienceLS}. Interactive methods have also been explored to enhance comprehension, like augmented reading interfaces~\cite{Head2020AugmentingSP}, multi-touch interaction techniques~\cite{Tashman2011LiquidTextAF}, and animated graphics in academic papers~\cite{Grossman2015YourPI}. Solutions like automated systems for mathematical texts~\cite{Pagel2014MathematicalLP} and e-Proofs with audio and visual aids~\cite{Alcock2009eProofsSE} have been introduced. Furthermore, human-AI collaborative approaches are also investigated in tools designed to support reading articles~\cite{chen2023marvista}. \yuansong{In addition to reading academic content, review comment analysis requires organizing and evaluating specific feedback.} Research on online reviews focuses on achieving a holistic understanding and delving into specific details of comments. For instance, Zhu et al. introduced a visual design to address self-selection bias in online product reviews~\cite{Zhu2022BiasAwareDF}. Review Spotlight provides an overarching view of online reviews~\cite{Yatani2011ReviewSA}, while the ReCloud platform leverages word cloud visualizations from online reviews~\cite{Wang2014ReCloudSW}. \yuansong{Despite assisting users in analyzing comments and making informed decisions, these studies are distinct from the scenarios we focus on, as review comment analysis also requires contextual understanding and preparation for revisions.} Interconnected textual materials, especially in academic exploration using citations and hyperlinks, have also been explored. Chang et al. presented a paper reading tool with visual enhancements near citations~\cite{Chang2023CiteSeeAC}, and Tayeh et al. introduced a dynamic link service for creating bi- and multi-directional hyperlinks across documents~\cite{Tayeh2014OpenCL, Tayeh2015ADE}. 

\par \yuansong{While these works have shown significant advantages of AI in text comprehension, its application in review comment analysis remains underexplored. Considering how AI can assist users in the review analysis process, such as organizing and summarizing comments or aligning them with the original text, provides opportunities to improve both the effectiveness and efficiency of the task. To address this, our study examines AI's role in supporting the review process within a human-AI collaboration framework, introducing an interactive system to aid authors in review digestion.}

\subsection{Text Segmentation, Sentence Textual Similarity, and Sentiment Analysis Techniques} 
\label{sec:3}
\par \yuansong{To help authors comprehend reviews tailored to their needs, \textit{ReviseMate} uses techniques to enhance review understanding. The first step is extracting comments from unstructured review text, where NLP methods like text segmentation are crucial. For example, Koshorek et al.\cite{Koshorek2018TextSA} treated text segmentation as a supervised learning task, while Li et al.\cite{Li2022NeuralTS} proposed a neural model for segmentation at various granularities, such as the sentence level.} These approaches provide practical solutions for extracting comments from raw text.

\par Recognizing that comments from different reviewers may address similar aspects \yuansong{with varying viewpoints}, it's crucial to identify and relate these comments. Recent research employs sentence embeddings to capture the semantic core and measure semantic similarities between sentences~\cite{Cer2017SemEval2017T1}. Knowledge-based and corpus-based approaches exist for this task~\cite{Chandrasekaran2020ComparativeAO}, \yuansong{but corpus-based methods struggle with word meaning distinctions~\cite{Chandrasekaran2020EvolutionOS}. In this study, we used the \textit{GPT-3.5-turbo-16k} model, which excels in capturing semantic nuances for similarity assessment.}

\par Determining sentiment in peer review text is also important. Sentiment analysis methods can be divided into three categories: lexicon-based, traditional machine learning, and deep learning. \yuansong{Lexicon-based methods estimate sentiment using predefined phrases~\cite{Turney2002ThumbsUO} but struggle with domain variability~\cite{Pang2008OpinionMA}. Traditional machine learning applies statistical techniques~\cite{Sharma2012ACS} but faces challenges with complex datasets.}

\par Based on prior research, we utilized LLMs, including \textit{GPT-3.5-turbo-16k} and \textit{bart-large-mnli}, to support review text analysis. \yuansong{We leverage these models to facilitate information extraction and organization, integrating user interactions to support informed decisions and improve analysis quality and efficiency.}

\section{Formative Study}
\par To address the significant challenges authors face in comprehending reviews, our formative study seeks to answer the following research inquiries: \textbf{RQ1}: \textit{What are the factors that authors are most concerned about when digesting reviews?} \textbf{RQ2}: \textit{What methods are acceptable for authors to enhance their comprehension while digesting the reviews?}

\subsection{Preliminary Interview}
\subsubsection{Methods}
\par To address \textbf{RQ1}, we conducted 20-minute semi-structured interviews to explore the challenges and concerns authors face while digesting review feedback. We interviewed 12 participants (7 male, 5 female) from a local university, selecting them through a snowball sampling method. The participants had varied research backgrounds and experiences, with all participants having undergone the complete submission process as first authors at least once. The detailed information of the participants is shown in \autoref{tab:participants_info} in the Appendix. The interviews began by gathering demographic details from the participants and inquiring about their personal experiences with the submission process. Participants expressed their challenges and pain points while digesting reviews. Subsequently, the interview sessions delved into the participants' usual approaches to reviewing and analyzing feedback. This exploration encompassed methods employed for review analysis, as well as preferences about the preparation of subsequent revisions. Concurrently, the participants detailed specific challenges they encountered throughout this process, furnishing intricate details. As the interviews concluded, participants were prompted to share their perspectives and expectations regarding the potential implementation of an interactive system aimed at facilitating their review analysis endeavors.

\subsubsection{Results}
\par Following the receipt of Institutional Review Board (IRB) approval, the interviews were meticulously recorded and analyzed by two members of our research group. 
The acquired findings were systematically structured in alignment with the aforementioned \textbf{Preprocessing Phase} and the \textbf{Analysis Phase}. 

\par \mybox{Preprocessing Phase} \textbf{It could be challenging when extracting comments from raw review text.} Certain participants encountered difficulties during this endeavor. Particularly, P5 (male, age: 23) articulated that ``\textit{The comments in the review text cover a variety of things like viewpoints, explanations, and suggestions. This mix of information makes it take a while to pull out what's important.}'' P9 (male, age: 24) pointed out the characteristic of the review analysis process and its inherent challenges: ``\textit{The whole review analysis and revision thing is like a never-ending ride. You're constantly going back and forth, digging for information. But it's tough to find the right bits with unstructured review text and no standard [review] template.}'' These statements highlight the need for a more streamlined method for comment extraction.

\par \mybox{Analysis Phase} \textbf{It could be laborious and time-intensive when mapping reviewers' comments to the corresponding paragraphs of the paper.} Most participants emphasized the importance of matching comments with their relevant paper paragraphs, \yuansong{including both the novice and experienced authors. As one of them,} P1 (female, age: 40) mentioned that ``\textit{The first thing I do when analyzing reviews is matching comments to the right sections of the paper they're talking about.}'' Nonetheless, this process often involves considerable time and effort, necessitating frequent navigation between different sections. P10 (male, age: 28), \yuansong{an author with some paper revision experience}, expressed: ``\textit{It's quite a task to match each review comment with the original text, especially when there are a lot of comments, and they're complex. Spending too long on analysis in one go can tire me out, which makes my analysis less effective and efficient. And that means I end up spending even more time trying to understand things and planning out revisions.}'' Evidently, participants presented the requirement for a tool that streamlines this process due to the challenges and time required for manual alignment.

\par \mybox{Analysis Phase} \textbf{It could be complicated and challenging when analyzing useful and accurate insights from comments.} Understanding review comments is essential for effective revisions and responses. To gain a thorough insight into the feedback, many participants use strategies such as categorization or quantification. P10 (male, age: 28) elucidated, ``\textit{When I'm going through review comments, I dive deep into each one, analyzing it carefully and considering its context. Personally, I like to categorize the comments based on their content, separating writing problems from system issues, and then I prioritize them based on how urgent they are. This helps me plan out my revisions in a more organized way.}'' However, participants also emphasized the difficulties encountered during the execution of these strategies. P2 (male, age: 28) articulated, ``\textit{Even though I understand the importance of doing a thorough analysis of the comments, sometimes I feel like I'm lacking the right skills and support to do it effectively. For instance, I try to organize the comments into tables and sort them by different categories, but when I get into it, I realize that I'm not always sure about which category to put certain comments in or how to prioritize them. It makes me question the accuracy of my analysis.}'' These insights highlight the difficulties in comprehensive comment analysis. Recognizing users' diverse categorization criteria, we suggest customized features to enhance the analysis.


\par \mybox{Analysis Phase} \textbf{It could be cumbersome when \yuansong{organizing analyzed insights of review comments} for subsequent revisions.} Analyzing review text primarily aims to facilitate subsequent revisions. After the extraction and analysis of review comments, users still face challenges in integrating comments and insights from their analysis. This step is vital for understanding the overall picture and planning effective revisions. P4 (male, age: 25) expressed the complexity, saying,``\textit{Before, it was tough to keep track of and update my thoughts on review comments because there weren't good ways to organize and present them. Sometimes, I'd forget things or change my mind, which made it hard to stay on top of everything.}'' Moreover, P10 (\revision{male}, age: 28) emphasized the difficulties in systematic organization of these insights:``\textit{It's hard to merge my thoughts with the comments because they're scattered across [different] sections of the paper. This mix of comments and my own reflections makes it tricky to see the big picture and figure out how to revise effectively.}'' Acknowledging the necessity of structured insights for future revisions, we integrated this requirement into our design and implemented relevant features accordingly.


\subsection{Design Process}
\subsubsection{Design of Storyboard}
\par
\begin{figure*}

 \centering 
 \includegraphics[width=\columnwidth]{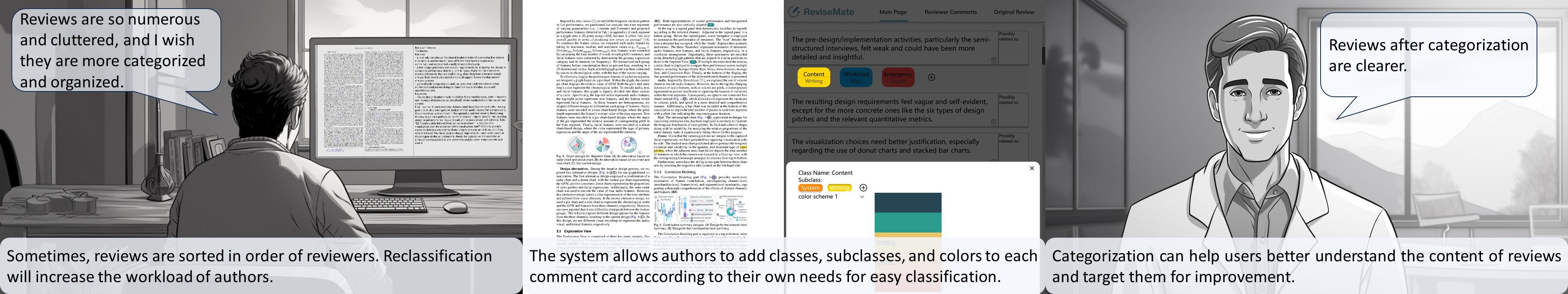}
 \caption{Storyboard2: Categorize the comments with customized criteria. \yuansong{This figure was created with the collaboration of an AI-powered image generation tool \textit{Midjourney}}.}
 \label{fig:storyboard}
\end{figure*}

 To answer \textbf{RQ2}, we designed scenarios using storyboards to explore the features to facilitate users' digestion of the reviews. \yuansong{Storyboarding is a common HCI technique for visually presenting system interfaces and usage contexts to gather feedback, as shown in prior research~\cite{luria2020social,chen2023meetscript}. By illustrating specific scenarios, it helps users understand problems and targeted experiences~\cite{truong2006storyboarding,peng2018requirements}. This approach is particularly useful in the early stages to convey system concepts with minimal production effort.} \yuansong{Based on challenges identified in preliminary interviews, we brainstormed scenarios where the system could address review digestion issues. Initially, 18 storyboards were created and refined into 10 by grouping similar ideas to represent common challenges.} These challenges included comment extraction, alignment with the original paper, categorization, and information integration. Each storyboard presents a user problem, our system's solution, and a hypothesized outcome (An example is mentioned in \autoref{fig:storyboard}). \yuansong{The features and interaction in the storyboards were inspired by previous work on document visualization and reading tools~\cite{Head2020AugmentingSP,lo2023semantic,Chang2023CiteSeeAC,Yuan2023CriTrainerAA}.}
A brief version is shown in \autoref{tab:storyboards} in the Appendix.

\subsubsection{Procedures}
\par We recruited 18 participants for our study (10 males, 8 females) through snowball sampling and university mailing lists. We ensured diverse feedback by maintaining a mix of both original and new participants. Specifically, 5 out of the participants (3 males, 2 females) were from the preliminary interview. The average age was 26.89 (SD = 5.5), and all had submitted a paper as the primary author at least once. The detailed information of participants is shown in \autoref{tab:participants_info_new} in the Appendix. Each participant was shown the storyboards individually in random order and asked to read the scenario and share their thoughts. During the interview, participants individually viewed the storyboards in a random sequence, taking around 3-5 minutes each to understand the presented problem and the system's solution. The average time spent by each participant was 45.5 minutes. After the interview, recordings were transcribed and analyzed using the affinity diagram method~\cite{Lucero2015UsingAD}.

\subsubsection{Findings}
\par The findings are derived from various storyboard scenarios such as comment extraction, comment analysis, and integration of the analyzed data. We structured these findings into two main phases: \textbf{Preprocessing} and \textbf{Analysis}.

\par \mybox{Preprocessing Phase} \textbf{Both automatic and user-driven comment extraction methods have the potential to improve overall user experience.}
Considering users' varied preferences during initial analysis, our storyboards offer multiple methods for comment extraction. Users can utilize the system's automatic extraction feature or opt to select and manually input comments from the review text. Most participants appreciated the diverse options, emphasizing their enhanced usability. For instance, P7 (male, age: 24) mentioned, ``\textit{Picking out text from the [original] reviews is a good way to get myself familiar with the content.}'' P15 (male, age:29) liked the flexibility, saying, ``\textit{At first, I'd probably add comments manually, but once I get the hang of the system, I might switch to automatic generation to save time.}''

\par \mybox{Analysis Phase} \textbf{Mapping each review comment to its corresponding content in the original paper can increase my efficiency.} The storyboards depict scenarios where \textit{ReviseMate} assists users in mapping comments with related paragraphs in the original paper. Instead of merely direct linking, our system suggests potentially relevant paragraphs to engage user interaction. Participants favored this interactive linking process. P11 (male, age:25) mentioned, ``\textit{I wouldn't just blindly trust the system and let it handle everything. Being involved in the process makes me double-check the connections for accuracy, using the system's [suggestions] as guidance.}'' Participants also expressed a potential improvement in their efficiency. P7 (male, age:24) added, ``\textit{The existing text analysis tools don't perfectly support linking comments with my paper, but having this feature really makes it easier for me to analyze and navigate through the content.}'' 

\par \mybox{Analysis Phase} \textbf{Customized categorization fits my various demands during different stages of comprehension.} From preliminary interview insights, we designed our system to support customized comment categorization. \yuansong{In the storyboard design, we explored using ratings, such as an \textit{Emergency} criterion, to indicate the urgency of comments. However, user feedback revealed that precise judgments through ratings were challenging, with classification proving to be a more intuitive and suitable approach. P1 (female, age:40) indicated, ``\textit{Categorizing comments by level allows us to quickly familiarize and classify them according to revision needs while leaving flexibility for subsequent analysis and mapping.}''} While categorizing comments using predefined criteria like content type or urgency ranking, users can introduce their own criteria. Participants valued this flexibility. For instance, P8 (female, age: 27) mentioned, ``\textit{The system's ability to handle different classification rules is really handy for me. It suits my needs perfectly and is way more practical than just theoretical advice.}'' Meanwhile, P17 (female, age: 40) emphasized that customizing criteria elevates the user experience, adding, ``\textit{Personally, I'd prefer a more detailed classification of comments to make things clearer. Having the option to add new criteria for categorization really helps me understand the comments better.}'' 

\par \mybox{Analysis Phase} \textbf{The integration of comments could facilitate me in planning the revision.} The storyboard showcases an integrated display of comments with the ``group by'' feature to streamline users' revision plans. Participant feedback was instrumental in refining this aspect. P13 (female, age: 35) highlighted the importance of such an overview, stating, ``\textit{As I go through the analysis, I naturally gather a lot of information. It's important for me to have an overview to keep track of everything.}'' We also explored tools like the \textit{Gantt} chart for planning. However, feedback indicated that users rarely set detailed timelines for each comment, given the difficulty in accurately estimating revision duration. Some participants also highlighted the need for assistance during the response stage. P3 (male, age: 28) said, ``\textit{I'm really looking forward to getting some help from the system, especially when it comes to drafting a cover letter during the response [phase].}'' This feedback guides us to refine and enhance the system's workflow.

\subsubsection{Design Goals}
\par Drawing from insights gathered during the initial interviews, we put forward three design goals (DG) that an interactive system should fulfill to effectively assist authors in digesting paper reviews:
\begin{itemize}
 \item \DG{DG1} The system should assist users in efficiently extracting comments from review text.
 \item \DG{DG2} The system should support users to organize reviews according to customized criteria and map them with relevant paragraphs in the paper, thereby facilitating the contextual and comprehensive analysis.
 \item \DG{DG3} The system should enable the integration of comments with analyzed insights, aiding users in organizing their thoughts for planning the revision and formulating responses.
\end{itemize}

\section{System}

\begin{figure}[h]
    \centering
    \includegraphics[width=\columnwidth]{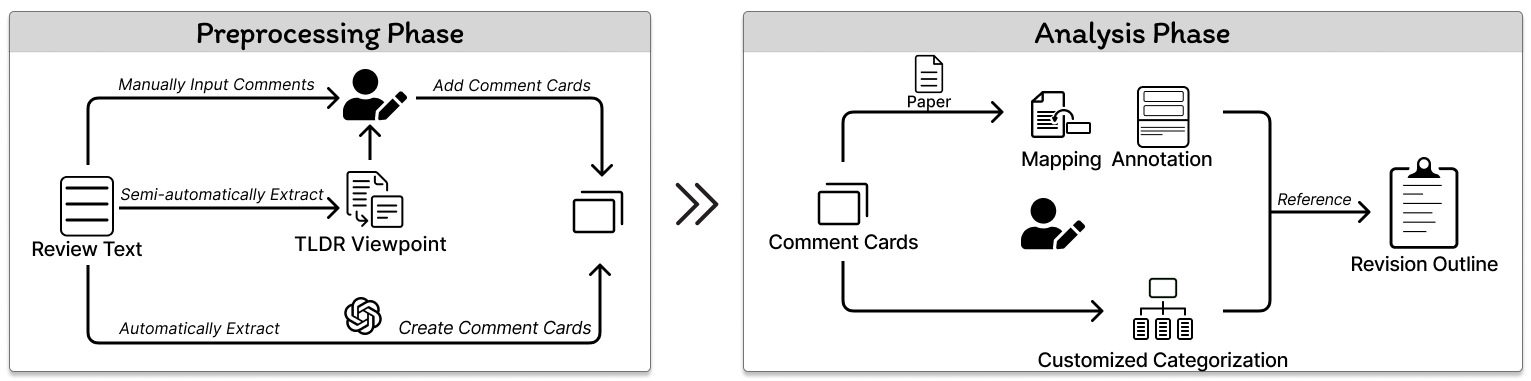}
    \caption{The review text analysis process facilitated by \textit{ReviseMate} comprises two main phases. Initially, in the preprocessing phase, users extract valuable information from raw review text. Subsequently, in the analysis phase, users map comments to relevant paragraphs of the paper, annotate their insights, categorize comments based on custom criteria, and organize the analyzed data into a revision outline to formulate revision plans.}
    \label{fig:workflow}
\end{figure}

\par To facilitate authors' academic reading of paper reviews, we proposed \textit{ReviseMate}, an interactive system with contextual and customized information analysis. The system workflow is divided into two main phases: the \textbf{Preprocessing} phase and the \textbf{Analysis} phase (\autoref{fig:workflow}), informed by formative study results. In the \textbf{Preprocessing} phase, \textit{ReviseMate} incorporates the \textit{GPT-3.5-turbo-16k} model to preprocess raw review text. This aids in extracting comments and generating comment cards for subsequent analysis. In the \textbf{Analysis} phase, \textit{ReviseMate} equips users with diverse features for a comprehensive understanding of individual and collective review comments. These include customized comment categorization, comment-to-paper mapping and annotation, and analyzed information organization. \autoref{fig:workflow} illustrates the holistic workflow using \textit{ReviseMate} for review text digestion. A user scenario guiding through the analysis process can be found in \autoref{Usage Scenario}.

\subsection{Preprocessing Phase}
\par In the \textbf{Preprocessing} phase, users begin by extracting comments from the raw review text with the support of the \textit{ReviseMate} system.

\begin{figure}[h]
    \centering
    \includegraphics[width=\columnwidth]{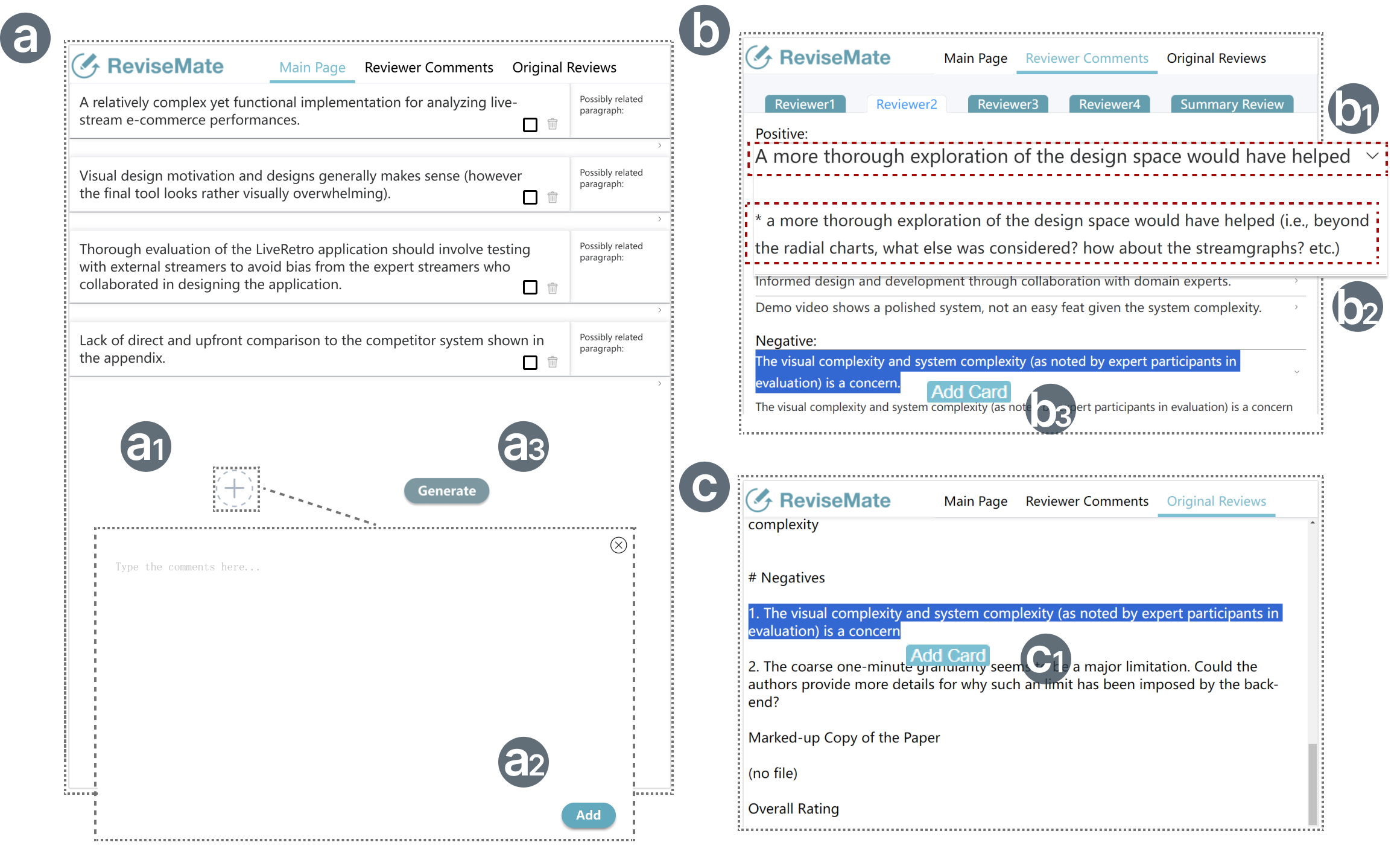}
    \caption{Interface of \textit{ReviseMate} in the \textbf{Preprocessing Phase}. Users can extract reviews through multiple options and form them into a comment card: (1) automatic generation \protect\darkcircled{a3} (2) manual input \protect\darkcircled{a2} (3) from system-organized reviews \protect\darkcircled{b3} or the original review text \protect\darkcircled{c1}.}
    \label{fig:Preprocessing}
\end{figure}

\subsubsection{Interface and Interaction}
\par The \textit{ReviseMate} interface (\autoref{fig:Categorization_Mapping}) is divided into two parts: the left section displays the original paper, while the right section is for the extraction and analysis of review text. Specifically, the right section (\autoref{fig:Preprocessing}) includes the \textit{Main} page, \textit{Reviewer Comments} page, and \textit{Original Review} page.

\par The process involves extracting individual comments from unstructured reviews provided by multiple reviewers. \textit{ReviseMate} constructs each comment as a comment card to support subsequent analysis. \yuansong{We offer different options for users during the extraction phase, considering varied preferences based on insights from the formative study.} Users have the option to extract and add comment cards through three methods: \textit{manual extraction}, \textit{semi-automatic extraction}, and \textit{automatic extraction.}

\begin{itemize}
    \item \yuansong{[\textit{manual extraction}]} The \textit{Main} page features has an \raisebox{-0.8ex}{\includegraphics[height=2.7ex]{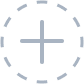}} button (\autoref{fig:Preprocessing}-\darkcircled{a1}), allowing manual comment card creation. A text input area (\autoref{fig:Preprocessing}-\darkcircled{a2}) appears on clicking \raisebox{-0.6ex}{\includegraphics[height=2.5ex]{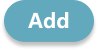}}, enabling direct addition of comment cards.
    \item \yuansong{[\textit{semi-automatic extraction}]} The \textit{Reviewer Comments} (\autoref{fig:Preprocessing}-\darkcircled{b}) and \textit{Original Review} (\autoref{fig:Preprocessing}-\darkcircled{c}) pages also facilitate comment card addition. On the \textit{Reviewer Comments} page, organized comments by reviewers and summary reviews are displayed (\autoref{fig:Preprocessing}-\darkcircled{b1}). Users can explore these and choose content for semi-automatic comment card creation (\autoref{fig:Preprocessing}-\darkcircled{b3}). They can also navigate to the \textit{Original Review} page and directly add comment cards based on raw review text (\autoref{fig:Preprocessing}-\darkcircled{c1}). 
    \item \yuansong{[\textit{automatic extraction}]} An automatic option exists: by clicking the \raisebox{-0.6ex}{\includegraphics[height=2.5ex]{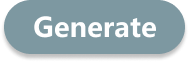}} button (\autoref{fig:Preprocessing}-\darkcircled{a3}) on the \textit{Main} page, the system uses the \textit{GPT-3.5-turbo-16k} model to analyze review text, extract and summarize distinct comments, and form comment cards automatically.
\end{itemize}

\par \yuansong{This design consideration provides users with various options to meet their analysis needs. Automatic extraction ensures comprehensive retrieval of review comments for user examination and editing. Semi-automatic extraction provides comments organized by reviewers, while manual extraction allows direct access to and interaction with the original comments.} Additionally, users can delete a comment card by clicking the \raisebox{-0.4ex}{\includegraphics[height=2.2ex]{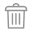}} button located at the bottom-right of the card when it is no longer needed.

\subsubsection{Implementation details}
\par \yuansong{For the extraction of reviews from their original review texts, we utilize OpenAI's \textit{GPT-3.5-turbo-16k} model~\cite{Radford2018ImprovingLU}, with a token capacity of 16,000 to generate comprehensive outputs without excessive truncation.} In our practice, we separate reviews contributed by distinct reviewers and subsequently subject them to the \textit{GPT-3.5-turbo-16k} model for review extraction. Subsequently, to establish alignment between the extracted comments and their corresponding source reviews, \textit{ReviseMate} employs the \textit{bert-base-uncased} model~\cite{Devlin2019BERTPO}. \yuansong{We chose the BERT model instead of traditional methods like TF-IDF for its advances in text data representation in NLP tasks~\cite{gonzalez2020comparing, subakti2022performance}.} This model is instrumental in encoding both the original texts and the extracted reviews into vector representations, \yuansong{enabling the computation of cosine similarities to match each original comment with its most similar extracted text}. The outcome of this process yields the transformation of the original, often elaborate, reviewer-contributed reviews into more succinct and concise language. This affords users the advantage of accessing both abridged summaries and the unabridged, original versions of the reviews for cross-reference and analysis.

\subsection{Analysis Phase}
\par In the \textbf{Analysis} phase, users can attain a profound understanding of the reviews, characterized by an individualized and comprehensive understanding of each review comment.

\subsubsection{Interface and Interaction}
\par Following the results of the formative study, the design of \textbf{Analysis} phase takes a proactive approach by involving users in categorizing comments and aligning them with the content of the original papers. This approach ensures that users can gain a customized and contextual understanding. Additionally, \textit{ReviseMate} supports users in structuring the analyzed information, assisting in their further revision planning.

\begin{figure*}[h]
    \centering
    \includegraphics[width=\textwidth]{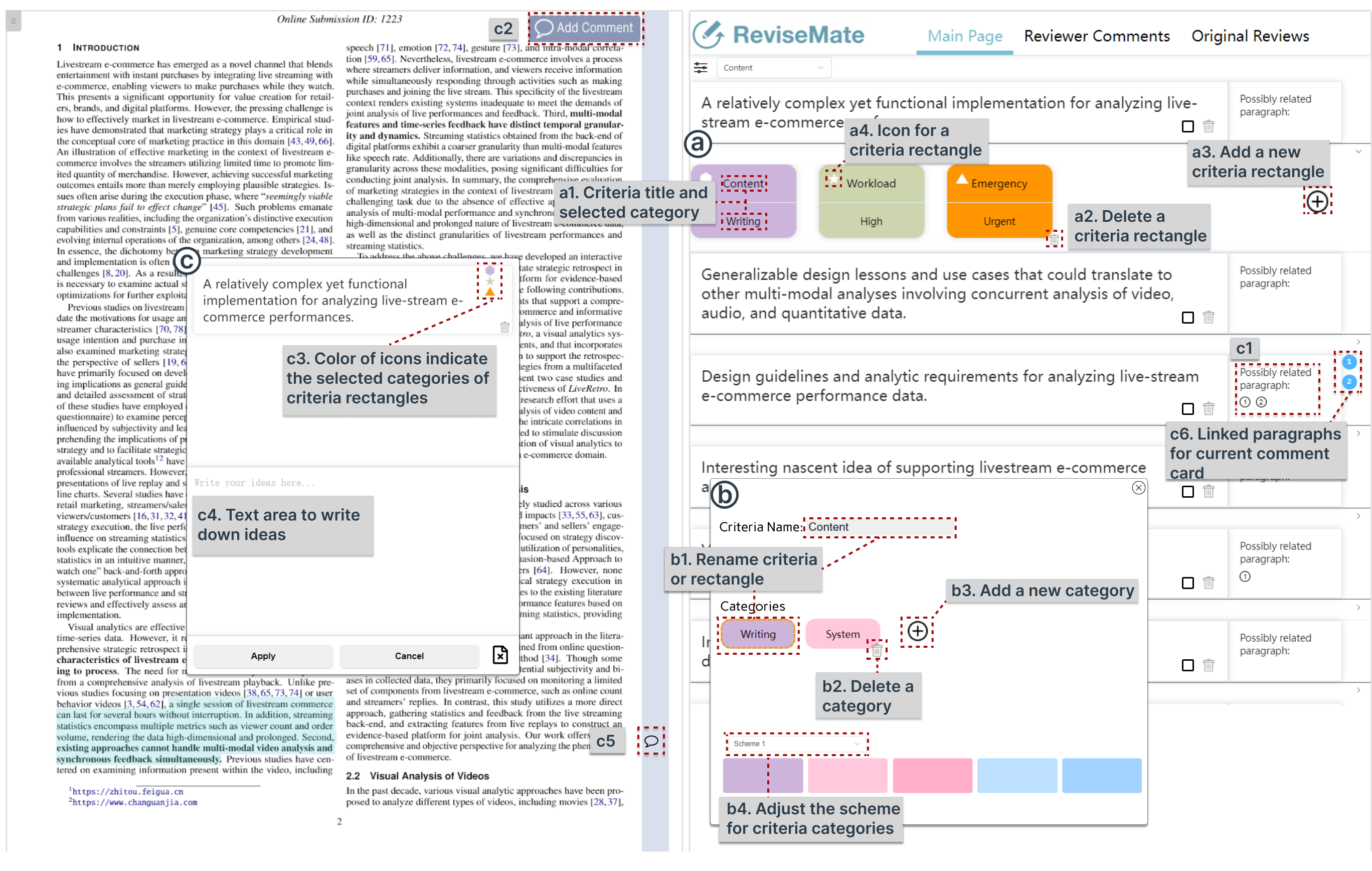}
    \caption{Interface and interactions in \textbf{Customized categorization} and \textbf{Mapping with original paper}. (a) Users can conduct customized categorization of comments using different criteria. (b) Users can edit the categories of a criterion. (c) Users can establish links between comments and relevant paragraphs in the paper, with the annotation of thoughts for revision.}
    \label{fig:Categorization_Mapping}
\end{figure*}

\par \textbf{Customized categorization.} Upon creating comment cards in the \textbf{Preprocessing} phase, users can modify attributes within collapsible sections beneath each card (\autoref{fig:Categorization_Mapping}-\transparentcircled{a}). Within these collapsible sections, there exist distinct rectangular representations denoting different criteria (\autoref{fig:Categorization_Mapping}-a1). \yuansong{The predefined set of criteria and categories are derived from the analysis of our storyboard interviews. For each comment card created through all three options in the \textit{Preprocessing} phase, it contains three predefined criteria as attributes with the heading: \textit{Content}, \textit{Workload}, and \textit{Emergency}.} Each criterion has selectable categories for classification. The selected categories are displayed at the bottom of these criteria rectangles. For example, under the \textit{Content} criterion, users can choose labels like \textit{writing} or \textit{system} to categorize comments.

\par Users possess the capability to further customize these criterion rectangles by clicking on them, which initiates the \textit{Editing Modal} (\autoref{fig:Categorization_Mapping}-\transparentcircled{b}). Within this modal, they can rename criteria or categories (\autoref{fig:Categorization_Mapping}-b1), remove specific criteria using the dedicated \raisebox{-0.4ex}{\includegraphics[height=2.2ex]{figure/Delete.png}} button (\autoref{fig:Categorization_Mapping}-b2), introduce new categories using the \raisebox{-0.4ex}{\includegraphics[height=2.2ex]{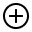}} button (\autoref{fig:Categorization_Mapping}-b3), and adjust color schemes (\autoref{fig:Categorization_Mapping}-b4). When a color scheme is selected for a criterion, the card color will match the chosen category, with the first category serving as the default.

\par Additionally, users can also tailor criteria by utilizing the \raisebox{-0.4ex}{\includegraphics[height=2.2ex]{figure/Delete.png}} (\autoref{fig:Categorization_Mapping}-a2) or \raisebox{-0.4ex}{\includegraphics[height=2.2ex]{figure/Add_criteria.png}} (\autoref{fig:Categorization_Mapping}-a3) button situated on the right side of the collapsible section. By customizing criteria and their associated categories, users can establish adaptable categorizations for comments, thereby gaining insights into each comment from diverse perspectives. To alleviate the potential visual overload, we have incorporated distinct icons (\autoref{fig:Categorization_Mapping}-a4) to symbolize each criterion. These icons are intended to represent the selected categories of each criterion rectangle in the subsequent process of mapping with the content of the original paper.


\par \textbf{Mapping with the original paper content.} \yuansong{For each comment card created through all three options in the \textit{Preprocessing} phase}, we provide a display of mapping cards on the right side of comment cards (\autoref{fig:Categorization_Mapping}-c1), containing potentially relevant paragraphs. These paragraphs are indicated by numbered circular buttons. Clicking any of these buttons results in a scroll to the corresponding paragraph in the original paper. Users can thoroughly examine the paragraph's content and determine its relevance to the current comment. When users assess the underlying connections between a paragraph and the given comment, they can select the relevant text within the paper. Subsequently, a \raisebox{-0.4ex}{\includegraphics[height=2.2ex]{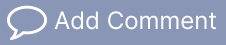}} button (\autoref{fig:Categorization_Mapping}-c2) becomes visible at the upper right corner of the paper view. By clicking this button, users can link the chosen comment card to the selected text through a drag-and-drop action, which takes place in the \textit{Editing Modal} (\autoref{fig:Categorization_Mapping}-\transparentcircled{c}). Icons (\autoref{fig:Categorization_Mapping}-c3) located on the right side of each comment card provide a simplified representation of the criteria rectangles. The color filled within these icons indicates the current selected category for each criterion. It's important to note that our approach recognizes that the automated model-based links between paragraphs and comments may not always be flawless. To address this, we present these linking results as suggestions for users to review, rather than as definitive links. To enhance accuracy and alignment, we introduce a dragging mechanism for comment cards into the \textit{Editing Modal}. This step requires users to manually confirm the links they create, which empowers them to correct any potential misalignments. Once the relationship between comments and corresponding paragraphs is established, users have the option to jot down their initial thoughts and revision ideas in the text area below (\autoref{fig:Categorization_Mapping}-c4). The dragged comment cards serve as a reference, aiding users in guiding and contextualizing their thought processes. Users can click the apply button within the \textit{Editing Modal} saves the linked comment cards and annotations in the right side \textit{Annotation Bar} with a \raisebox{-0.4ex}{\includegraphics[height=2.2ex]{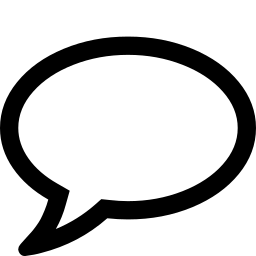}} icon (\autoref{fig:Categorization_Mapping}-c5), with the previously selected text highlighted. Additionally, linked paragraphs are presented as bubbles next to the current comment card on the right (\autoref{fig:Categorization_Mapping}-c6). Users can easily click this icon to view or edit the annotation content associated with the highlighted text in the \textit{Editing Modal}.

\begin{wrapfigure}{r}{0.7\textwidth}
  \centering
  \includegraphics[width=0.7\textwidth]
  {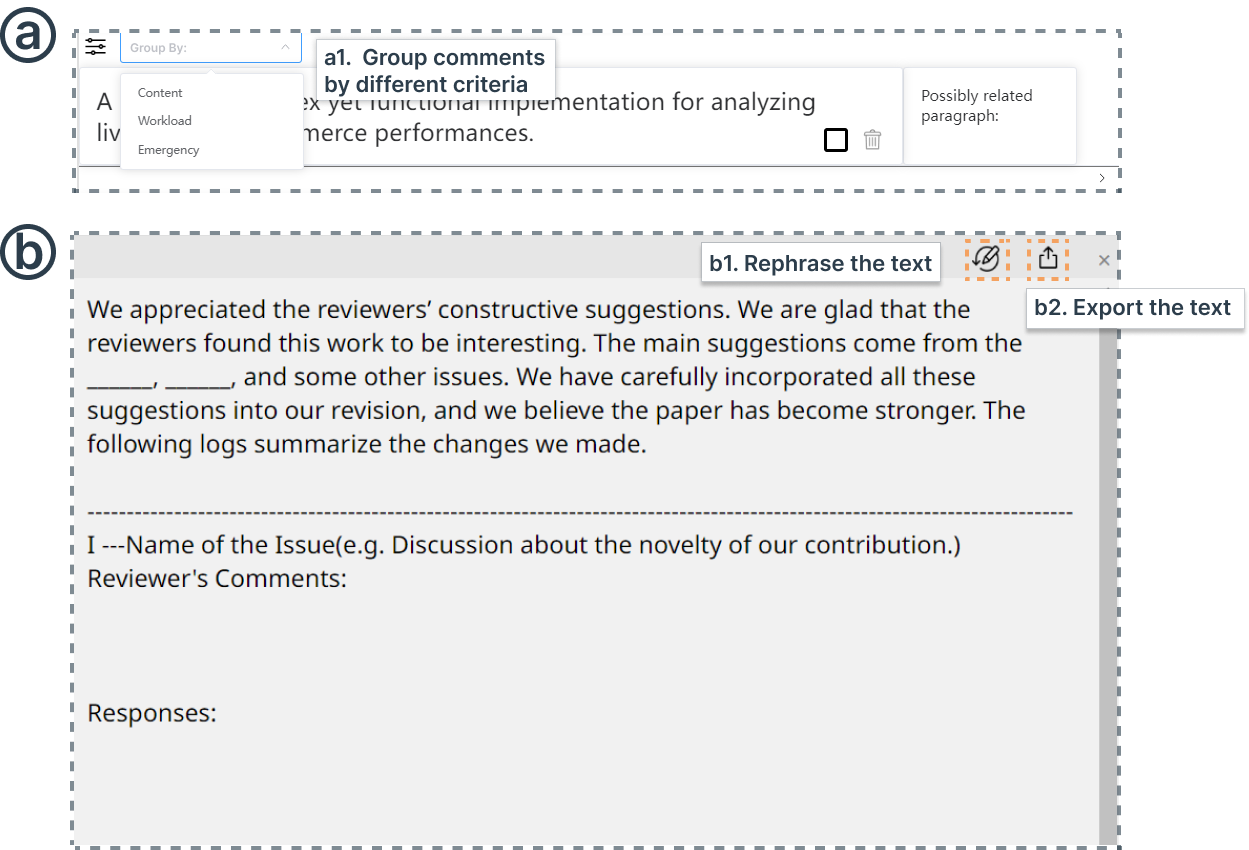}
  \caption{Interface and interactions in \textbf{Integrated organization}. (a) Users can group comments based on different criteria. (b) Users can integrate the analyzed information into the revision outline for making revision plans.}
  \label{fig:Organization}
\end{wrapfigure}

\par \textbf{Integrated organization.} The preceding analysis focuses on the individual attributes of comments and their corresponding paragraphs. While this approach gathers analysis information, it lacks cohesion and results in a fragmented understanding. Recognizing this gap, we address the need for a consolidated presentation of the analyzed data, affording users an overarching understanding of reviews and enabling them to strategize informed revisions. For comment cards grouped by category, users can click the \raisebox{-0.4ex}{\includegraphics[height=2.2ex]{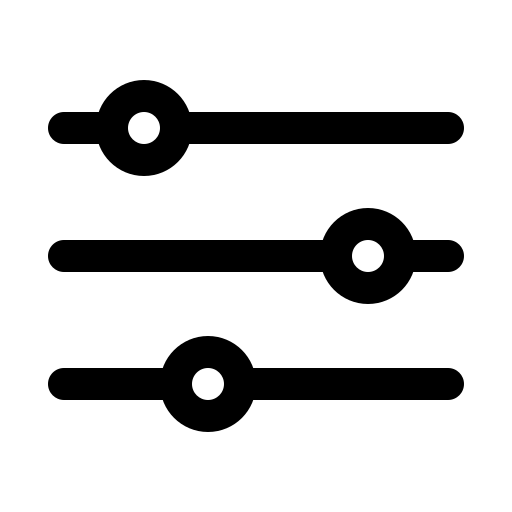}} button (\autoref{fig:Organization}-a1) to arrange them based on designated criteria. For instance, selecting \textit{Workload} as the criterion leads to the categorization of the comment cards under that specific aspect. In the case of predefined \textit{Workload} categories, comment cards marked as \textit{High} are clustered together, while those labeled \textit{Low} form a separate grouping. Insights gathered from interviews underscore the necessity of establishing links between comprehending reviews and the subsequent revision process. To address this, we introduce a \textit{Revision Editing} sidebar (\autoref{fig:Organization}-\transparentcircled{b}). Here, users can incorporate comments alongside their revision concepts into an initial template. This process involves referencing grouped comment cards and the original paper annotated with relevant notes. The \textit{Revision Editing} page supports the integration of comments and revisions, arranging them according to distinct aspects. Each aspect encompasses a collection of interconnected comments. This organized compilation facilitates planning revisions and formulating responses. Upon completing the process, users have the option to rephrase their text using the \raisebox{-0.4ex}{\includegraphics[height=2.2ex]{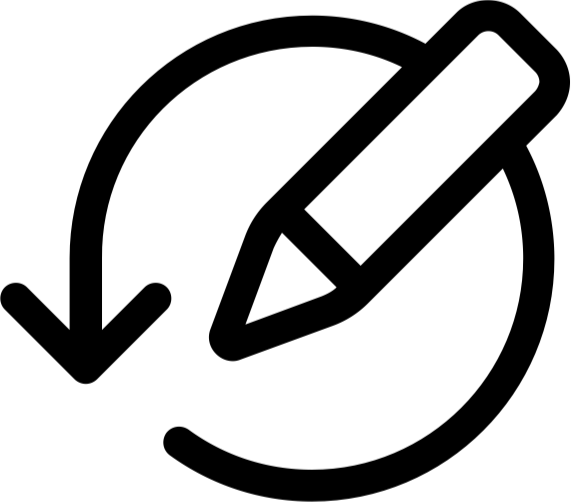}} button (\autoref{fig:Organization}-b1) or export the content utilizing the \raisebox{-0.4ex}{\includegraphics[height=2.2ex]{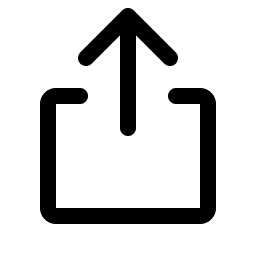}} button (\autoref{fig:Organization}-b2). Notably, we do not use automated model-based generation of revision plans in the \textit{Revision Editing} process, as the system focuses on enhancing the comprehension of scientific paper reviews. Users are encouraged to devise their own revision plans, reflecting their personal thoughts and ideas.

\subsubsection{Implementation details}
\par The central aspect of the \textbf{Analysis} phase revolves around mapping with the content of the original paper. To achieve this, we utilize the \textit{bart-large-mnli} model~\cite{Lewis2019BARTDS}, which has undergone fine-tuning for the \textit{MultiNLI} task\yuansong{~\cite{N18-1101}}. This task is designed to determine the relationship between two sentences, categorizing them as \textit{contradictory}, \textit{entailing}, or \textit{neutral}. In our approach, we segment the original paper into distinct paragraphs. Subsequently, for each extracted review \revision{comment, we employ the \textit{bart-large-mnli} model to rank these relationships based on their confidence scores and select the top 10 matching paragraphs presented as the potentially related paragraphs for each review comment.} This presentation serves to facilitate additional analysis by users.

\section{Controlled User Study}
\par
\begin{figure*}[h]
    \centering
    \includegraphics[width=\textwidth]{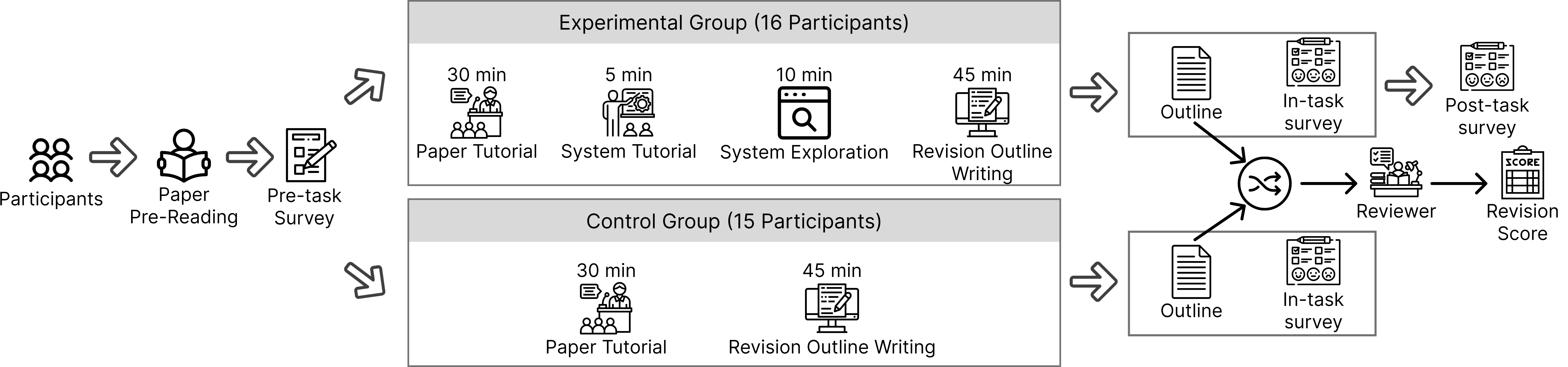}
    \caption{Participants were randomly divided into two groups: one used \textit{ReviseMate} for review text analysis, while the other group worked with raw review text. All participants underwent pre-training for the paper, which included a tutorial to familiarize them with the paper's content. The group using \textit{ReviseMate} received a tutorial on how to use the system. Throughout the study, participants completed pre-task surveys for demographic and background information, in-task surveys to assess the usability and effectiveness of the system, and post-task surveys to evaluate their acceptance of and collaboration with the system.}
    \label{fig:UserStudy}
\end{figure*}

 To address \textbf{RQ3-RQ4}, we conducted a controlled between-subject user study. Participants were divided into two groups: the \textit{ReviseMate} condition as the experimental group and the baseline condition as the control group. Based on findings from the formative study, there are no widely adopted tools specifically designed for authors to analyze review text. Therefore, for the control group, we developed a baseline system with an interface similar to \textit{ReviseMate}, \yuansong{which displayed the original manuscript and the review text side by side but did not include any additional interactive features.} \yuansong{The interface of the baseline system is shown in \autoref{fig:baseline} in Appendix.} This study was conducted with institutional IRB approval. During the study, participants were tasked with analyzing the review text alongside the provided paper and creating a revision outline that presented the review comments and their revision thoughts. \yuansong{We decided to conduct the between-subjects study design to prevent cross-condition effects, such as prior exposure to the paper and reviews, and to avoid learning or fatigue from multiple experimental conditions~\cite{kirk2013experimental}.} We collected both quantitative and qualitative data through pre-task, in-task, and post-task surveys, as well as evaluations of the submitted revision outlines. The study procedure is illustrated in \autoref{fig:UserStudy}.

\subsection{Participants and Procedure}
\par We recruited 31 participants (18 male, 13 female) through email and snowball sampling within the university community. Their average age was 25.9 (SD=5.6). These participants held various research roles, including assistant professors, Ph.D. students, graduate students, and undergraduate students, and represented diverse research domains such as Human-Computer Interaction (HCI), Machine Learning (ML), Biomedical Engineering (BME), Computer Vision (CV), and Software Engineering (SE). All participants had experience with the complete manuscript submission process, although specific experience as first authors was not required for this study. Detailed participant information is provided in \autoref{tab:participants1} in the Appendix. Participants were randomly assigned to two groups: 16 to the \textit{ReviseMate} condition and 15 to the raw review text condition.

\par To mitigate potential evaluation bias stemming from variations in papers and review text, we selected a specific paper from the Visualization domain, which discussed an interactive visualization system in a practical context, and its corresponding review feedback from a conference submission. This choice was based on the relatively low complexity of Visualization papers in terms of technical terminology and concepts compared to papers in other domains, making it easier for participants to comprehend. To further mitigate difficulties related to paper familiarity, participants received the \textit{PDF} version of the paper one week before the controlled user study and were instructed to thoroughly read it. Additionally, a 30-minute tutorial was conducted before the actual study to ensure participants understood the paper's structure and content. Participants completed a pre-task survey to provide demographic and background information. In the \textit{ReviseMate} condition, participants received a 5-minute system tutorial followed by a 10-minute interactive session to explore the system and ask questions. For both conditions, the review text analysis process was screen-recorded with participants' consent and lasted for 45 minutes. Participants were required to create a revision outline using a provided template (as shown in \autoref{fig:Organization}-\transparentcircled{b}), summarizing issues identified in the review comments and proposing revision thoughts. After the analysis, participants filled out an in-task questionnaire to rate their user experience during the review text analysis. Participants in the \textit{ReviseMate} condition also completed a post-task questionnaire and participated in semi-structured interviews to delve into their interaction and acceptance of the system. All participants received compensation at a rate of \$20 per hour.

\subsection{Measurement}
\subsubsection{RQ3: Usability and effectiveness}
\par We assessed usability and effectiveness using a 7-point Likert scale (1: Strongly disagree, 7: Strongly agree) through an in-task questionnaire. To evaluate \textbf{usability}, we examined factors such as ease of use, support for review text analysis, system-related distractions, user satisfaction, likelihood of recommending the system, future use intent, and awareness of system feature usefulness, with reference to SUS~\cite{Brooke1996SUSA}. To evaluate \textbf{effectiveness}, we considered criteria such as users' perceived review comment extraction accuracy, accessibility of detailed comment analysis, the efficiency of organizing analyzed information, perceived analysis confidence, quality of revision outline, time efficiency, and cost. We also collected users' revision outlines and obtained ratings from two experienced researchers with over 15 years of research experience in the VIS domain. \yuansong{They rated each outline blindly, without knowledge of the corresponding condition.} Based on the literature on information quality evaluation~\cite{miller1996multiple,cai2003evaluating,pierce2008evaluating} and review guidance on HCI/CSCW papers~\cite{hinckley2015so,chi_review_criteria,cscw_student_mentor_program,cscw_reviewing_for_cscw}, we developed evaluation criteria for revision outlines from three dimensions: completeness, accuracy, and organization. The specific evaluation criteria are shown in \autoref{tab:evaluation_criteria}. We did not use the time spent on review analysis as a metric to measure the system's effectiveness because when users are told that faster analysis is better, they tend to complete the task in a shorter time to achieve better performance. This approach compromises the quality of the analysis and contradicts the true intention of the review comment analysis process. We also conducted a field deployment study to obtain user feedback in real-world scenarios as a supplement.

\begin{table}[h]
\centering
\begin{tabular}{p{0.2\textwidth} p{0.7\textwidth}}
\toprule
\textbf{Criteria} & \textbf{Description} \\
\midrule
\textbf{Completeness} & \\
0-3 & The outline is missing many key points from the review comments, with only a few addressed. \\
4-6 & The outline addresses some of the key points, but several important comments are overlooked. \\
7-9 & Most of the key points are addressed, with only minor omissions. \\
10 & All key points from the review comments are comprehensively addressed. \\
\midrule
\textbf{Accuracy} & \\
0-3 & The outline contains significant inaccuracies or misinterpretations of the review comments. \\
4-6 & The outline has some inaccuracies, with several points misinterpreted or incorrectly addressed. \\
7-9 & The outline is mostly accurate, with only minor misinterpretations. \\
10 & The outline accurately reflects all review comments without any misinterpretations. \\
\midrule
\textbf{Organization} & \\
0-3 & The outline is poorly organized, with a lack of logical structure and coherence. \\
4-6 & The outline has some organizational issues, making it difficult to follow in parts. \\
7-9 & The outline is well-organized, with a clear structure and logical flow, though minor improvements could be made. \\
10 & The outline is exceptionally well-organized, with a clear, logical, and coherent structure. \\
\bottomrule
\end{tabular}
\caption{Evaluation Criteria for Revision Outline}
\label{tab:evaluation_criteria}
\end{table}

\subsubsection{RQ4: Interaction and Acceptance}
\par Screen recordings of user interactions with the system were collected with user consent. Additionally, semi-structured interviews in the post-task survey gathered insights into users' interactions with the system. In the post-task survey, we included a questionnaire to measure participants' acceptance and collaboration with the system. Specifically, we measured users' perceived system accuracy, overall trust in the system, and trust in specific system-generated results, including comments extraction and suggestions for related paragraphs. Semi-structured interviews were also conducted to capture participants' thoughts on acceptance and collaboration with the system. Details of the questionnaire and interview questions shown in \autoref{tab:Pre_task_questions}, \autoref{tab:In_task_questions}, \autoref{tab:Post_task_questions}, and \autoref{tab:Controlled_study_questions}.


\subsection{Results and Analysis}
\par We conducted both quantitative and qualitative analyses to address research questions \textbf{RQ3-RQ4}. \textbf{Quantitative data} from questionnaires and revision outline scores were analyzed using the Mann-Whitney U test~\cite{Mann1947OnAT}, a method commonly used for comparing independent conditions~\cite{Chandrasekharan2017YouCS, Kim2022StyletteST, Umar2019DetectionAA}. Descriptive statistics were also employed to examine central tendencies and variabilities in the results. \textbf{Qualitative data}, focusing on \textbf{RQ4: Interaction and Acceptance} were gathered through open-ended questions in post-surveys. Participants' thoughts and motivations behind their interactions and acceptance were captured through audio recordings, transcribed into text, and analyzed using thematic analysis~\cite{braun2006using, braun2012thematic}.

\begin{figure}[h]
    \centering
    \begin{subfigure}[b]{0.8\textwidth}
        \includegraphics[width=\textwidth]{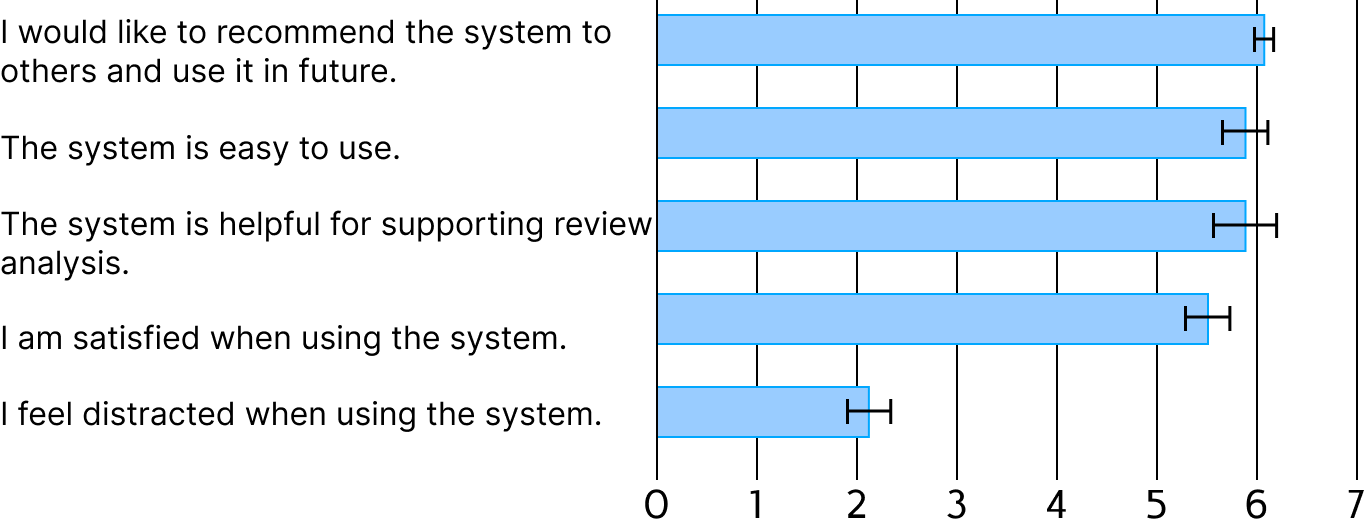}
    \caption{}
    \label{fig:in-task1}
    \end{subfigure}
    \begin{subfigure}[b]{0.8\textwidth}
        \includegraphics[width=\textwidth]{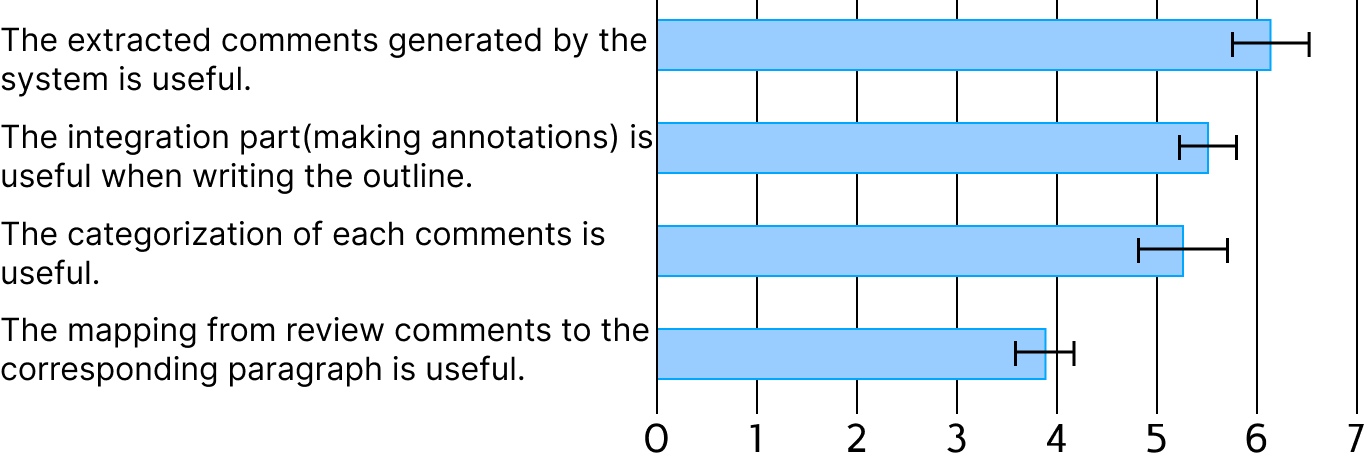}
        \caption{}
        \label{fig:in-task2}
    \end{subfigure}
    \caption{The usability of the system. (a) The usability of \textit{ReviseMate} perceived by users. (b) The perceived usefulness of different features in \textit{ReviseMate}.}
\end{figure}

\subsubsection{\textbf{RQ3: How are the usability and effectiveness of the system in review text analysis}}
\par \textbf{(1) Usability.} In the \textit{ReviseMate} condition, the in-task questionnaire included specific questions regarding system usability. The ratings for these questions are presented in descending order in \autoref{fig:in-task1}. The results indicate high scores for ease of use, helpfulness in supporting review analysis, user satisfaction, and the likelihood of recommending and using the system in the future. Additionally, participants reported relatively low levels of distraction while using the system, suggesting that the system is perceived as focused and user-friendly. In \autoref{fig:in-task2}, we present the perceived usefulness of different \textit{ReviseMate} features. While most features were deemed useful, the ratings for the feature ``mapping from review comments to corresponding paragraphs'' were comparatively lower than other features.

\par \textbf{Summary of findings from RQ3 (usability):} The findings indicate that \textit{ReviseMate} is generally perceived as useful with minimal distractions. While most features are considered beneficial, the ``mapping from review comments to the corresponding paragraph'' received relatively lower ratings, which will be further explored in RQ4 below.

\par
\begin{figure}[h]
    \centering
    \includegraphics[width=\textwidth]{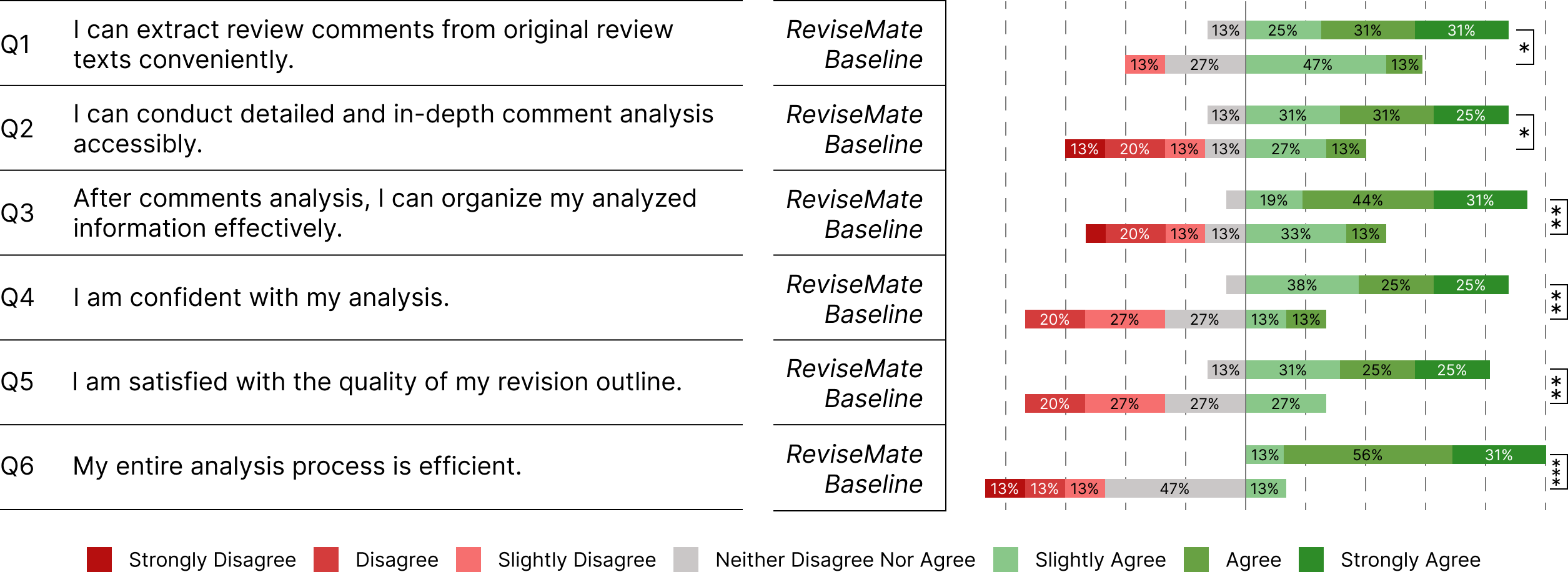}
    \caption{Effectiveness-related questions in \textit{ReviseMate} and Baseline conditions. (*: p<.05; **: p<.01; ***: p<.001)}
    \label{fig:effectiveness}
\end{figure}

 \textbf{(2) Effectiveness.} We evaluated the effectiveness of \textit{ReviseMate} by combining user ratings from the in-task questionnaire with domain expert evaluations of the submitted revision outlines. This dual approach allowed us to capture both user perceptions and an objective assessment of performance. For the in-task questionnaire data, we employed the Mann-Whitney U test~\cite{Mann1947OnAT} to compare the effectiveness of \textit{ReviseMate} with the baseline. The results, as shown in \autoref{fig:effectiveness}, indicated that participants in the \textit{ReviseMate} condition reported significantly higher ratings for effectiveness. The review text analysis process in \textit{ReviseMate} consists of two phases: 1) comment extraction during the \textit{Preprocessing} phase, and 2) in-depth comment analysis and the structuring of analyzed data with revision plans during the \textit{Analysis} phase. Users provided feedback at each stage, and the results revealed that participants in the \textit{ReviseMate} condition performed better in various stages:
\begin{itemize}
\item \textit{Comment extraction:} \textit{ReviseMate} significantly enhanced the convenience of comment extraction from review text (U=14, p<0.05).
\item \textit{In-depth analysis:} Users found \textit{ReviseMate} more accessible for conducting an in-depth analysis of review comments (U=11, p <0.05).
\item \textit{Analyzed information organization and revision outline creation:} Participants perceived \textit{ReviseMate} as more efficient in organizing previously analyzed information (U=7, p<0.01).
\item \textit{Confidence and satisfaction:} Users reported significantly higher confidence in their analyzed results (U=8.5, p<0.01) and greater satisfaction with their finished revision outlines (U=5, p<0.01).
\item \textit{Efficiency:} The \textit{ReviseMate} condition group also rated the review text analysis process as significantly more efficient (U=0.5, p<0.001).
\end{itemize}

\par We conducted a Mann-Whitney U test to compare the ratings of submitted revision outlines between the two groups. Each revision outline is evaluated based on the criteria of \textit{completeness}, \textit{accuracy}, and \textit{organization}, with each criterion rated on a scale from 0 to 10. As shown in \autoref{tab:scores}, the results indicate that users in the \textit{ReviseMate} condition outperformed those in the control condition, with significantly high scores for completeness (U=46, p<0.01), accuracy (U=29.5, p<0.001), and organization (U=46.5, p<0.01).

\begin{table}[h]
\begin{tabular}{@{}ccccc@{}}
\toprule
 & Control Group & Experiment Group & U & p\\ \midrule
Completeness & 4.53 & 5.88 & 46.00 & 0.003 \\
Accuracy & 3.60 & 5.31 & 29.50 & <0.001\\
Organization & 3.73 & 4.88 & 46.50 & 0.003\\
\bottomrule
\end{tabular}
\caption{Comparison of average scores in Completeness, Accuracy, and Organization between Control and Experiment Groups. ``U'' indicates Mann-Whitney U test results, and ``p'' indicates p-values for statistical significance.}
\label{tab:scores}
\end{table}

\par \textbf{Summary of findings from RQ3 (effectiveness):} Participants using \textit{ReviseMate} gave significantly higher ratings for their review analysis tasks compared to those in the raw text condition. Moreover, the revision outlines submitted in the \textit{ReviseMate} condition demonstrated superior performance in terms of \textit{completeness}, \textit{accuracy}, and \textit{organization} compared to the control condition.

\subsubsection{\textbf{RQ4: How do authors interact and accept the system during their review analysis?}}
\par We employed thematic analysis to analyze the qualitative data obtained from screen recordings and open-ended questions regarding \textbf{interaction and acceptance} during semi-structured interviews. The first author transcribed the data and generated initial open codes, which were then reviewed and discussed to resolve disagreement and develop the high-level themes~\cite{braun2012thematic, braun2006using}. 

\par (1) \textbf{Interaction.} The resulting insights are presented based on the recurring themes from thematic analysis.

\par \textbf{ReviseMate enables users to engage in diverse review comment extraction and make contextual adjustments.} Our analysis revealed that users adopt various methods for review comment extraction and comment card creation. Among the 16 participants in the \textit{ReviseMate} condition, 4 opted to use the automated generation feature provided by \textit{ReviseMate} to create comment cards entirely. In contrast, 9 users preferred to examine the original review text and manually select comments to construct their cards. Additionally, 13 users chose to explore the \textit{Reviewer comments} page, which organizes comments by reviewers and provides summarized titles for each comment. Users acknowledge that their choices are based on personal preferences and habits. T22 (male, age: 23) expressed a preference for starting with the original review comments, stating, ``\textit{I do this because I like to read the original review comments first before coming up with my own extracted ideas. It helps me catch all the important details and truly grasp what the reviewers are trying to convey.}'' Besides, T5 (male, age: 25) mentioned a personal habit of exploring comments categorized by reviewers, explaining, ``\textit{I start by browsing through the comments categorized by the reviewer because I have a personal habit of getting to know the perspectives of different reviewers first. This helps me to better grasp the diversity and focus of the comments.}'' These findings demonstrate that \textit{ReviseMate} meets the needs of researchers with diverse preferences for analyzing review comments.

 \par Users also frequently made contextual adjustments to comments after initially creating comment cards. Some users (9 out of 16) revisited the \textit{original} page following the extraction of comment cards. They read through the original review and added new comments to supplement their comment cards. T15 (female, age: 24) mentioned, ``\textit{I revisit the original review comments to make sure the extracted comment cards are accurate and that nothing important has been left out.}'' Additionally, users (12 out of 16) checked the \textit{Reviewer Comment} page and deleted existing comment cards, typically because these cards contained positive comments that were deemed unnecessary. T21 (male, age: 20) pointed out, ``\textit{As I went through the reviewer comments page, I noticed that some of these comment cards were actually positive. Currently, I don't think they would be all that important.}'' Users recognized the system's flexibility in supporting the editing of extracted comment cards. T4 (female, age: 29) highlighted this flexibility, stating, ``\textit{Sometimes, I want to get rid of cards that don't matter or aren't relevant so I can concentrate on the important stuff. This feature gives me the freedom to tailor my responses to fit exactly what I need.}'' These findings are consistent with our formative study observations and design goals, which emphasize offering various comment extraction options and flexible editing to accommodate different user needs.

\par \textbf{ReviseMate supports users for custom categorization of extracted comments and contextual analysis with the original text.} \textit{ReviseMate} provides predefined and customized categorization features to support in-depth comment organization and analysis. Users demonstrated two distinct usage patterns for these categorization features: some users opted to add additional criteria and numerous categories to facilitate comprehensive analysis, while others preferred to work within the confines of predefined criteria and categories. Users' rationales for their choices varied, encompassing factors such as personal preferences and levels of experience. \yuansong{The motivation to add additional criteria and categories primarily derives from users' consideration for a personalized analysis in understanding the review.} \revision{For instance, users may create categories like 'Introduction' or 'Results' based on the paper’s sections for contextual analysis, making the feedback more structured and easier to review.} As explained by T14 (male, age: 27), ``\textit{I prefer to introduce extra categories because it offers a more comprehensive perspective on the paper, especially when it deals with complex topics or issues. It allows me to pinpoint the information relevant to my research and respond more effectively to the review comments.}'' Notably, less experienced users tended to rely on predefined standards and classifications for their analysis, as exemplified by T3, T24, and T30, who expressed their inclination to adhere to existing criteria and categories during the categorization process.

\par An intriguing finding was that users' analysis processes did not conform precisely to our anticipated workflow of ``first categorizing all comments and then conducting a unified contextual analysis with the original paper''. Instead, users seemed to strike a balance between comment categorization and contextual analysis \yuansong{in the task-driven interaction of analyzing reviews and finishing revision outlines within the controlled study. \revision{We recognize that since participants are not the paper authors, this contextual analysis may fail to capture the actual reflections and practices of the authors. However, the interview results} indicate this is consistent with the actual review analysis process: examining the review comments, identifying correspondences with the original paper, and making informed comprehensive analysis.} Specifically, a majority of users (11 out of 16) engaged in a stepwise approach, categorizing comments individually and subsequently establishing connections with corresponding sections of the original paper. Users recognized that combining categorization with contextual analysis of the original text enhanced their ability to comprehensively evaluate each comment. T26 (male, age: 28) explained, ``\textit{I go through comments one by one, linking them back to the original text. This helps me grasp each comment better, ensuring I have the right context. It also makes it easier for me to adapt my analysis as needed.}'' In addition, users expressed concerns about potentially losing spontaneous ideas and thoughts if they deferred a task until the completion of the previous one. T5 (male, age: 25) articulated this concern, stating, ``\textit{I worry that waiting until the end might make me forget some instant ideas and thoughts. So, I prefer adding new criteria or categories as I go along with the categorization and analysis to ensure I don't miss important points.}'' Although inconsistent from our initially anticipated workflow, we found that when users analyze the extracted comment cards against the original text, the system still supports flexible adjustments to criteria or categories. These findings inspired us to revise our envisioned system workflow, suggesting that users categorize and analyze the extracted comments against the original text one by one.

\par \textbf{ReviseMate supports users in real-time recording thoughts from each review comment and organization for revision outlines.} Users demonstrated diverse approaches to incorporating their analyzed information into the revision outlines. Surprisingly, only a minority of users (5 out of 16) strictly followed the sequential process of completing all previous stages before writing the revision outline. In contrast, the majority (11 out of 16) merged information integration with comment analysis, simultaneously integrating insights into the evolving revision outline as they analyzed each comment. Although users demonstrated diverse approaches for incorporating analyzed information into their revision outlines, they all expressed the convenience brought by \textit{ReviseMate} for supporting real-time recording of thoughts in the corresponding sections of the original text. T25 (male, age: 25) offered advantages such as enhanced productivity and maintaining clarity of thought throughout the process, ``\textit{When I check the corresponding sections of the review comments in the original text, I can record my thoughts as side-annotations conveniently, it helped me stay on track and think more clearly.}'' Additionally, users generally recognized the organization of analyzed insights during the preparation for revision. As T12 stated, ``\textit{When I have a structured template, I only need to find the previously recorded issues and related thoughts in the original text sections and organize them one by one into the revision outline. This structured organization is very helpful for generating the subsequent cover letter or making manuscript modifications according to that.}''

\par \textbf{Summary of findings from RQ4 (interaction):} \textit{ReviseMate} facilitates analysis across various stages of the review comment process, allowing for flexible transitions between different features. Users exhibit diverse behaviors during actual usage, frequently navigating between original comments, the manuscript, and recording analytical insights to ensure comprehensive comment extraction and analysis. This fluid interaction suggests blurred boundaries between different stages of the system's workflow. User feedback prompted adjustments to clarify and refine the interaction workflow in subsequent studies.

\par (2) \textbf{Acceptance.} We assessed users' acceptance and trust in the system by collecting their ratings through questionnaires in the post-task survey and gathered insights through open-ended questions during system usage.

\par
\begin{figure}[h]
    \centering
    \includegraphics[width=0.8\textwidth]{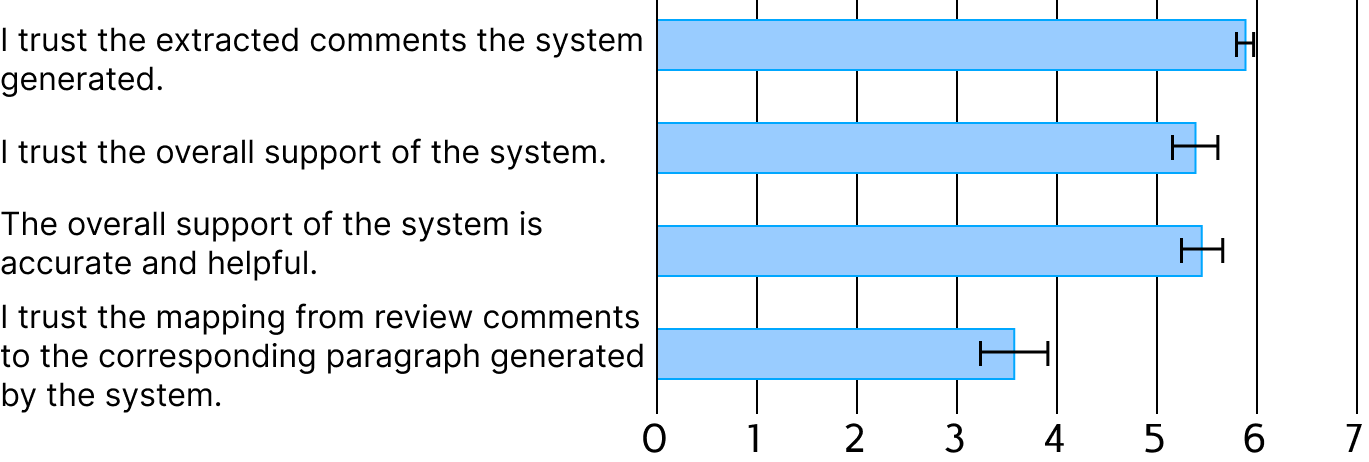}
    \caption{User trust in \textit{ReviseMate}: This includes users' perceptions of the system's accuracy, their overall trust in the system, and the average level of trust attributed to each individual feature.}
    \label{fig:post-task}
\end{figure}

 \autoref{fig:post-task} shows users' overall acceptance and trust in the system, along with the trust levels associated with each feature within the system. Regarding the system's acceptance, users perceived it as accurate and helpful overall (mean = 5.88, SD = 0.33). In terms of trust, users generally trust most of the system's features and its overall functionality. However, there was notably lower trust in the generated suggested related paragraphs of comments, which received an average score of 3.56 (SD = 0.7). We delve into the qualitative findings from interviews to uncover the underlying reasons for this trust disparity.

\par \textbf{Finding 1. Users recognize the system's value:} Users consistently rated the system highly in terms of its usefulness. All participants provided scores above the midpoint for questions related to usefulness. This can be attributed to the relief of having information presented and integrated during the analysis, a significant improvement compared to working with \yuansong{a system presenting raw review text and original paper without additional interaction features (baseline condition is shown in \autoref{fig:baseline})}. For example, T6 (female, age: 31) expressed satisfaction, stating, ``\textit{I gave it a high score because having a system is obviously better than not having one \yuansong{<system with features for review analysis>}. It provided more information and improved the integration of information, making it easier for me to analyze and fully understand the comments.}'' T17 (female, age: 30) stressed the importance of additional references when dealing with the massive information in unstructured reviews, saying ``\textit{I'm rating it high because having extra references is key when sifting through a bunch of reviews. It speeds up the analysis, adds accuracy, and lets me dig deeper into the comments.}''

\par \textbf{Finding 2. Evolving trust levels during system exploration:} Users initially approached the system with a cautious mindset, reserving judgment until they had explored its capabilities further. The majority of users (10 out of 16) acknowledged that the system's accuracy in summarizing extracted comments played a pivotal role in building trust, particularly in the associated features. However, trust levels fluctuated when users engaged with the system's feature, suggesting possibly related paragraphs in the original paper. The primary challenge stemmed from the presence of multiple suggestions for a single comment. Users found it tedious to click and check each suggestion individually, which negatively impacted their trust in the system. For instance, T10 (female, age: 20) shared their experience, stating ``\textit{I discovered that there were many suggested paragraphs, which led me to click and check them one by one, which was not convenient.}'' From these findings, we choose to limit the presentation of suggested related paragraphs to only the top 5 for each comment. Another noteworthy observation was that once users had an unsatisfactory experience with a particular feature, it became challenging to regain their trust. Some users indicated that they were hesitant to explore suggested paragraphs further after encountering situations where the initial suggestions did not align with their expectations. As explained by T25 (male, age: 25), ``\textit{I checked the first few relevant paragraphs suggested by the system, but they were not very accurate. I guessed that the next ones might also be inaccurate, so I would rather find the relevant content myself.}''

\par \textbf{Finding 3. Variability in system acceptance among users:} We found that users displayed varying levels of acceptance when initially using the system. Notably, the extent of critical thinking and exploration correlated with users' self-confidence and research experience. More experienced users tended to approach the system with a degree of skepticism and engaged in comprehensive assessments, considering high-level perspectives. T18 (male, age: 29), for instance, expressed this sentiment, stating, ``\textit{I've gained some experience in this field, and I won't be easily convinced by the system. For example, I like to read original review texts first to determine whether the system's extracted information is accurate or not.}'' Conversely, users with less experience or those less familiar with the specific research domain exhibited a more observant and accepting attitude. T20 (female, age: 22) exemplified this perspective, sharing, ``\textit{As a novice with less submission experience, I choose to trust and strictly follow the system's processes to reduce making mistakes.}'' This finding underscored the role of users' background and experience in shaping their initial acceptance of the system.

\par \textbf{Summary of findings from RQ4 (acceptance):} Users generally acknowledged the value of \textit{ReviseMate}, noting that its features enhanced the efficiency and quality of their analysis. However, trust in the system evolved during the exploration process. When presented with multiple choices for a review comment, users tended to assess trust based on the accuracy of initially clicked paragraphs. Excessive recommendations could lead to inaccuracies and increase review burdens, thereby impacting trust and feature usage. Consequently, we opted to provide only the top 5 most relevant paragraphs per comment. Acceptance of the system varied among users: experienced researchers approached it cautiously from a higher-level perspective, while novices tended to accept results more readily and follow instructions.

\subsection{Discussion of Controlled Study}
\par In the controlled study, participants were assigned to either the \textit{ReviseMate} condition or the Baseline condition to evaluate \textit{ReviseMate}'s usability and effectiveness in assisting with review comment analysis, and to collect feedback on system interaction and acceptance. We acknowledge the limitations of the controlled study: participants were not necessarily the first authors of the papers, and the actual analysis process may involve more time and discussion than completing tasks within a limited experimental timeframe. \yuansong{We also acknowledge that the between-subjects study may introduce biases due to individual differences, especially when dealing with subjective and personalized tasks~\cite{kirk2013experimental}. Therefore, the primary goal of our controlled study is to assess users' understanding and analysis of the review content within a specified time frame. This assessment is reflected through user feedback and the ratings of the revision outline evaluated based on three criteria: completeness, accuracy, and organization, as shown in \autoref{tab:evaluation_criteria}. \revision{We compared our system to a baseline that provides only the original paper and review comments, mirroring the authors' current approach to processing reviewer feedback. While this design enables a direct comparison with \textit{ReviseMate} and ensures evaluation under controlled conditions, we recognize its limitations. The baseline system, akin to a ``strawman'' comparison, may be overly simplistic and not fully representative of real-world revision workflows. Therefore,} the primary concern of the controlled study} was to simulate real-world review comment analysis scenarios under controlled conditions, validate the system's usability and effectiveness preliminarily, and use feedback to refine the system's interaction workflow and features. Subsequently, we conducted a field deployment study to gather additional user feedback in real-world settings, complementing and extending findings from the controlled study.

\section{Field Deployment Study}
\par We further conducted a field deployment study (N=6) to assess the usage of \textit{ReviseMate} in authors' digestions on the review of their own papers, following up with semi-structured interviews to gather qualitative data. \yuansong{In this study, we aim to first evaluate the effectiveness of \textit{ReviseMate} in practical review digestions, while also gathering insights into system usability through factors such as AI acceptance and collaboration arising from the usage processes.} Based on the insights from the controlled study, we make the following adjustments to \textit{ReviseMate} workflow and features.

\begin{itemize}
\item We reconsidered the user interaction workflow and encouraged diverse and personalized review analysis. After extracting review comments and creating comment cards using various system-supported methods, users are no longer confined to a step-by-step process. Instead, they can flexibly switch between the original text, comment cards, and revision outlines. This flexibility allows for either complete analysis and organization of individual comments or batch analysis followed by collective organization.
\item We modified the comment-to-text mapping feature to show only the top 5 paragraphs when multiple sections are identified, reducing choice overload and increasing result reliability.
\end{itemize}

\subsection{Participants and Procedure}
\par We recruited participants from the controlled user study to evaluate their paper's reviews using \textit{ReviseMate}. The group included six individuals, averaging 29.8 years old, with four males and two females. \yuansong{We acknowledge that the ideal setting is recruiting participants to be the first authors in their papers and having unread reviews. However, due to the constraints of fixed review cycles, it is challenging to find participants who are both first authors and have unread reviews. Therefore, we recruited the first authors who had not read the review comments for over three months and the second authors who had not yet read the review comments to participate in our study. By allowing participants to perform tasks in their own familiar environment, we simulate the real-world scenario of authors reading review comments for the first time to maintain ecological validity in the study.} Participant details are in \autoref{tab:participants2}. Each participant used \textit{ReviseMate} on their own computer to assess their review text. We asked participants to document their experiences and provide screen recordings with their consent.

\begin{wraptable}{r}{9cm} 
\centering
\begin{tabular}{@{}ccccc p{2cm} @{}}
\toprule
ID & Gender/Age & Exp & Role & Major\\ \midrule
P1 & F/25 & 1/3 & Graduate & VIS \\
P2 & F/39 & 20/39 & Asst.Prof & HCI \\
P3 & M/29 & 3/8 & PhD & ML \\
P4 & M/25 & 1/2 & Graduate & BME \\
P5 & M/27 & 4/11 & PhD & VIS \\
P6 & M/34 & 13/37 & Asst.Prof & HCI \\
\bottomrule
\end{tabular}
\caption{Participant data (F = Female, M = Male) included their experience level (Exp), denoted as A/B, where A represents the number of papers each participant authored as the first author, and B represents the total number of papers per participant. The participants' professional titles included Assistant Professor (Asst.Prof.), and their academic disciplines covered Machine Learning (ML), Biomedical Engineering (BME), Human-Computer Interaction (HCI), and Visualization (VIS).}
\label{tab:participants2}
\vspace{-15pt}
\end{wraptable}


\par Following the review analysis, we held 40-minute semi-structured interviews with each participant. During these interviews, participants were instructed to systematically reflect on their analysis process using their notes and screen recordings. They were also encouraged to share their thoughts and provide suggestions for improvements. Each participant received a \$30 compensation for their participation. With participant consent, we gathered data from these interviews. Two researchers then used thematic analysis methods~\cite{braun2006using, braun2012thematic} to study the interview results. The interview questions are listed in \autoref{tab:Qualitative_study_questions}.

\subsection{Results and Analysis}
\par In the interview, participants began by sharing their overall analysis experience. Each participant had reviewed at least one paper, with some participants (P1 and P3) analyzing two review texts for their respective papers. These original papers covered diverse research topics, aligning with the participants' areas of expertise, including fields such as VIS, BME, ML, and HCI. The findings are based on the questions of \textit{usability and effectiveness} and \textit{acceptance and collaboration with the system}.


\par \textbf{Finding 1. ReviseMate assists users in extracting review comments and enhances the efficiency of organizing and analyzing them.} Regarding the usability of specific system features, users held a positive perception of the multi-choice comment extraction, viewing it as a valuable tool for both assistance and reference. P5 emphasized the flexibility offered by the multiple comment extraction options, contrasting it favorably with the traditional, laborious process of manually extracting information from raw review text. P5 (male, age: 27) explained, ``\textit{I typically begin with the fully automated comment extraction because it captures all the essential comments at once. If I need to make adjustments to the automated comments, I use the semi-automatic method by copying key sentences and making edits. And when the automated method misses specific details, I manually enter them. These three comment extraction methods cater perfectly to my varied needs, ensuring comprehensive coverage while also providing flexibility.}'' P1 also expressed a positive perspective on comment cards, highlighting their ability to present both summarized comments and their corresponding original text from the reviews. P1 (female, age: 25) remarked, ``\textit{Comment cards offer more than just a quick summary of review comments; they also display the original comment text. For instance, if a comment card states, `The discussion of visualization options is not comprehensive enough', and I see the original text that reads, `The selection of some tables, such as pie charts, is not comprehensive enough', I instantly know what needs to be revised. Having this reference to the original text helps me ensure I don't overlook crucial details.}''

\par Users also consistently expressed highly positive views of the features related to comment card categorization. This positive sentiment can be attributed to the system's ability to meet the general need for organizing unstructured comments, as well as its support for flexible, personalized categorization based on user preferences. For example, as mentioned by P2 (female, age: 39), ``\textit{In the past, when I got review comments, I had to spend much time organizing them on my own. But now, I can easily group them based on predefined criteria such as content and workload. This makes my revisions more focused and also facilitates my comprehension of the review comments.}'' Furthermore, P6 (male, age: 34) emphasized the importance of supporting categorization with customizable criteria to facilitate a comprehensive analysis of comments. P6 explained, ``\textit{When I'm going through the review comments and thinking about what needs to be revised, I always like adding new criteria based on the given reviews, which would make it easier for me to connect and respond to the reviews later on. It is also a good practice for my organization of reviews.}'' Additionally, P6 highlighted that authors often deal with multiple rounds of submissions and reviews for the same article. We plan to address this point in our discussion on the generalizability of \textit{ReviseMate}.

\par \textbf{Finding 2. Users exhibit a generally positive attitude towards the AI results in \textit{ReviseMate}.} Users first acknowledged that AI involvement in human-AI collaboration is an ongoing trend, with AI increasingly being used to assist in the organization and analysis of complex raw data, thereby improving work efficiency. For example, P5 (male, age: 27) stated, ``\textit{Using AI to assist us in understanding long chunks of text seems to be an unstoppable trend. It's like a double win since it improves efficiency and gives me a hand in making decisions}.'' Additionally, some experienced scholars emphasized that the fear of AI making mistakes should not deter its use. As long as AI demonstrates sufficient accuracy, it can be used under human supervision and with final decisions made by humans, ensuring the quality of the analysis is maintained. ``\textit{I am open to human-AI collaboration,'' P6 explained. ``Mistakes can happen in any process, but we should not reject the involvement of AI for fear of possible errors. AI enhances the efficiency of comment extraction and mapping to the original text, while I still make the analysis and final revision plans, ensuring both the quality of analysis and greater convenience.}''


\par \textbf{Finding 3. Consideration of collaboration factors during Analysis.} Several users (P1, P3, P4) mentioned the topic of collaborative work when analyzing review texts. They indicated that there is a combination of individual analysis and discussions among authors in the actual review comment analysis process. Initially, individuals independently complete the organization and analysis of review comments, recording their thoughts. Subsequently, the authors discuss the involved comments and determine the revision plans. Using \textit{ReviseMate} ensures that each point in the final revision outline is traceable through the extraction and analysis of review comments, therefore facilitating collaboration. As proposed by P4 (male, age: 25), ``\textit{My analysis process with ReviseMate ensures that each comment undergoes customized categorization, corresponding text analysis, and recording of analytical thoughts. Therefore, when I organize the analytical thoughts into the revision outline, the previous analysis trace remains intact, facilitating review and modification during subsequent discussions.}'' Users also present the expectation of supporting simultaneous edits like \textit{overleaf}, which would be considered a future improvement.


\subsection{Discussion of Field Deployment Study}
\par \yuansong{Our field deployment study aimed to explore how users from diverse backgrounds interact with the system in real-world scenarios for digesting paper reviews. \revision{Compared to the controlled study, which focuses more on findings related to effectiveness and usability, the field deployment study reveals more practical usage insights. Specifically, the field deployment study} evaluated the system’s usability and effectiveness in practical contexts, gaining insights into user perceptions of AI-generated results and multi-user collaboration dynamics. Unlike controlled studies with predefined tasks, this approach provided feedback with minimal intervention, offering valuable insights for system optimization and future improvements.}

\section{Discussion}
\subsection{Design Implications on Review Comment Analysis}
\subsubsection{Enhance the Interaction between Different Features}
\par \textit{ReviseMate} divides the review comment analysis process into the \textbf{Preprocessing} and \textbf{Analysis} phases. According to the prescribed workflow, users proceed through comments extraction, categorization, and mapping with the original paper sequentially, ultimately yielding the final analysis results, which are then structured on the revision outline. However, in practice, users tend to exhibit a ``flexible analysis across different features'' characteristics. For example, influenced by personal experiences and preferences, users may not strictly adhere to the anticipated process but instead, navigate between various features during analysis. To alleviate potential inconveniences arising from this switching behavior, one solution is to enhance the interaction between different features. This can be achieved by implementing features such as previewing the content of comment cards in a hovering manner, facilitating interaction between comments categorization and mapping with the original paper. Additionally, marking the categorized information of each comment in the revision outline enhances the interaction between comment categorization and integrated information organization. These approaches enable users to conduct a seamless analysis across features based on their preferences, thereby improving the user experience and efficiency.

\subsubsection{Support Integrated Analysis across Review Comments}
\par While ensuring the diversity of user interactions with the system, the features provide predefined settings such as \textit{Content}, \textit{Workload}, and \textit{Emergency} within \textit{Customized categorization}. However, while these classifications establish a general standard for categorizing comments at the individual comment level, they lack an analysis of relationships between different comments. Specifically, users often find the need to analyze each comment individually to identify consistencies or discrepancies, assisting in consolidating feedback on similar topics. However, as T12 mentioned in the controlled study, this process can be inconvenient, especially when dealing with a large number of comments, without inter-comment level analysis. To enhance the user's analysis process of comments, a recommended solution is to analyze the topics of each comment and integrate them accordingly. Therefore, future system improvements could include refining the backend model to support review analysis across different comments and presenting cross-comment interaction features.

\subsection{Extending Construction-Integration Theory from Reading Comprehension to Feedback Processing}
\par The Construction-Integration (CI) theory~\cite{Kintsch1991TheCM} is used to explain and study the cognitive mechanisms involved in reading comprehension and text processing. It provides guidance on how people construct meaning from texts and integrate these meanings into their existing knowledge structures during academic reading. While the analysis of review text also involves the construction and integration of information based on comments, there are distinctions between traditional reading comprehension and the review comment analysis process. The primary purpose of reading comprehension is to acquire and understand new knowledge, with an emphasis on literature review, information absorption, and critical thinking~\cite{chen2023marvista, Yuan2023CriTrainerAA, tashman2011active}. In contrast, the analysis of review comments consists of feedback processing, focuses on addressing feedback on existing work, with an emphasis on problem-solving and developing specific revision plans~\cite{shute2008focus, kelly2014peer}. Compared to academic reading, review comment analysis places greater importance on the completeness and effectiveness of the analysis.

\par Our work, based on these insights, extends the CI theory to the domain of review comment analysis grounded in feedback processing. We divide the analysis process into preprocessing (extraction and organization of comments) and analysis (categorization, contextual analysis, and recording and integration of insights) stages to align with the CI theoretical framework. \yuansong{By separating the stages, the system can focus on different tasks at each step, optimizing the flow of information and enhancing the flexibility of the design. This division aligns with the cognitive processes described in CI theory and informs the design of system features and iterative workflows for review comment analysis.} More generally, the review text analysis process can be seen as feedback processing, involving the receipt, integration, and action on feedback. \yuansong{Beyond academic review, this framework can be adapted to textual feedback processing in other domains, such as categorizing feedback in customer reviews and analyzing sentiment in social media comments.} \revision{Although comments from other domains could differ from those in the review analysis, the system offers a space to explore design implications for feedback-oriented systems.} 

\subsection{Reducing Overreliance on AI in ReviseMate}
\par \yuansong{Research on AI can be primarily divided into two goals: simulating and understanding human abilities to automate tasks or applying AI capabilities to assist humans in a mixed-initiative human-AI workflow~\cite{shneiderman2020design}.}
Although AI-assisted systems are believed to enhance human analysis efficiency and quality significantly, some studies indicate that \yuansong{mixed-initiative human-AI workflows} can often be inferior to standalone AI~\cite{green2019principles, vaccaro2019effects, bansal2021does, liu2021understanding}. A key reason for this is that humans tend to over-rely on AI results, even in cases where humans would make better decisions and AI may be wrong. We acknowledge that excessive reliance on automation introduces the risk of misinterpretations and biases~\cite{Schemmer2022OnTI}, therefore, \textit{ReviseMate} primarily aims to \yuansong{incorporate human effort in the workflow,} assisting users in extracting and organizing information for analysis instead of making AI-driven decisions about revision strategies or responses. Therefore, it is crucial for how AI-based results are presented to encourage appropriate reliance~\cite{cao2022understanding}. \yuansong{Despite Explainable AI (XAL) being leveraged, research indicates that this can only increase the chances of accepting AI's recommendation without substantially reducing overreliance~\cite{bansal2021does,buccinca2021trust}.} While adding cognitive forcing functions is considered a promising approach~\cite{buccinca2021trust}. This includes having humans make decisions before seeing AI results and allowing users to view AI recommendations selectively. \yuansong{In designing \textit{ReviseMate}, we adopted a consistent approach. In the \textbf{Preprocessing Phase}, we retained the semi-automatic extraction method to help users extract comments while performing cognitive effort. At the same time, the fully automatic comment cards also preserved the correspondence with the original text to help users check the content. During the \textbf{Analysis Phase}, \textit{ReviseMate} provides correspondence between comments and the original paper by demonstrating} ``possibly related paragraphs'' for each comment as an optional reference, emphasizing the value of users' active cognitive effort in the analysis and decision-making phases. \yuansong{We also plan to update \textit{ReviseMate}'s interaction workflow to encourage users to think independently before reviewing AI-provided results during the comment extraction and text correspondence processes as a future direction}.

\subsection{Generalizability of ReviseMate}
\par \yuansong{We are currently focused on analyzing academic paper reviews in the STEM field, streamlining the digestion of received review text for authors. However, as mentioned in the field deployment study, users may experience multiple submissions, each receiving its own set of reviews, which is common in areas such as the humanities or social sciences. Analyzing newly received reviews often involves referencing and comparing with prior feedback. Recognizing this, we plan to expand \textit{ReviseMate} in the future to facilitate comparisons and analyses across multiple reviews.}


\par \yuansong{Although our controlled study primarily involved participants with a background in computer science, the workflow developed in ReviseMate—consisting of comment organization, contextual analysis, and integration—can be adapted to broader disciplinary contexts.} For instance, during the field deployment study, P2 suggested potential applications of \textit{ReviseMate} in the education domain, where students submit their work and receive feedback from instructors. We envision that \textit{ReviseMate} could be employed by students to analyze and \yuansong{organize} teachers' feedback, thereby aiding in more effective revision and refinement processes. While differences in content and primary concerns may necessitate refinements to features like predefined criteria and categories in comment categorization for specific contexts, these observations highlight the system's potential adaptability and compatibility \yuansong{for diverse feedback-driven processes} beyond academia. \yuansong{These observations suggest that the \textit{ReviseMate}'s design, which combines structured feedback digestion with flexible and personalized analysis, provides opportunities for extension into other contexts requiring textual feedback integration.}

\subsection{Limitation and Future Work}
\par This study has several limitations. \yuansong{First, we acknowledge that the design process intentionally prioritizes authors' perspectives. \revision{Therefore, \textit{ReviseMate} provides features such as aligning comments with relevant paragraphs to facilitate authors' analysis and comprehension.} We consider this a starting point in exploring support for STEM paper review digestion. \revision{Through establishing a cognitive scaffolding for review digestion grounded in authors' perspectives, the system enables future structured collaborations between reviewers and authors across iterative design cycles.} Future directions include integrating different aspects of this process, such as involving reviewers' perspectives into \textit{ReviseMate}'s workflow, enabling reviewers to write comments linked to specific parts of the original text, and promoting authors to perform subsequent analysis. This approach aims to balance diverse needs and foster value co-creation~\cite{gronroos2013critical}.} 

\par Second, the controlled user study utilized the same review for all participants, differing from \textit{ReviseMate}'s primary use case of helping authors analyze their received reviews. We acknowledge this compromise was made to initially evaluate \textit{ReviseMate}'s usability and effectiveness in a controlled environment. To address this limitation, we subsequently conducted a field deployment study for practical review analysis. Third, participants in the field deployment study included first authors who hadn't reviewed their paper's comments for over three months and second authors who hadn't thoroughly reviewed their feedback, which may not fully represent real-world scenarios. To address this, we plan to conduct a longitudinal study to collect continuous usage feedback. 

\par In our between-subjects study, we assessed the system's effectiveness by rating the quality of revision outlines submitted by users within a limited time, using an objective standard. A more comprehensive approach would be to evaluate the time savings \textit{ReviseMate} provides during analysis. Therefore, we plan to track each user's usage and obtain statistical average time improvements in the longitudinal study. \revision{Additionally}, our participants were recruited through snowball sampling within the university community, predominantly from the field of Computer Science. We plan to increase the sample size and recruit diverse individuals from various research domains. Lastly, our prototype has encountered usability issues such as time delays and inaccurate results. We recognize these problems and intend to address them through iterative enhancements of the backend model to improve the system's usability.

\section{Conclusion}
\par We introduce \textit{ReviseMate}, a system designed to streamline \yuansong{STEM} review analysis by dividing the process into \textbf{Preprocessing} and \textbf{Analysis} stages. This system aids users in analyzing review texts and facilitating collaborative article revisions and responses to reviewers. \textit{ReviseMate} generates comment cards from review comments, which users can manually create or construct with system assistance based on the original feedback. The system supports the customized organization and categorization of comments, along with contextual analysis linked to the original text. Additionally, \textit{ReviseMate} allows for real-time recording during the analysis and structured organization of final insights. We initially validated the system's usability and effectiveness through a controlled study with 31 participants, gaining valuable insights into user interaction and collaboration. Based on these experimental results, we redesigned the workflow and modified certain features. Subsequently, we conducted a field deployment study to evaluate \textit{ReviseMate}'s performance in real-world review comment analysis, obtaining qualitative feedback. Our work discusses design implications for supporting cross-interaction between analysis stages and the integrated analysis of review comments. It extends the Construction-Integration theory to the feedback-processing domain and explores strategies to reduce overreliance on AI results. Our findings offer insights into applying feedback processing in academic review analysis.


\begin{acks}
\par We thank anonymous reviewers for their valuable feedback. This work is supported by grants from the National Natural Science Foundation of China (No. 62372298), Shanghai Engineering Research Center of Intelligent Vision and Imaging, Shanghai Frontiers Science Center of Human-centered Artificial Intelligence (ShangHAI), and MoE Key Laboratory of Intelligent Perception and Human-Machine Collaboration (KLIP-HuMaCo).
\end{acks}


\bibliographystyle{ACM-Reference-Format}
\bibliography{sample-base}


\begin{thebibliography}{94}


\ifx \showCODEN    \undefined \def \showCODEN     #1{\unskip}     \fi
\ifx \showISBNx    \undefined \def \showISBNx     #1{\unskip}     \fi
\ifx \showISBNxiii \undefined \def \showISBNxiii  #1{\unskip}     \fi
\ifx \showISSN     \undefined \def \showISSN      #1{\unskip}     \fi
\ifx \showLCCN     \undefined \def \showLCCN      #1{\unskip}     \fi
\ifx \shownote     \undefined \def \shownote      #1{#1}          \fi
\ifx \showarticletitle \undefined \def \showarticletitle #1{#1}   \fi
\ifx \showURL      \undefined \def \showURL       {\relax}        \fi
\providecommand\bibfield[2]{#2}
\providecommand\bibinfo[2]{#2}
\providecommand\natexlab[1]{#1}
\providecommand\showeprint[2][]{arXiv:#2}

\bibitem[(1885)(2013)]%
        {Ebbinghaus18852013MemoryAC}
\bibfield{author}{\bibinfo{person}{Hermann~Ebbinghaus (1885)}.}
  \bibinfo{year}{2013}\natexlab{}.
\newblock \showarticletitle{Memory: A Contribution to Experimental Psychology}.
\newblock \bibinfo{journal}{\emph{Annals of Neurosciences}}
  \bibinfo{volume}{20} (\bibinfo{year}{2013}), \bibinfo{pages}{155 -- 156}.
\newblock
\urldef\tempurl%
\url{https://api.semanticscholar.org/CorpusID:31173730}
\showURL{%
\tempurl}


\bibitem[{ACM CHI Conference}({[n.\,d.]})]%
        {chi_review_criteria}
\bibfield{author}{\bibinfo{person}{{ACM CHI Conference}}.}
  \bibinfo{year}{[n.\,d.]}\natexlab{}.
\newblock \bibinfo{title}{Guide to Reviewing Papers}.
\newblock
  \bibinfo{howpublished}{\url{https://chi2024.acm.org/submission-guides/guide-to-reviewing-papers/}}.
\newblock
\newblock
\shownote{Accessed: 2024-06-29}.


\bibitem[{ACM CSCW Conference}({[n.\,d.]})]%
        {cscw_student_mentor_program}
\bibfield{author}{\bibinfo{person}{{ACM CSCW Conference}}.}
  \bibinfo{year}{[n.\,d.]}\natexlab{}.
\newblock \bibinfo{title}{Instructions for CSCW Student Mentor Program}.
\newblock
  \bibinfo{howpublished}{\url{https://cscw.acm.org/2016/volunteer/InstructionsCSCWStudentMentorProgram.pdf}}.
\newblock
\newblock
\shownote{Accessed: 2024-06-29}.


\bibitem[Alcock(2009)]%
        {Alcock2009eProofsSE}
\bibfield{author}{\bibinfo{person}{Lara Alcock}.}
  \bibinfo{year}{2009}\natexlab{}.
\newblock \showarticletitle{e-Proofs: Student Experience of Online Resources to
  Aid Understanding of Mathematical Proofs}.
\newblock
\urldef\tempurl%
\url{https://api.semanticscholar.org/CorpusID:211019993}
\showURL{%
\tempurl}


\bibitem[Anderson and Pearson(1988)]%
        {Anderson1988ASV}
\bibfield{author}{\bibinfo{person}{Richard~C. Anderson} {and}
  \bibinfo{person}{P.~David Pearson}.} \bibinfo{year}{1988}\natexlab{}.
\newblock \showarticletitle{A Schema-Theoretic View of Basic Processes in
  Reading Comprehension. Technical Report No. 306.}
\newblock
\urldef\tempurl%
\url{https://api.semanticscholar.org/CorpusID:47005501}
\showURL{%
\tempurl}


\bibitem[Bansal et~al\mbox{.}(2021)]%
        {bansal2021does}
\bibfield{author}{\bibinfo{person}{Gagan Bansal}, \bibinfo{person}{Tongshuang
  Wu}, \bibinfo{person}{Joyce Zhou}, \bibinfo{person}{Raymond Fok},
  \bibinfo{person}{Besmira Nushi}, \bibinfo{person}{Ece Kamar},
  \bibinfo{person}{Marco~Tulio Ribeiro}, {and} \bibinfo{person}{Daniel Weld}.}
  \bibinfo{year}{2021}\natexlab{}.
\newblock \showarticletitle{Does the whole exceed its parts? the effect of ai
  explanations on complementary team performance}. In
  \bibinfo{booktitle}{\emph{Proceedings of the 2021 CHI conference on human
  factors in computing systems}}. \bibinfo{pages}{1--16}.
\newblock


\bibitem[Baumeister et~al\mbox{.}(2001)]%
        {Baumeister2001BadIS}
\bibfield{author}{\bibinfo{person}{Roy~F. Baumeister}, \bibinfo{person}{Ellen
  Bratslavsky}, \bibinfo{person}{Catrin Finkenauer}, {and}
  \bibinfo{person}{Kathleen~D. Vohs}.} \bibinfo{year}{2001}\natexlab{}.
\newblock \showarticletitle{Bad is Stronger than Good}.
\newblock \bibinfo{journal}{\emph{Review of General Psychology}}
  \bibinfo{volume}{5} (\bibinfo{year}{2001}), \bibinfo{pages}{323 -- 370}.
\newblock
\urldef\tempurl%
\url{https://api.semanticscholar.org/CorpusID:262038677}
\showURL{%
\tempurl}


\bibitem[Biolkov{\'a} et~al\mbox{.}(2023)]%
        {Biolkov2023InvestigationOP}
\bibfield{author}{\bibinfo{person}{Marie Biolkov{\'a}}, \bibinfo{person}{Tom
  Moore}, \bibinfo{person}{Karen Schindler}, \bibinfo{person}{Karl Swann},
  \bibinfo{person}{Andrew Vail}, \bibinfo{person}{Lindsay Flook},
  \bibinfo{person}{Helen Dick}, \bibinfo{person}{Greg Fitzharris},
  \bibinfo{person}{Christopher~A. Price}, {and} \bibinfo{person}{Norah
  Spears}.} \bibinfo{year}{2023}\natexlab{}.
\newblock \showarticletitle{Investigation of potential gender bias in the peer
  review system at Reproduction}.
\newblock \bibinfo{journal}{\emph{Learned Publishing}}  \bibinfo{volume}{36}
  (\bibinfo{year}{2023}).
\newblock


\bibitem[Braun and Clarke(2006)]%
        {braun2006using}
\bibfield{author}{\bibinfo{person}{Virginia Braun} {and}
  \bibinfo{person}{Victoria Clarke}.} \bibinfo{year}{2006}\natexlab{}.
\newblock \showarticletitle{Using thematic analysis in psychology}.
\newblock \bibinfo{journal}{\emph{Qualitative research in psychology}}
  \bibinfo{volume}{3}, \bibinfo{number}{2} (\bibinfo{year}{2006}),
  \bibinfo{pages}{77--101}.
\newblock


\bibitem[Braun and Clarke(2012)]%
        {braun2012thematic}
\bibfield{author}{\bibinfo{person}{Virginia Braun} {and}
  \bibinfo{person}{Victoria Clarke}.} \bibinfo{year}{2012}\natexlab{}.
\newblock \bibinfo{booktitle}{\emph{Thematic analysis.}}
\newblock \bibinfo{publisher}{American Psychological Association}.
\newblock


\bibitem[Brooke(1996)]%
        {Brooke1996SUSA}
\bibfield{author}{\bibinfo{person}{J.~B. Brooke}.}
  \bibinfo{year}{1996}\natexlab{}.
\newblock \showarticletitle{SUS: A 'Quick and Dirty' Usability Scale}.
\newblock
\urldef\tempurl%
\url{https://api.semanticscholar.org/CorpusID:107686571}
\showURL{%
\tempurl}


\bibitem[Bu{\c{c}}inca et~al\mbox{.}(2021)]%
        {buccinca2021trust}
\bibfield{author}{\bibinfo{person}{Zana Bu{\c{c}}inca},
  \bibinfo{person}{Maja~Barbara Malaya}, {and} \bibinfo{person}{Krzysztof~Z
  Gajos}.} \bibinfo{year}{2021}\natexlab{}.
\newblock \showarticletitle{To trust or to think: cognitive forcing functions
  can reduce overreliance on AI in AI-assisted decision-making}.
\newblock \bibinfo{journal}{\emph{Proceedings of the ACM on Human-computer
  Interaction}} \bibinfo{volume}{5}, \bibinfo{number}{CSCW1}
  (\bibinfo{year}{2021}), \bibinfo{pages}{1--21}.
\newblock


\bibitem[Cachola et~al\mbox{.}(2020)]%
        {Cachola2020TLDRES}
\bibfield{author}{\bibinfo{person}{Isabel Cachola}, \bibinfo{person}{Kyle Lo},
  \bibinfo{person}{Arman Cohan}, {and} \bibinfo{person}{Daniel~S. Weld}.}
  \bibinfo{year}{2020}\natexlab{}.
\newblock \showarticletitle{TLDR: Extreme Summarization of Scientific
  Documents}. In \bibinfo{booktitle}{\emph{Findings}}.
\newblock


\bibitem[Cai et~al\mbox{.}(2019)]%
        {cai2019hello}
\bibfield{author}{\bibinfo{person}{Carrie~J Cai}, \bibinfo{person}{Samantha
  Winter}, \bibinfo{person}{David Steiner}, \bibinfo{person}{Lauren Wilcox},
  {and} \bibinfo{person}{Michael Terry}.} \bibinfo{year}{2019}\natexlab{}.
\newblock \showarticletitle{" Hello AI": uncovering the onboarding needs of
  medical practitioners for human-AI collaborative decision-making}.
\newblock \bibinfo{journal}{\emph{Proceedings of the ACM on Human-computer
  Interaction}} \bibinfo{volume}{3}, \bibinfo{number}{CSCW}
  (\bibinfo{year}{2019}), \bibinfo{pages}{1--24}.
\newblock


\bibitem[Cai and Ziad(2003)]%
        {cai2003evaluating}
\bibfield{author}{\bibinfo{person}{Yu Cai} {and} \bibinfo{person}{Mostapha
  Ziad}.} \bibinfo{year}{2003}\natexlab{}.
\newblock \showarticletitle{Evaluating completeness of an information product}.
\newblock \bibinfo{journal}{\emph{AMCIS 2003 Proceedings}}
  (\bibinfo{year}{2003}), \bibinfo{pages}{294}.
\newblock


\bibitem[Cao and Huang(2022)]%
        {cao2022understanding}
\bibfield{author}{\bibinfo{person}{Shiye Cao} {and} \bibinfo{person}{Chien-Ming
  Huang}.} \bibinfo{year}{2022}\natexlab{}.
\newblock \showarticletitle{Understanding user reliance on AI in assisted
  decision-making}.
\newblock \bibinfo{journal}{\emph{Proceedings of the ACM on Human-Computer
  Interaction}} \bibinfo{volume}{6}, \bibinfo{number}{CSCW2}
  (\bibinfo{year}{2022}), \bibinfo{pages}{1--23}.
\newblock


\bibitem[Cer et~al\mbox{.}(2017)]%
        {Cer2017SemEval2017T1}
\bibfield{author}{\bibinfo{person}{Daniel~Matthew Cer},
  \bibinfo{person}{Mona~T. Diab}, \bibinfo{person}{Eneko Agirre},
  \bibinfo{person}{I{\~n}igo Lopez-Gazpio}, {and} \bibinfo{person}{Lucia
  Specia}.} \bibinfo{year}{2017}\natexlab{}.
\newblock \showarticletitle{SemEval-2017 Task 1: Semantic Textual Similarity
  Multilingual and Crosslingual Focused Evaluation}. In
  \bibinfo{booktitle}{\emph{International Workshop on Semantic Evaluation}}.
\newblock
\urldef\tempurl%
\url{https://api.semanticscholar.org/CorpusID:4421747}
\showURL{%
\tempurl}


\bibitem[Chandrasekaran and Mago(2020a)]%
        {Chandrasekaran2020EvolutionOS}
\bibfield{author}{\bibinfo{person}{Dhivya Chandrasekaran} {and}
  \bibinfo{person}{Vijay Mago}.} \bibinfo{year}{2020}\natexlab{a}.
\newblock \showarticletitle{Evolution of Semantic Similarity—A Survey}.
\newblock \bibinfo{journal}{\emph{ACM Computing Surveys (CSUR)}}
  \bibinfo{volume}{54} (\bibinfo{year}{2020}), \bibinfo{pages}{1 -- 37}.
\newblock
\urldef\tempurl%
\url{https://api.semanticscholar.org/CorpusID:216641688}
\showURL{%
\tempurl}


\bibitem[Chandrasekaran and Mago(2020b)]%
        {Chandrasekaran2020ComparativeAO}
\bibfield{author}{\bibinfo{person}{Dhivya Chandrasekaran} {and}
  \bibinfo{person}{Vijay~K. Mago}.} \bibinfo{year}{2020}\natexlab{b}.
\newblock \showarticletitle{Comparative analysis of word embeddings in
  assessing semantic similarity of complex sentences}.
\newblock \bibinfo{journal}{\emph{IEEE Access}}  \bibinfo{volume}{PP}
  (\bibinfo{year}{2020}), \bibinfo{pages}{1--1}.
\newblock
\urldef\tempurl%
\url{https://api.semanticscholar.org/CorpusID:235795286}
\showURL{%
\tempurl}


\bibitem[Chandrasekharan et~al\mbox{.}(2017)]%
        {Chandrasekharan2017YouCS}
\bibfield{author}{\bibinfo{person}{Eshwar Chandrasekharan},
  \bibinfo{person}{Umashanthi Pavalanathan}, \bibinfo{person}{Anirudh
  Srinivasan}, \bibinfo{person}{Adam Glynn}, \bibinfo{person}{Jacob
  Eisenstein}, {and} \bibinfo{person}{Eric Gilbert}.}
  \bibinfo{year}{2017}\natexlab{}.
\newblock \showarticletitle{You Can't Stay Here: The Efficacy of Reddit's 2015
  Ban Examined Through Hate Speech}.
\newblock \bibinfo{journal}{\emph{Proc. ACM Hum. Comput. Interact.}}
  \bibinfo{volume}{1} (\bibinfo{year}{2017}), \bibinfo{pages}{31:1--31:22}.
\newblock
\urldef\tempurl%
\url{https://api.semanticscholar.org/CorpusID:221315649}
\showURL{%
\tempurl}


\bibitem[Chang et~al\mbox{.}(2023)]%
        {Chang2023CiteSeeAC}
\bibfield{author}{\bibinfo{person}{Joseph~Chee Chang}, \bibinfo{person}{Amy~X.
  Zhang}, \bibinfo{person}{Jonathan Bragg}, \bibinfo{person}{Andrew Head},
  \bibinfo{person}{Kyle Lo}, \bibinfo{person}{Doug Downey}, {and}
  \bibinfo{person}{Daniel~S. Weld}.} \bibinfo{year}{2023}\natexlab{}.
\newblock \showarticletitle{CiteSee: Augmenting Citations in Scientific Papers
  with Persistent and Personalized Historical Context}.
\newblock \bibinfo{journal}{\emph{Proceedings of the 2023 CHI Conference on
  Human Factors in Computing Systems}} (\bibinfo{year}{2023}).
\newblock
\urldef\tempurl%
\url{https://api.semanticscholar.org/CorpusID:256868353}
\showURL{%
\tempurl}


\bibitem[Chen et~al\mbox{.}(2023a)]%
        {chen2023meetscript}
\bibfield{author}{\bibinfo{person}{Xinyue Chen}, \bibinfo{person}{Shuo Li},
  \bibinfo{person}{Shipeng Liu}, \bibinfo{person}{Robin Fowler}, {and}
  \bibinfo{person}{Xu Wang}.} \bibinfo{year}{2023}\natexlab{a}.
\newblock \showarticletitle{Meetscript: designing transcript-based interactions
  to support active participation in group video meetings}.
\newblock \bibinfo{journal}{\emph{Proceedings of the ACM on Human-Computer
  Interaction}} \bibinfo{volume}{7}, \bibinfo{number}{CSCW2}
  (\bibinfo{year}{2023}), \bibinfo{pages}{1--32}.
\newblock


\bibitem[Chen et~al\mbox{.}(2023b)]%
        {chen2023marvista}
\bibfield{author}{\bibinfo{person}{Xiang~“Anthony” Chen},
  \bibinfo{person}{Chien-Sheng Wu}, \bibinfo{person}{Lidiya Murakhovs’~ka},
  \bibinfo{person}{Philippe Laban}, \bibinfo{person}{Tong Niu},
  \bibinfo{person}{Wenhao Liu}, {and} \bibinfo{person}{Caiming Xiong}.}
  \bibinfo{year}{2023}\natexlab{b}.
\newblock \showarticletitle{Marvista: exploring the design of a human-AI
  collaborative news reading tool}.
\newblock \bibinfo{journal}{\emph{ACM Transactions on Computer-Human
  Interaction}} \bibinfo{volume}{30}, \bibinfo{number}{6}
  (\bibinfo{year}{2023}), \bibinfo{pages}{1--27}.
\newblock


\bibitem[Day(1980)]%
        {Day1980HowTW}
\bibfield{author}{\bibinfo{person}{Robert~A. Day}.}
  \bibinfo{year}{1980}\natexlab{}.
\newblock \showarticletitle{How to write and publish a scientific paper}.
\newblock
\urldef\tempurl%
\url{https://api.semanticscholar.org/CorpusID:56622110}
\showURL{%
\tempurl}


\bibitem[Devlin et~al\mbox{.}(2019)]%
        {Devlin2019BERTPO}
\bibfield{author}{\bibinfo{person}{Jacob Devlin}, \bibinfo{person}{Ming-Wei
  Chang}, \bibinfo{person}{Kenton Lee}, {and} \bibinfo{person}{Kristina
  Toutanova}.} \bibinfo{year}{2019}\natexlab{}.
\newblock \showarticletitle{BERT: Pre-training of Deep Bidirectional
  Transformers for Language Understanding}.
\newblock \bibinfo{journal}{\emph{ArXiv}}  \bibinfo{volume}{abs/1810.04805}
  (\bibinfo{year}{2019}).
\newblock
\urldef\tempurl%
\url{https://api.semanticscholar.org/CorpusID:52967399}
\showURL{%
\tempurl}


\bibitem[Gonz{\'a}lez-Carvajal and Garrido-Merch{\'a}n(2020)]%
        {gonzalez2020comparing}
\bibfield{author}{\bibinfo{person}{Santiago Gonz{\'a}lez-Carvajal} {and}
  \bibinfo{person}{Eduardo~C Garrido-Merch{\'a}n}.}
  \bibinfo{year}{2020}\natexlab{}.
\newblock \showarticletitle{Comparing BERT against traditional machine learning
  text classification}.
\newblock \bibinfo{journal}{\emph{arXiv preprint arXiv:2005.13012}}
  (\bibinfo{year}{2020}).
\newblock


\bibitem[Green and Chen(2019)]%
        {green2019principles}
\bibfield{author}{\bibinfo{person}{Ben Green} {and} \bibinfo{person}{Yiling
  Chen}.} \bibinfo{year}{2019}\natexlab{}.
\newblock \showarticletitle{The principles and limits of algorithm-in-the-loop
  decision making}.
\newblock \bibinfo{journal}{\emph{Proceedings of the ACM on Human-Computer
  Interaction}} \bibinfo{volume}{3}, \bibinfo{number}{CSCW}
  (\bibinfo{year}{2019}), \bibinfo{pages}{1--24}.
\newblock


\bibitem[Gr{\"o}nroos and Voima(2013)]%
        {gronroos2013critical}
\bibfield{author}{\bibinfo{person}{Christian Gr{\"o}nroos} {and}
  \bibinfo{person}{P{\"a}ivi Voima}.} \bibinfo{year}{2013}\natexlab{}.
\newblock \showarticletitle{Critical service logic: making sense of value
  creation and co-creation}.
\newblock \bibinfo{journal}{\emph{Journal of the academy of marketing science}}
   \bibinfo{volume}{41} (\bibinfo{year}{2013}), \bibinfo{pages}{133--150}.
\newblock


\bibitem[Grossman et~al\mbox{.}(2015)]%
        {Grossman2015YourPI}
\bibfield{author}{\bibinfo{person}{Tovi Grossman}, \bibinfo{person}{Fanny
  Chevalier}, {and} \bibinfo{person}{Rubaiat~Habib Kazi}.}
  \bibinfo{year}{2015}\natexlab{}.
\newblock \showarticletitle{Your Paper is Dead!: Bringing Life to Research
  Articles with Animated Figures}.
\newblock \bibinfo{journal}{\emph{Proceedings of the 33rd Annual ACM Conference
  Extended Abstracts on Human Factors in Computing Systems}}
  (\bibinfo{year}{2015}).
\newblock
\urldef\tempurl%
\url{https://api.semanticscholar.org/CorpusID:14312410}
\showURL{%
\tempurl}


\bibitem[Gu et~al\mbox{.}(2021)]%
        {gu2021lessons}
\bibfield{author}{\bibinfo{person}{Hongyan Gu}, \bibinfo{person}{Jingbin
  Huang}, \bibinfo{person}{Lauren Hung}, {and} \bibinfo{person}{Xiang'Anthony'
  Chen}.} \bibinfo{year}{2021}\natexlab{}.
\newblock \showarticletitle{Lessons learned from designing an AI-enabled
  diagnosis tool for pathologists}.
\newblock \bibinfo{journal}{\emph{Proceedings of the ACM on Human-computer
  Interaction}} \bibinfo{volume}{5}, \bibinfo{number}{CSCW1}
  (\bibinfo{year}{2021}), \bibinfo{pages}{1--25}.
\newblock


\bibitem[Happell(2011)]%
        {happell2011responding}
\bibfield{author}{\bibinfo{person}{Brenda Happell}.}
  \bibinfo{year}{2011}\natexlab{}.
\newblock \showarticletitle{Responding to reviewers’ comments as part of
  writing for publication}.
\newblock \bibinfo{journal}{\emph{Nurse Researcher}} \bibinfo{volume}{18},
  \bibinfo{number}{4} (\bibinfo{year}{2011}).
\newblock


\bibitem[Head et~al\mbox{.}(2020)]%
        {Head2020AugmentingSP}
\bibfield{author}{\bibinfo{person}{Andrew Head}, \bibinfo{person}{Kyle Lo},
  \bibinfo{person}{Dongyeop Kang}, \bibinfo{person}{Raymond Fok},
  \bibinfo{person}{Sam Skjonsberg}, \bibinfo{person}{Daniel~S. Weld}, {and}
  \bibinfo{person}{Marti~A. Hearst}.} \bibinfo{year}{2020}\natexlab{}.
\newblock \showarticletitle{Augmenting Scientific Papers with Just-in-Time,
  Position-Sensitive Definitions of Terms and Symbols}.
\newblock \bibinfo{journal}{\emph{Proceedings of the 2021 CHI Conference on
  Human Factors in Computing Systems}} (\bibinfo{year}{2020}).
\newblock
\urldef\tempurl%
\url{https://api.semanticscholar.org/CorpusID:222066998}
\showURL{%
\tempurl}


\bibitem[Hidouri et~al\mbox{.}(2024)]%
        {hidouri2024key}
\bibfield{author}{\bibinfo{person}{Saida Hidouri}, \bibinfo{person}{Hela
  Kamoun}, \bibinfo{person}{Sana Salah}, \bibinfo{person}{Anis Jellad}, {and}
  \bibinfo{person}{Helmi~Ben Saad}.} \bibinfo{year}{2024}\natexlab{}.
\newblock \showarticletitle{Key Guidelines for Responding to Reviewers}.
\newblock \bibinfo{journal}{\emph{F1000Research}}  \bibinfo{volume}{13}
  (\bibinfo{year}{2024}).
\newblock


\bibitem[Hinckley(2015)]%
        {hinckley2015so}
\bibfield{author}{\bibinfo{person}{Ken Hinckley}.}
  \bibinfo{year}{2015}\natexlab{}.
\newblock \showarticletitle{So you’re a program committee member now: On
  excellence in reviews and meta-reviews and championing submitted work that
  has merit}.
\newblock  (\bibinfo{year}{2015}).
\newblock


\bibitem[Jansen et~al\mbox{.}(2016)]%
        {Jansen2016WhatDA}
\bibfield{author}{\bibinfo{person}{Yvonne Jansen}, \bibinfo{person}{Kasper
  Hornb{\ae}k}, {and} \bibinfo{person}{Pierre Dragicevic}.}
  \bibinfo{year}{2016}\natexlab{}.
\newblock \showarticletitle{What Did Authors Value in the CHI'16 Reviews They
  Received?}
\newblock \bibinfo{journal}{\emph{Proceedings of the 2016 CHI Conference
  Extended Abstracts on Human Factors in Computing Systems}}
  (\bibinfo{year}{2016}).
\newblock


\bibitem[Kelly et~al\mbox{.}(2014)]%
        {kelly2014peer}
\bibfield{author}{\bibinfo{person}{Jacalyn Kelly}, \bibinfo{person}{Tara
  Sadeghieh}, {and} \bibinfo{person}{Khosrow Adeli}.}
  \bibinfo{year}{2014}\natexlab{}.
\newblock \showarticletitle{Peer review in scientific publications: benefits,
  critiques, \& a survival guide}.
\newblock \bibinfo{journal}{\emph{Ejifcc}} \bibinfo{volume}{25},
  \bibinfo{number}{3} (\bibinfo{year}{2014}), \bibinfo{pages}{227}.
\newblock


\bibitem[Kim et~al\mbox{.}(2022)]%
        {Kim2022StyletteST}
\bibfield{author}{\bibinfo{person}{Tae~Soo Kim}, \bibinfo{person}{Yoonseo
  Choi}, \bibinfo{person}{Daeun Choi}, {and} \bibinfo{person}{Juho Kim}.}
  \bibinfo{year}{2022}\natexlab{}.
\newblock \showarticletitle{Stylette: Styling the Web with Natural Language}.
\newblock \bibinfo{journal}{\emph{Proceedings of the 2022 CHI Conference on
  Human Factors in Computing Systems}} (\bibinfo{year}{2022}).
\newblock
\urldef\tempurl%
\url{https://api.semanticscholar.org/CorpusID:248378697}
\showURL{%
\tempurl}


\bibitem[Kim et~al\mbox{.}(2016)]%
        {Kim2016SimpleScienceLS}
\bibfield{author}{\bibinfo{person}{Yea-Seul Kim}, \bibinfo{person}{Jessica~R.
  Hullman}, \bibinfo{person}{Matthew Burgess}, {and} \bibinfo{person}{Eytan
  Adar}.} \bibinfo{year}{2016}\natexlab{}.
\newblock \showarticletitle{SimpleScience: Lexical Simplification of Scientific
  Terminology}. In \bibinfo{booktitle}{\emph{Conference on Empirical Methods in
  Natural Language Processing}}.
\newblock
\urldef\tempurl%
\url{https://api.semanticscholar.org/CorpusID:8784583}
\showURL{%
\tempurl}


\bibitem[Kintsch and Welsch(1991)]%
        {Kintsch1991TheCM}
\bibfield{author}{\bibinfo{person}{Walter Kintsch} {and}
  \bibinfo{person}{David~M. Welsch}.} \bibinfo{year}{1991}\natexlab{}.
\newblock \showarticletitle{The construction-integration model: A framework for
  studying memory for text.}
\newblock
\urldef\tempurl%
\url{https://api.semanticscholar.org/CorpusID:69884604}
\showURL{%
\tempurl}


\bibitem[Kirk(2013)]%
        {kirk2013experimental}
\bibfield{author}{\bibinfo{person}{R.~E. Kirk}.}
  \bibinfo{year}{2013}\natexlab{}.
\newblock \bibinfo{booktitle}{\emph{Experimental Design: Procedures for the
  Behavioral Sciences}}.
\newblock \bibinfo{publisher}{SAGE Publications, Inc.}
\newblock
\href{https://doi.org/10.4135/9781483384733}{doi:\nolinkurl{10.4135/9781483384733}}


\bibitem[Koshorek et~al\mbox{.}(2018)]%
        {Koshorek2018TextSA}
\bibfield{author}{\bibinfo{person}{Omri Koshorek}, \bibinfo{person}{Adir
  Cohen}, \bibinfo{person}{Noam Mor}, \bibinfo{person}{Michael Rotman}, {and}
  \bibinfo{person}{Jonathan Berant}.} \bibinfo{year}{2018}\natexlab{}.
\newblock \showarticletitle{Text Segmentation as a Supervised Learning Task}.
  In \bibinfo{booktitle}{\emph{North American Chapter of the Association for
  Computational Linguistics}}.
\newblock
\urldef\tempurl%
\url{https://api.semanticscholar.org/CorpusID:4411469}
\showURL{%
\tempurl}


\bibitem[Lane and Pearson(1982)]%
        {Lane1982TheDO}
\bibfield{author}{\bibinfo{person}{David~M. Lane} {and}
  \bibinfo{person}{Deborah~A. Pearson}.} \bibinfo{year}{1982}\natexlab{}.
\newblock \showarticletitle{The Development of Selective Attention.}
\newblock \bibinfo{journal}{\emph{Merrill-palmer Quarterly}}
  \bibinfo{volume}{28} (\bibinfo{year}{1982}).
\newblock
\urldef\tempurl%
\url{https://api.semanticscholar.org/CorpusID:140989080}
\showURL{%
\tempurl}


\bibitem[Lewis et~al\mbox{.}(2019)]%
        {Lewis2019BARTDS}
\bibfield{author}{\bibinfo{person}{Mike Lewis}, \bibinfo{person}{Yinhan Liu},
  \bibinfo{person}{Naman Goyal}, \bibinfo{person}{Marjan Ghazvininejad},
  \bibinfo{person}{Abdel rahman Mohamed}, \bibinfo{person}{Omer Levy},
  \bibinfo{person}{Veselin Stoyanov}, {and} \bibinfo{person}{Luke
  Zettlemoyer}.} \bibinfo{year}{2019}\natexlab{}.
\newblock \showarticletitle{BART: Denoising Sequence-to-Sequence Pre-training
  for Natural Language Generation, Translation, and Comprehension}. In
  \bibinfo{booktitle}{\emph{Annual Meeting of the Association for Computational
  Linguistics}}.
\newblock
\urldef\tempurl%
\url{https://api.semanticscholar.org/CorpusID:204960716}
\showURL{%
\tempurl}


\bibitem[Li et~al\mbox{.}(2022)]%
        {Li2022NeuralTS}
\bibfield{author}{\bibinfo{person}{J. Li}, \bibinfo{person}{Billy Chiu},
  \bibinfo{person}{Shuo Shang}, {and} \bibinfo{person}{Ling Shao}.}
  \bibinfo{year}{2022}\natexlab{}.
\newblock \showarticletitle{Neural Text Segmentation and its Application to
  Sentiment Analysis}.
\newblock \bibinfo{journal}{\emph{IEEE Transactions on Knowledge and Data
  Engineering}}  \bibinfo{volume}{34} (\bibinfo{year}{2022}),
  \bibinfo{pages}{828--842}.
\newblock
\urldef\tempurl%
\url{https://api.semanticscholar.org/CorpusID:216430804}
\showURL{%
\tempurl}


\bibitem[Liu et~al\mbox{.}(2021)]%
        {liu2021understanding}
\bibfield{author}{\bibinfo{person}{Han Liu}, \bibinfo{person}{Vivian Lai},
  {and} \bibinfo{person}{Chenhao Tan}.} \bibinfo{year}{2021}\natexlab{}.
\newblock \showarticletitle{Understanding the effect of out-of-distribution
  examples and interactive explanations on human-ai decision making}.
\newblock \bibinfo{journal}{\emph{Proceedings of the ACM on Human-Computer
  Interaction}} \bibinfo{volume}{5}, \bibinfo{number}{CSCW2}
  (\bibinfo{year}{2021}), \bibinfo{pages}{1--45}.
\newblock


\bibitem[Lo et~al\mbox{.}(2023)]%
        {lo2023semantic}
\bibfield{author}{\bibinfo{person}{Kyle Lo}, \bibinfo{person}{Joseph~Chee
  Chang}, \bibinfo{person}{Andrew Head}, \bibinfo{person}{Jonathan Bragg},
  \bibinfo{person}{Amy~X Zhang}, \bibinfo{person}{Cassidy Trier},
  \bibinfo{person}{Chloe Anastasiades}, \bibinfo{person}{Tal August},
  \bibinfo{person}{Russell Authur}, \bibinfo{person}{Danielle Bragg},
  {et~al\mbox{.}}} \bibinfo{year}{2023}\natexlab{}.
\newblock \showarticletitle{The semantic reader project: Augmenting scholarly
  documents through ai-powered interactive reading interfaces}.
\newblock \bibinfo{journal}{\emph{arXiv preprint arXiv:2303.14334}}
  (\bibinfo{year}{2023}).
\newblock


\bibitem[Lucero(2015)]%
        {Lucero2015UsingAD}
\bibfield{author}{\bibinfo{person}{Andr{\'e}s Lucero}.}
  \bibinfo{year}{2015}\natexlab{}.
\newblock \showarticletitle{Using Affinity Diagrams to Evaluate Interactive
  Prototypes}. In \bibinfo{booktitle}{\emph{IFIP TC13 International Conference
  on Human-Computer Interaction}}.
\newblock
\urldef\tempurl%
\url{https://api.semanticscholar.org/CorpusID:45566772}
\showURL{%
\tempurl}


\bibitem[Luria et~al\mbox{.}(2020)]%
        {luria2020social}
\bibfield{author}{\bibinfo{person}{Michal Luria}, \bibinfo{person}{Rebecca
  Zheng}, \bibinfo{person}{Bennett Huffman}, \bibinfo{person}{Shuangni Huang},
  \bibinfo{person}{John Zimmerman}, {and} \bibinfo{person}{Jodi Forlizzi}.}
  \bibinfo{year}{2020}\natexlab{}.
\newblock \showarticletitle{Social boundaries for personal agents in the
  interpersonal space of the home}. In \bibinfo{booktitle}{\emph{Proceedings of
  the 2020 CHI conference on human factors in computing systems}}.
  \bibinfo{pages}{1--12}.
\newblock


\bibitem[Mackeprang et~al\mbox{.}(2019)]%
        {mackeprang2019discovering}
\bibfield{author}{\bibinfo{person}{Maximilian Mackeprang},
  \bibinfo{person}{Claudia M{\"u}ller-Birn}, {and}
  \bibinfo{person}{Maximilian~Timo Stauss}.} \bibinfo{year}{2019}\natexlab{}.
\newblock \showarticletitle{Discovering the sweet spot of human-computer
  configurations: A case study in information extraction}.
\newblock \bibinfo{journal}{\emph{Proceedings of the ACM on Human-Computer
  Interaction}} \bibinfo{volume}{3}, \bibinfo{number}{CSCW}
  (\bibinfo{year}{2019}), \bibinfo{pages}{1--30}.
\newblock


\bibitem[Mann and Whitney(1947)]%
        {Mann1947OnAT}
\bibfield{author}{\bibinfo{person}{Henry~B. Mann} {and}
  \bibinfo{person}{Douglas~R. Whitney}.} \bibinfo{year}{1947}\natexlab{}.
\newblock \showarticletitle{On a Test of Whether one of Two Random Variables is
  Stochastically Larger than the Other}.
\newblock \bibinfo{journal}{\emph{Annals of Mathematical Statistics}}
  \bibinfo{volume}{18} (\bibinfo{year}{1947}), \bibinfo{pages}{50--60}.
\newblock
\urldef\tempurl%
\url{https://api.semanticscholar.org/CorpusID:14328772}
\showURL{%
\tempurl}


\bibitem[Miller(1996)]%
        {miller1996multiple}
\bibfield{author}{\bibinfo{person}{Holmes Miller}.}
  \bibinfo{year}{1996}\natexlab{}.
\newblock \showarticletitle{The multiple dimensions of information quality}.
\newblock \bibinfo{journal}{\emph{Information Systems Management}}
  \bibinfo{volume}{13}, \bibinfo{number}{2} (\bibinfo{year}{1996}),
  \bibinfo{pages}{79--82}.
\newblock


\bibitem[{Nature}(2024)]%
        {nature_peer_review}
\bibfield{author}{\bibinfo{person}{{Nature}}.} \bibinfo{year}{2024}\natexlab{}.
\newblock \showarticletitle{{Peer review policies | Humanities and Social
  Sciences Communications}}.
\newblock \bibinfo{journal}{\emph{{Humanities and Social Sciences
  Communications}}} (\bibinfo{year}{2024}).
\newblock
\urldef\tempurl%
\url{https://www.nature.com/palcomms/journal-policies}
\showURL{%
\tempurl}


\bibitem[Nickerson(1998)]%
        {Nickerson1998ConfirmationBA}
\bibfield{author}{\bibinfo{person}{Raymond~S. Nickerson}.}
  \bibinfo{year}{1998}\natexlab{}.
\newblock \showarticletitle{Confirmation Bias: A Ubiquitous Phenomenon in Many
  Guises}.
\newblock \bibinfo{journal}{\emph{Review of General Psychology}}
  \bibinfo{volume}{2} (\bibinfo{year}{1998}), \bibinfo{pages}{175 -- 220}.
\newblock
\urldef\tempurl%
\url{https://api.semanticscholar.org/CorpusID:8508954}
\showURL{%
\tempurl}


\bibitem[Noble(2017)]%
        {Noble2017TenSR}
\bibfield{author}{\bibinfo{person}{William~Stafford Noble}.}
  \bibinfo{year}{2017}\natexlab{}.
\newblock \showarticletitle{Ten simple rules for writing a response to
  reviewers}.
\newblock \bibinfo{journal}{\emph{PLoS Computational Biology}}
  \bibinfo{volume}{13} (\bibinfo{year}{2017}).
\newblock
\urldef\tempurl%
\url{https://api.semanticscholar.org/CorpusID:2815851}
\showURL{%
\tempurl}


\bibitem[Pagel and Schubotz(2014)]%
        {Pagel2014MathematicalLP}
\bibfield{author}{\bibinfo{person}{Robert Pagel} {and} \bibinfo{person}{Moritz
  Schubotz}.} \bibinfo{year}{2014}\natexlab{}.
\newblock \showarticletitle{Mathematical Language Processing Project}.
\newblock \bibinfo{journal}{\emph{ArXiv}}  \bibinfo{volume}{abs/1407.0167}
  (\bibinfo{year}{2014}).
\newblock
\urldef\tempurl%
\url{https://api.semanticscholar.org/CorpusID:1723652}
\showURL{%
\tempurl}


\bibitem[Pang and Lee(2008)]%
        {Pang2008OpinionMA}
\bibfield{author}{\bibinfo{person}{Bo Pang} {and} \bibinfo{person}{Lillian
  Lee}.} \bibinfo{year}{2008}\natexlab{}.
\newblock \showarticletitle{Opinion Mining and Sentiment Analysis}.
\newblock \bibinfo{journal}{\emph{Found. Trends Inf. Retr.}}
  \bibinfo{volume}{2} (\bibinfo{year}{2008}), \bibinfo{pages}{1--135}.
\newblock


\bibitem[Park et~al\mbox{.}(2022)]%
        {park2022exploring}
\bibfield{author}{\bibinfo{person}{Soya Park}, \bibinfo{person}{Jonathan
  Bragg}, \bibinfo{person}{Michael Chang}, \bibinfo{person}{Kevin Larson},
  {and} \bibinfo{person}{Danielle Bragg}.} \bibinfo{year}{2022}\natexlab{}.
\newblock \showarticletitle{Exploring Team-Sourced Hyperlinks to Address
  Navigation Challenges for Low-Vision Readers of Scientific Papers}.
\newblock \bibinfo{journal}{\emph{Proceedings of the ACM on Human-Computer
  Interaction}} \bibinfo{volume}{6}, \bibinfo{number}{CSCW2}
  (\bibinfo{year}{2022}), \bibinfo{pages}{1--23}.
\newblock


\bibitem[Peng and Martens(2018)]%
        {peng2018requirements}
\bibfield{author}{\bibinfo{person}{Qiong Peng} {and}
  \bibinfo{person}{Jean~Bernard Martens}.} \bibinfo{year}{2018}\natexlab{}.
\newblock \showarticletitle{Requirements gathering for tools in support of
  storyboarding in user experience design}. In \bibinfo{booktitle}{\emph{32nd
  International BCS Human Computer Interaction Conference, HCI 2018}}. British
  Computer Society (BCS), \bibinfo{pages}{1--10}.
\newblock


\bibitem[Picarra et~al\mbox{.}(2015)]%
        {picarra2015open}
\bibfield{author}{\bibinfo{person}{Mafalda Picarra}, \bibinfo{person}{EKT
  Victoria~Tsoukala}, {and} \bibinfo{person}{Alma Swan}.}
  \bibinfo{year}{2015}\natexlab{}.
\newblock \showarticletitle{Open Access to scientific information: facilitating
  knowledge transfer and technological innovation from the academic to the
  private sector}.
\newblock \bibinfo{journal}{\emph{Recuperado de http://www. pasteur4oa.
  eu/sites/pasteur4oa/files/resource/Brief\_OA\% 20and\% 20knowledge\%
  20transfer\% 20to\% 20the\% 20private\% 20sector. pdf (consultado el 4 de
  marzo de 2018)}} (\bibinfo{year}{2015}).
\newblock


\bibitem[Pierce(2008)]%
        {pierce2008evaluating}
\bibfield{author}{\bibinfo{person}{Roger Pierce}.}
  \bibinfo{year}{2008}\natexlab{}.
\newblock \showarticletitle{Evaluating information: Validity, reliability,
  accuracy, triangulation}.
\newblock \bibinfo{journal}{\emph{Res Methods Polit}} (\bibinfo{year}{2008}),
  \bibinfo{pages}{79--99}.
\newblock


\bibitem[Radford and Narasimhan(2018)]%
        {Radford2018ImprovingLU}
\bibfield{author}{\bibinfo{person}{Alec Radford} {and} \bibinfo{person}{Karthik
  Narasimhan}.} \bibinfo{year}{2018}\natexlab{}.
\newblock \showarticletitle{Improving Language Understanding by Generative
  Pre-Training}.
\newblock
\urldef\tempurl%
\url{https://api.semanticscholar.org/CorpusID:49313245}
\showURL{%
\tempurl}


\bibitem[Randall({[n.\,d.]})]%
        {cscw_reviewing_for_cscw}
\bibfield{author}{\bibinfo{person}{Dave Randall}.}
  \bibinfo{year}{[n.\,d.]}\natexlab{}.
\newblock \bibinfo{title}{Reviewing for CSCW}.
\newblock
  \bibinfo{howpublished}{\url{https://cscw.acm.org/2016/volunteer/DaveRandallReviewingforCSCW.pdf}}.
\newblock
\newblock
\shownote{Accessed: 2024-06-29}.


\bibitem[Rodr{\'i}guez-Bravo et~al\mbox{.}(2017)]%
        {RodrguezBravo2017PeerRT}
\bibfield{author}{\bibinfo{person}{Blanca Rodr{\'i}guez-Bravo},
  \bibinfo{person}{David Nicholas}, \bibinfo{person}{Eti Herman},
  \bibinfo{person}{Ch{\'e}rifa Boukacem-Zeghmouri}, \bibinfo{person}{Anthony
  Watkinson}, \bibinfo{person}{Jie Xu}, \bibinfo{person}{Abdullah Abrizah},
  {and} \bibinfo{person}{Marzena Świgoń}.} \bibinfo{year}{2017}\natexlab{}.
\newblock \showarticletitle{Peer review: The experience and views of early
  career researchers}.
\newblock \bibinfo{journal}{\emph{Learned Publishing}}  \bibinfo{volume}{30}
  (\bibinfo{year}{2017}).
\newblock
\urldef\tempurl%
\url{https://api.semanticscholar.org/CorpusID:44145736}
\showURL{%
\tempurl}


\bibitem[Ross(2010)]%
        {Ross2010ThePO}
\bibfield{author}{\bibinfo{person}{Brian~H. Ross}.}
  \bibinfo{year}{2010}\natexlab{}.
\newblock \showarticletitle{The Psychology of Learning and Motivation: Advances
  in Research and Theory}.
\newblock
\urldef\tempurl%
\url{https://api.semanticscholar.org/CorpusID:17140139}
\showURL{%
\tempurl}


\bibitem[Schemmer et~al\mbox{.}(2022)]%
        {Schemmer2022OnTI}
\bibfield{author}{\bibinfo{person}{Maximilian Schemmer},
  \bibinfo{person}{Niklas K{\"u}hl}, \bibinfo{person}{Carina Benz}, {and}
  \bibinfo{person}{Gerhard Satzger}.} \bibinfo{year}{2022}\natexlab{}.
\newblock \showarticletitle{On the Influence of Explainable AI on Automation
  Bias}.
\newblock \bibinfo{journal}{\emph{ArXiv}}  \bibinfo{volume}{abs/2204.08859}
  (\bibinfo{year}{2022}).
\newblock
\urldef\tempurl%
\url{https://api.semanticscholar.org/CorpusID:248239725}
\showURL{%
\tempurl}


\bibitem[Sharma and Dey(2012)]%
        {Sharma2012ACS}
\bibfield{author}{\bibinfo{person}{Anuj Sharma} {and}
  \bibinfo{person}{Shubhamoy Dey}.} \bibinfo{year}{2012}\natexlab{}.
\newblock \showarticletitle{A comparative study of feature selection and
  machine learning techniques for sentiment analysis}. In
  \bibinfo{booktitle}{\emph{Research in Adaptive and Convergent Systems}}.
\newblock


\bibitem[Shneiderman(2020)]%
        {shneiderman2020design}
\bibfield{author}{\bibinfo{person}{Ben Shneiderman}.}
  \bibinfo{year}{2020}\natexlab{}.
\newblock \showarticletitle{Design lessons from AI’s two grand goals: human
  emulation and useful applications}.
\newblock \bibinfo{journal}{\emph{IEEE Transactions on Technology and Society}}
  \bibinfo{volume}{1}, \bibinfo{number}{2} (\bibinfo{year}{2020}),
  \bibinfo{pages}{73--82}.
\newblock


\bibitem[Shute(2008)]%
        {shute2008focus}
\bibfield{author}{\bibinfo{person}{Valerie~J Shute}.}
  \bibinfo{year}{2008}\natexlab{}.
\newblock \showarticletitle{Focus on formative feedback}.
\newblock \bibinfo{journal}{\emph{Review of educational research}}
  \bibinfo{volume}{78}, \bibinfo{number}{1} (\bibinfo{year}{2008}),
  \bibinfo{pages}{153--189}.
\newblock


\bibitem[Squazzoni et~al\mbox{.}(2021)]%
        {squazzoni2021peer}
\bibfield{author}{\bibinfo{person}{Flaminio Squazzoni},
  \bibinfo{person}{Giangiacomo Bravo}, \bibinfo{person}{Mike Farjam},
  \bibinfo{person}{Ana Marusic}, \bibinfo{person}{Bahar Mehmani},
  \bibinfo{person}{Michael Willis}, \bibinfo{person}{Aliaksandr Birukou},
  \bibinfo{person}{Pierpaolo Dondio}, {and} \bibinfo{person}{Francisco
  Grimaldo}.} \bibinfo{year}{2021}\natexlab{}.
\newblock \showarticletitle{Peer review and gender bias: A study on 145
  scholarly journals}.
\newblock \bibinfo{journal}{\emph{Science advances}} \bibinfo{volume}{7},
  \bibinfo{number}{2} (\bibinfo{year}{2021}), \bibinfo{pages}{eabd0299}.
\newblock


\bibitem[Stanovich(1980)]%
        {stanovich1980toward}
\bibfield{author}{\bibinfo{person}{Keith~E Stanovich}.}
  \bibinfo{year}{1980}\natexlab{}.
\newblock \showarticletitle{Toward an interactive-compensatory model of
  individual differences in the development of reading fluency}.
\newblock \bibinfo{journal}{\emph{Reading research quarterly}}
  (\bibinfo{year}{1980}), \bibinfo{pages}{32--71}.
\newblock


\bibitem[Stelmakh et~al\mbox{.}(2019)]%
        {stelmakh2019peerreview4all}
\bibfield{author}{\bibinfo{person}{Ivan Stelmakh}, \bibinfo{person}{Nihar~B
  Shah}, {and} \bibinfo{person}{Aarti Singh}.} \bibinfo{year}{2019}\natexlab{}.
\newblock \showarticletitle{PeerReview4All: Fair and accurate reviewer
  assignment in peer review}. In \bibinfo{booktitle}{\emph{Algorithmic Learning
  Theory}}. PMLR, \bibinfo{pages}{828--856}.
\newblock


\bibitem[Stelmakh et~al\mbox{.}(2020)]%
        {Stelmakh2020PriorAP}
\bibfield{author}{\bibinfo{person}{Ivan Stelmakh}, \bibinfo{person}{Nihar~B.
  Shah}, \bibinfo{person}{Aarti Singh}, {and} \bibinfo{person}{Hal Daum'e}.}
  \bibinfo{year}{2020}\natexlab{}.
\newblock \showarticletitle{Prior and Prejudice: The Novice Reviewers' Bias
  against Resubmissions in Conference Peer Review}.
\newblock \bibinfo{journal}{\emph{ArXiv}}  \bibinfo{volume}{abs/2011.14646}
  (\bibinfo{year}{2020}).
\newblock
\urldef\tempurl%
\url{https://api.semanticscholar.org/CorpusID:227227578}
\showURL{%
\tempurl}


\bibitem[Subakti et~al\mbox{.}(2022)]%
        {subakti2022performance}
\bibfield{author}{\bibinfo{person}{Alvin Subakti}, \bibinfo{person}{Hendri
  Murfi}, {and} \bibinfo{person}{Nora Hariadi}.}
  \bibinfo{year}{2022}\natexlab{}.
\newblock \showarticletitle{The performance of BERT as data representation of
  text clustering}.
\newblock \bibinfo{journal}{\emph{Journal of big Data}} \bibinfo{volume}{9},
  \bibinfo{number}{1} (\bibinfo{year}{2022}), \bibinfo{pages}{15}.
\newblock


\bibitem[Tashman and Edwards(2011a)]%
        {tashman2011active}
\bibfield{author}{\bibinfo{person}{Craig~S Tashman} {and}
  \bibinfo{person}{W~Keith Edwards}.} \bibinfo{year}{2011}\natexlab{a}.
\newblock \showarticletitle{Active reading and its discontents: the situations,
  problems and ideas of readers}. In \bibinfo{booktitle}{\emph{Proceedings of
  the SIGCHI Conference on Human Factors in Computing Systems}}.
  \bibinfo{pages}{2927--2936}.
\newblock


\bibitem[Tashman and Edwards(2011b)]%
        {Tashman2011LiquidTextAF}
\bibfield{author}{\bibinfo{person}{Craig~S. Tashman} {and}
  \bibinfo{person}{W.~Keith Edwards}.} \bibinfo{year}{2011}\natexlab{b}.
\newblock \showarticletitle{LiquidText: a flexible, multitouch environment to
  support active reading}.
\newblock \bibinfo{journal}{\emph{Proceedings of the SIGCHI Conference on Human
  Factors in Computing Systems}} (\bibinfo{year}{2011}).
\newblock
\urldef\tempurl%
\url{https://api.semanticscholar.org/CorpusID:12836641}
\showURL{%
\tempurl}


\bibitem[Tayeh and Signer(2014)]%
        {Tayeh2014OpenCL}
\bibfield{author}{\bibinfo{person}{Ahmed A.~O. Tayeh} {and} \bibinfo{person}{B.
  Signer}.} \bibinfo{year}{2014}\natexlab{}.
\newblock \showarticletitle{Open Cross-Document Linking and Browsing Based on a
  Visual Plug-in Architecture}. In \bibinfo{booktitle}{\emph{WISE}}.
\newblock
\urldef\tempurl%
\url{https://api.semanticscholar.org/CorpusID:1980030}
\showURL{%
\tempurl}


\bibitem[Tayeh and Signer(2015)]%
        {Tayeh2015ADE}
\bibfield{author}{\bibinfo{person}{Ahmed A.~O. Tayeh} {and} \bibinfo{person}{B.
  Signer}.} \bibinfo{year}{2015}\natexlab{}.
\newblock \showarticletitle{A Dynamically Extensible Open Cross-Document Link
  Service}. In \bibinfo{booktitle}{\emph{WISE}}.
\newblock
\urldef\tempurl%
\url{https://api.semanticscholar.org/CorpusID:8084189}
\showURL{%
\tempurl}


\bibitem[{The British Academy}(2007)]%
        {british_academy_peer_review}
\bibfield{author}{\bibinfo{person}{{The British Academy}}.}
  \bibinfo{year}{2007}\natexlab{}.
\newblock \bibinfo{booktitle}{\emph{{Peer Review: the challenges for the
  humanities and social sciences}}}.
\newblock \bibinfo{type}{{T}echnical {R}eport}. \bibinfo{institution}{{The
  British Academy}}.
\newblock
\urldef\tempurl%
\url{https://www.thebritishacademy.ac.uk/documents/197/Peer-review-challenges-for-humanities-social-sciences.pdf}
\showURL{%
\tempurl}


\bibitem[Truong et~al\mbox{.}(2006)]%
        {truong2006storyboarding}
\bibfield{author}{\bibinfo{person}{Khai~N Truong}, \bibinfo{person}{Gillian~R
  Hayes}, {and} \bibinfo{person}{Gregory~D Abowd}.}
  \bibinfo{year}{2006}\natexlab{}.
\newblock \showarticletitle{Storyboarding: an empirical determination of best
  practices and effective guidelines}. In \bibinfo{booktitle}{\emph{Proceedings
  of the 6th conference on Designing Interactive systems}}.
  \bibinfo{pages}{12--21}.
\newblock


\bibitem[Tschandl et~al\mbox{.}(2020)]%
        {tschandl2020human}
\bibfield{author}{\bibinfo{person}{Philipp Tschandl},
  \bibinfo{person}{Christoph Rinner}, \bibinfo{person}{Zoe Apalla},
  \bibinfo{person}{Giuseppe Argenziano}, \bibinfo{person}{Noel Codella},
  \bibinfo{person}{Allan Halpern}, \bibinfo{person}{Monika Janda},
  \bibinfo{person}{Aimilios Lallas}, \bibinfo{person}{Caterina Longo},
  \bibinfo{person}{Josep Malvehy}, {et~al\mbox{.}}}
  \bibinfo{year}{2020}\natexlab{}.
\newblock \showarticletitle{Human--computer collaboration for skin cancer
  recognition}.
\newblock \bibinfo{journal}{\emph{Nature Medicine}} \bibinfo{volume}{26},
  \bibinfo{number}{8} (\bibinfo{year}{2020}), \bibinfo{pages}{1229--1234}.
\newblock


\bibitem[Turney(2002)]%
        {Turney2002ThumbsUO}
\bibfield{author}{\bibinfo{person}{Peter~D. Turney}.}
  \bibinfo{year}{2002}\natexlab{}.
\newblock \showarticletitle{Thumbs Up or Thumbs Down? Semantic Orientation
  Applied to Unsupervised Classification of Reviews}. In
  \bibinfo{booktitle}{\emph{Annual Meeting of the Association for Computational
  Linguistics}}.
\newblock


\bibitem[Tversky and Kahneman(1974)]%
        {Tversky1974JudgmentUU}
\bibfield{author}{\bibinfo{person}{Amos Tversky} {and} \bibinfo{person}{Daniel
  Kahneman}.} \bibinfo{year}{1974}\natexlab{}.
\newblock \showarticletitle{Judgment under Uncertainty: Heuristics and Biases}.
\newblock \bibinfo{journal}{\emph{Science}}  \bibinfo{volume}{185}
  (\bibinfo{year}{1974}), \bibinfo{pages}{1124 -- 1131}.
\newblock
\urldef\tempurl%
\url{https://api.semanticscholar.org/CorpusID:6196452}
\showURL{%
\tempurl}


\bibitem[Umar et~al\mbox{.}(2019)]%
        {Umar2019DetectionAA}
\bibfield{author}{\bibinfo{person}{Prasanna Umar}, \bibinfo{person}{Anna~Cinzia
  Squicciarini}, {and} \bibinfo{person}{Sarah~Michele Rajtmajer}.}
  \bibinfo{year}{2019}\natexlab{}.
\newblock \showarticletitle{Detection and Analysis of Self-Disclosure in Online
  News Commentaries}.
\newblock \bibinfo{journal}{\emph{The World Wide Web Conference}}
  (\bibinfo{year}{2019}).
\newblock
\urldef\tempurl%
\url{https://api.semanticscholar.org/CorpusID:86467407}
\showURL{%
\tempurl}


\bibitem[Vaccaro and Waldo(2019)]%
        {vaccaro2019effects}
\bibfield{author}{\bibinfo{person}{Michelle Vaccaro} {and} \bibinfo{person}{Jim
  Waldo}.} \bibinfo{year}{2019}\natexlab{}.
\newblock \showarticletitle{The effects of mixing machine learning and human
  judgment}.
\newblock \bibinfo{journal}{\emph{Commun. ACM}} \bibinfo{volume}{62},
  \bibinfo{number}{11} (\bibinfo{year}{2019}), \bibinfo{pages}{104--110}.
\newblock


\bibitem[Wang et~al\mbox{.}(2019)]%
        {wang2019human}
\bibfield{author}{\bibinfo{person}{Dakuo Wang}, \bibinfo{person}{Justin~D
  Weisz}, \bibinfo{person}{Michael Muller}, \bibinfo{person}{Parikshit Ram},
  \bibinfo{person}{Werner Geyer}, \bibinfo{person}{Casey Dugan},
  \bibinfo{person}{Yla Tausczik}, \bibinfo{person}{Horst Samulowitz}, {and}
  \bibinfo{person}{Alexander Gray}.} \bibinfo{year}{2019}\natexlab{}.
\newblock \showarticletitle{Human-AI collaboration in data science: Exploring
  data scientists' perceptions of automated AI}.
\newblock \bibinfo{journal}{\emph{Proceedings of the ACM on human-computer
  interaction}} \bibinfo{volume}{3}, \bibinfo{number}{CSCW}
  (\bibinfo{year}{2019}), \bibinfo{pages}{1--24}.
\newblock


\bibitem[Wang et~al\mbox{.}(2014)]%
        {Wang2014ReCloudSW}
\bibfield{author}{\bibinfo{person}{Ji Wang}, \bibinfo{person}{Jian Zhao},
  \bibinfo{person}{Sheng Guo}, \bibinfo{person}{Chris North}, {and}
  \bibinfo{person}{Naren Ramakrishnan}.} \bibinfo{year}{2014}\natexlab{}.
\newblock \showarticletitle{ReCloud: semantics-based word cloud visualization
  of user reviews}. In \bibinfo{booktitle}{\emph{Graphics Interface}}.
\newblock
\urldef\tempurl%
\url{https://api.semanticscholar.org/CorpusID:6038574}
\showURL{%
\tempurl}


\bibitem[Weber et~al\mbox{.}(2002)]%
        {Weber2002AuthorPO}
\bibfield{author}{\bibinfo{person}{Ellen~J Weber}, \bibinfo{person}{Patricia~P.
  Katz}, \bibinfo{person}{Joseph~F. Waeckerle}, {and}
  \bibinfo{person}{Michael~L. Callaham}.} \bibinfo{year}{2002}\natexlab{}.
\newblock \showarticletitle{Author perception of peer review: impact of review
  quality and acceptance on satisfaction.}
\newblock \bibinfo{journal}{\emph{JAMA}}  \bibinfo{volume}{287 21}
  (\bibinfo{year}{2002}), \bibinfo{pages}{2790--3}.
\newblock


\bibitem[Williams et~al\mbox{.}(2018)]%
        {N18-1101}
\bibfield{author}{\bibinfo{person}{Adina Williams}, \bibinfo{person}{Nikita
  Nangia}, {and} \bibinfo{person}{Samuel Bowman}.}
  \bibinfo{year}{2018}\natexlab{}.
\newblock \showarticletitle{A Broad-Coverage Challenge Corpus for Sentence
  Understanding through Inference}. In \bibinfo{booktitle}{\emph{Proceedings of
  the 2018 Conference of the North American Chapter of the Association for
  Computational Linguistics: Human Language Technologies, Volume 1 (Long
  Papers)}} (New Orleans, Louisiana). \bibinfo{publisher}{Association for
  Computational Linguistics}, \bibinfo{pages}{1112--1122}.
\newblock
\urldef\tempurl%
\url{http://aclweb.org/anthology/N18-1101}
\showURL{%
\tempurl}


\bibitem[Williams(2004)]%
        {williams2004reply}
\bibfield{author}{\bibinfo{person}{Hywel~C Williams}.}
  \bibinfo{year}{2004}\natexlab{}.
\newblock \showarticletitle{How to reply to referees' comments when submitting
  manuscripts for publication}.
\newblock \bibinfo{journal}{\emph{Journal of the American Academy of
  Dermatology}} \bibinfo{volume}{51}, \bibinfo{number}{1}
  (\bibinfo{year}{2004}), \bibinfo{pages}{79--83}.
\newblock


\bibitem[Wolters(1933)]%
        {Wolters1933RememberingAS}
\bibfield{author}{\bibinfo{person}{A.~W.~P. Wolters}.}
  \bibinfo{year}{1933}\natexlab{}.
\newblock \showarticletitle{Remembering: A Study in Experimental and Social
  Psychology . By F. C. Bartlett. (Cambridge University Press. 1932. Pp. x +
  317. Price 16s. net.)}.
\newblock \bibinfo{journal}{\emph{Philosophy}}  \bibinfo{volume}{8}
  (\bibinfo{year}{1933}), \bibinfo{pages}{374 -- 376}.
\newblock
\urldef\tempurl%
\url{https://api.semanticscholar.org/CorpusID:145357365}
\showURL{%
\tempurl}


\bibitem[Yatani et~al\mbox{.}(2011)]%
        {Yatani2011ReviewSA}
\bibfield{author}{\bibinfo{person}{Koji Yatani}, \bibinfo{person}{Michael
  Novati}, \bibinfo{person}{Andrew Trusty}, {and} \bibinfo{person}{Khai~Nhut
  Truong}.} \bibinfo{year}{2011}\natexlab{}.
\newblock \showarticletitle{Review spotlight: a user interface for summarizing
  user-generated reviews using adjective-noun word pairs}.
\newblock \bibinfo{journal}{\emph{Proceedings of the SIGCHI Conference on Human
  Factors in Computing Systems}} (\bibinfo{year}{2011}).
\newblock
\urldef\tempurl%
\url{https://api.semanticscholar.org/CorpusID:16393334}
\showURL{%
\tempurl}


\bibitem[Yuan et~al\mbox{.}(2023)]%
        {Yuan2023CriTrainerAA}
\bibfield{author}{\bibinfo{person}{Kangyu Yuan}, \bibinfo{person}{Hehai Lin},
  \bibinfo{person}{Shilei Cao}, \bibinfo{person}{Zhenhui Peng},
  \bibinfo{person}{Qingyu Guo}, {and} \bibinfo{person}{Xiaojuan Ma}.}
  \bibinfo{year}{2023}\natexlab{}.
\newblock \showarticletitle{CriTrainer: An Adaptive Training Tool for Critical
  Paper Reading}.
\newblock \bibinfo{journal}{\emph{Proceedings of the 36th Annual ACM Symposium
  on User Interface Software and Technology}} (\bibinfo{year}{2023}).
\newblock
\urldef\tempurl%
\url{https://api.semanticscholar.org/CorpusID:264350375}
\showURL{%
\tempurl}


\bibitem[Zhu et~al\mbox{.}(2022)]%
        {Zhu2022BiasAwareDF}
\bibfield{author}{\bibinfo{person}{Qian Zhu}, \bibinfo{person}{Leo Yu-Ho Lo},
  \bibinfo{person}{Meng Xia}, \bibinfo{person}{Zixin Chen}, {and}
  \bibinfo{person}{Xiaojuan Ma}.} \bibinfo{year}{2022}\natexlab{}.
\newblock \showarticletitle{Bias-Aware Design for Informed Decisions: Raising
  Awareness of Self-Selection Bias in User Ratings and Reviews}.
\newblock \bibinfo{journal}{\emph{Proceedings of the ACM on Human-Computer
  Interaction}}  \bibinfo{volume}{6} (\bibinfo{year}{2022}), \bibinfo{pages}{1
  -- 31}.
\newblock
\urldef\tempurl%
\url{https://api.semanticscholar.org/CorpusID:252367665}
\showURL{%
\tempurl}


\bibitem[Zou(2024)]%
        {zou2024chatgpt}
\bibfield{author}{\bibinfo{person}{James Zou}.}
  \bibinfo{year}{2024}\natexlab{}.
\newblock \showarticletitle{ChatGPT is transforming peer review—how can we
  use it responsibly?}
\newblock \bibinfo{journal}{\emph{Nature}} \bibinfo{volume}{635},
  \bibinfo{number}{8037} (\bibinfo{year}{2024}), \bibinfo{pages}{10--10}.
\newblock


\end{thebibliography}


\pagebreak
\appendix
\section{APPENDIX}

\subsection{Usage Scenario of the system}
\label{Usage Scenario}
\par Let's consider Emily, a graduate student specializing in the field of Human-Computer Interaction (HCI). Despite her relative inexperience as an author, Emily has embarked on the scholarly submission process as the first author twice. She has recently received valuable feedback from reviewers following her initial paper submission. Currently, Emily is diligently scrutinizing the reviewers' comments, with plans to engage in thorough revisions.

\par As she delves into the comments, her objective is to extract the essence of each point, align it with the relevant section of her manuscript, and subsequently proceed to the next set of comments. However, this seemingly straightforward process quickly evolves into a laborious task. Even though she possesses an intricate understanding of her manuscript, extracting specific feedback from the unstructured text provided by reviewers and then associating it with the corresponding content proves to be mentally taxing. Emily also encounters challenges when attempting to organize the comments into meaningful categories. These categories would serve a dual purpose: aiding her in comprehending the feedback and facilitating her revision planning. Initially, she classifies the comments based on their thematic content, distinguishing between feedback on writing aspects and those pertaining to technical elements like interface design. Nonetheless, Emily realizes that as she progresses through the process of understanding the reviews and strategizing her revisions, an additional need arises. She recognizes the importance of categorizing the comments based on the level of effort required to address each of them. This new aspect entails categorizing the comments by the workload they represent. Consequently, she finds herself revisiting her earlier categorizations and making adjustments, which adds extra effort and necessitates reorganization. Upon transitioning from the stage of comprehending the feedback to the actual process of revising her paper, Emily encounters another obstacle. The valuable insights and reflections she had during her initial analysis are not readily accessible. This situation prompts her to re-read and re-analyze the comments in order to grasp the fundamental issues before embarking on the revision phase.

\par Overwhelmed by the extensive review analysis process, Emily begins to seek a more efficient and less draining approach. She turns to \textit{ReviseMate} for assistance. After uploading her paper and the accompanying review text, Emily discovers that the tool offers three distinct methods for extracting comments from raw reviews and generating corresponding comment cards: \textit{manual} (\autoref{fig:Preprocessing}-\darkcircled{a2}), \textit{semi-automatic} (\autoref{fig:Preprocessing}-\darkcircled{b3}), and \textit{automatic} (\autoref{fig:Preprocessing}-\darkcircled{a3}) extraction. As she is still navigating the system, Emily initially starts by manually inputting comments. She also explores the \textit{Reviewer Comments} page (\autoref{fig:Preprocessing}-\darkcircled{b}), where she finds comments grouped by reviewers. Emily selects this approach to extract comments. Seeking comprehensive coverage, she eventually opts for automatic comment generation.

\par Emily finds the system's collapsible comment cards with categorization features to be particularly valuable (\autoref{fig:Categorization_Mapping}-\darkcircled{a}). She decides to customize categories within the content criteria rectangle (\autoref{fig:Categorization_Mapping}-\darkcircled{b}), adding new categories to address concerns related to her evaluations. Following the categorization based on \textit{Content} and \textit{Workload}, Emily shifts her attention to the \textit{possibly related paragraphs} displayed on the right side of each comment card (\autoref{fig:Categorization_Mapping}-c1). These paragraphs help her match comments with their corresponding sections in the original paper. Clicking on a suggested paragraph resulted in smooth scrolling, highlighting the relevant part of the paper for reference. Emily reviews the paragraph carefully and confirms its relevance. Subsequently, she selects the relevant text and clicks the \raisebox{-0.4ex}{\includegraphics[height=2.2ex]{figure/Addcomment.png}} button that appears in the top-right corner (\autoref{fig:Categorization_Mapping}-c2). This action establishes a link between the comment and the selected paragraph. Dragging the comment card into the \textit{Editing Modal} (\autoref{fig:Categorization_Mapping}-c), Emily notices icons on the card's right side (\autoref{fig:Categorization_Mapping}-c3). These icons depict shapes representing selected categories within each criterion. This visual aid significantly aids Emily in swiftly comprehending the categorization of each comment. Underneath the comment card, Emily takes the opportunity to jot down her revision ideas for comments related to the paragraph (\autoref{fig:Categorization_Mapping}-c4). Upon completion, she clicks the \raisebox{-0.4ex}{\includegraphics[height=2.7ex]{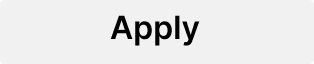}} button to save her changes. Emily observes a \raisebox{-0.4ex}{\includegraphics[height=2.2ex]{figure/annotation.png}} icon on the \textit{Annotation Bar} located to the right of the paper (\autoref{fig:Categorization_Mapping}-c5). This feature enables her to modify existing annotations and adjust associated comment cards as needed.

\par In the final stage of her analysis, Emily aims to streamline her evaluated data to craft a structured overview and formulate her revision plan. Clicking on the \raisebox{-0.4ex}{\includegraphics[height=2.2ex]{figure/Sort.png}} button (\autoref{fig:Organization}-a1), she chooses to organize the comments based on \textit{Workload}. This results in cards labeled as  \textit{High} being grouped separately from those labeled as \textit{Low}. Moving forward, Emily accesses the \textit{Revision Editing} sidebar (\autoref{fig:Organization}-b), utilizing the provided template to arrange her comments and analysis. This template includes fields for \textit{Name of the Issue}, \textit{Reviewer's Comments}, and \textit{Responses}. In the concluding stage, Emily rephrases her revision notes by clicking the \raisebox{-0.4ex}{\includegraphics[height=2.2ex]{figure/rephrase.png}} button (\autoref{fig:Organization}-b1). She subsequently exports these notes through \raisebox{-0.4ex}{\includegraphics[height=2.2ex]{figure/Export.png}} button (\autoref{fig:Organization}-b2) to serve as guidelines for her upcoming revisions and the foundation for her response cover letter. Drawing from her comprehensive experience with the review analysis process, Emily concludes that \textit{ReviseMate} proves invaluable in both understanding reviewer feedback and strategically planning her revisions.

\pagebreak
\subsection{Participants Information in Preliminary Interview}
\begin{table}[htbp]
\begin{tabular}{@{}ccccc p{2cm} @{}}
\toprule
ID & Gender/Age & Exp & Role & Major\\ \midrule
P1 & F/40 & 15/43 & Asst.Prof. & HCI\\
P2 & M/28 & 5/12 & PhD & HCI\\
P3 & F/25 & 2/4 & Graduate & CV\\
P4 & M/25 & 1/4 & Graduate & ML\\
P5 & M/23 & 1/2 & Graduate & ML\\
P6 & M/29 & 5/10 & PhD & CV\\
P7 & F/24 & 1/4 & Graduate & HCI\\
P8 & F/25 & 2/4 & Graduate & CV\\
P9 & M/24 & 2/3 & Graduate & HCI\\
P10 & M/28 & 5/11 & PhD & ML\\
P11 & F/25 & 2/4 & Graduate & HCI\\
P12 & M/37 & 15/41 & Asst.Prof. & HCI\\
\bottomrule
\end{tabular}
\caption{Participants information (F = Female; M = Male; The Experience (Exp), A/B, A means Number of Papers Per Participant as the First Author, B means Number of Papers Per Participant; Asst.Prof. = Assistant Professor; ML = Machine Learning, CV = Computer Vision, HCI = Human-Computer Interaction).}
\label{tab:participants_info}
\end{table}

\clearpage
\subsection{Participants Information in Storyboard Interview}
\begin{table}[htbp]
\begin{tabular}{@{}ccccc p{2cm} @{}}
\toprule
ID & Gender/Age & Exp & Role & Major\\ \midrule
P1 & F/25 & 2/4 & Graduate & CV\\
P2 & M/37 & 15/41 & Asst.Prof. & HCI\\
P3 & M/28 & 5/12 & PhD & HCI\\
P4 & M/24 & 2/3 & Graduate & HCI\\
P5 & F/25 & 2/4 & Graduate & CV\\
P6 & M/31 & 18/44 & Asst.Prof. & HCI\\
P7 & M/24 & 1/3 & Graduate & ML\\
P8 & F/27 & 4/12 & PhD & SE\\
P9 & F/25 & 2/3 & Graduate & CV\\
P10 & M/19 & 0/1 & Undergraduate & SE\\
P11 & M/25 & 1/3 & Graduate & VIS\\
P12 & M/21 & 1/2 & Undergraduate & ML\\
P13 & F/35 & 15/42 & Asst.Prof. & VIS\\
P14 & M/21 & 1/2 & Undergraduate & ML\\
P15 & M/29 & 6/10 & PhD & VIS\\
P16 & F/23 & 2/4 & Graduate & BME\\
P17 & F/40 & 15/50 & Asst.Prof. & ML\\
P18 & F/25 & 1/4 & Graduate & HCI\\
\bottomrule
\end{tabular}
\caption{Participants information (F = Female; M = Male; The Experience (Exp), A/B, A means Number of Papers Per Participant as the First Author, B means Number of Papers Per Participant; Asst.Prof. = Assistant Professor; ML = Machine Learning, CV = Computer Vision, SE = Software Engineering, BME = Biomedical Engineering, HCI = Human-Computer Interaction, VIS = Visualization).}
\label{tab:participants_info_new}
\end{table}

\clearpage
\subsection{Participants Information in Controlled User Study}
\begin{table}[h]
\begin{tabular}{@{}ccccc@{}}
\toprule
ID & Gender/Age & Exp & Role & Major\\ \midrule
T1 & F/19 & 0/1 & Undergraduate & VIS\\
T2 & M/22 & 1/1 & Undergraduate & CV\\
T3 & M/23 & 0/2 & Undergraduate & VIS\\
T4 & F/29 & 4/7 & PhD & HCI\\
T5 & M/25 & 2/3 & Graduate & NLP\\
T6 & F/31 & 3/8 & PhD & BME\\
T7 & M/39 & 20/39 & Asst.Prof. & HCI\\
T8 & M/29 & 3/8 & PhD & ML\\
T9 & M/19 & 1/1 & Undergraduate & CV\\
T10 & F/20 & 0/2 & Undergraduate & ML\\
T11 & F/23 & 2/3 & Graduate & SE\\
T12 & M/35 & 17/33 & Asst.Prof. & CV\\
T13 & F/24 & 1/2 & Graduate & BME\\
T14 & M/27 & 4/11 & PhD & VIS\\
T15 & F/24 & 1/3 & Graduate & CV\\
T16 & M/23 & 2/3 & Graduate & ML\\
T17 & F/30 & 5/9 & PhD & VIS\\
T18 & M/29 & 3/12 & PhD & ML\\
T19 & M/37 & 11/32 & Asst.Prof. & VIS\\
T20 & F/22 & 0/1 & Undergraduate & ML\\
T21 & M/20 & 0/2 & Undergraduate & SE\\
T22 & M/23 & 1/1 & Undergraduate & HCI\\
T23 & M/20 & 1/2 & Undergraduate & CV\\
T24 & F/19 & 0/1 & Undergraduate & HCI\\
T25 & M/25 & 2/4 & Graduate & CG\\
T26 & M/28 & 5/7 & PhD & BME\\
T27 & F/33 & 17/22 & Asst.Prof. & VIS\\
T28 & M/34 & 13/37 & Asst.Prof. & HCI\\
T29 & M/28 & 5/9 & PhD & CV\\
T30 & F/20 & 0/1 & Undergraduate & ML\\
T31 & F/24 & 1/2 & Graduate & VIS\\
\bottomrule
\end{tabular}
\caption{Participants information (F = Female; M = Male; The Experience (Exp), A/B, A means Number of Papers Per Participant as the First Author, B means Number of Papers Per Participant; Asst.Prof. = Assistant Professor; ML = Machine Learning, CV = Computer Vision, SE = Software Engineering, BME = Biomedical Engineering, HCI = Human-Computer Interaction, VIS = Visualization, CG = Computer Graphics, NLP = Natural Language Processing).}
\label{tab:participants1}
\end{table}

\clearpage
\subsection{Designed Questions in Questionnaires and Interviews}
\begin{table}[h]
\centering
\begin{tabular}{p{13cm}}
\hline\hline
1. Your Gender: \quad \textcircled{1} Male \quad \textcircled{2} Female \quad \textcircled{3} Others \\
2. Your current academic level: \quad \textcircled{1} Undergraduate \quad \textcircled{2} Graduate \quad \textcircled{3} PhD \quad \textcircled{4} Asst. Prof \\
3. Your field of study/Your Major: \quad \textcircled{1} HCI \quad \textcircled{2} VIS \quad \textcircled{3} CV \quad \textcircled{4} ML \quad \textcircled{5} SE \quad \textcircled{6} BME \quad \textcircled{7} Other \\
4. How many academic papers have you submitted as a researcher? \\
5. How many academic papers have you contributed to? \\
\hline\hline
\end{tabular}
\captionof{table}{Pre-task survey.}
\label{tab:Pre_task_questions}
\end{table}
\begin{table}[h]
\centering
\begin{tabular}{l p{0.8\linewidth}}
\hline\hline
\textbf{Category} & \textbf{Question} \\
\hline
\multirow{6}{*}{Effectiveness} & 1. I can extract review comments from original review texts conveniently. \\
 & 2. I can conduct detailed and in-depth comment analysis accessibly. \\
 & 3. After comments analysis, I can organize my analyzed information effectively. \\
 & 4. I am confident with my analysis. \\
 & 5. I am satisfied with the quality of my revision outline. \\
 & 6. My entire analysis process is efficient. \\
\hline
\multirow{9}{*}{Usability} & 1. The extracted comments generated by the system are useful. \\
 & 2. The integration part (making annotations) is useful when writing the outline. \\
 & 3. The categorization of each comment is useful. \\
 & 4. The mapping from review comments to the corresponding paragraph is useful. \\
 & 5. I would like to recommend the system to others and use it in the future. \\
 & 6. The system is easy to use. \\
 & 7. The system is helpful for supporting review analysis. \\
 & 8. I am satisfied when using the system. \\
 & 9. I feel distracted when using the system. \\
\hline\hline
\end{tabular}
\caption{In-task survey for participants in 7-point Likert scale(1: Strongly Disagree, 7: Strongly Agree).}
\label{tab:In_task_questions}
\end{table}
\begin{table}[h]
\centering
\begin{tabular}{l p{0.8\linewidth}}
\hline\hline
\textbf{Category} & \textbf{Question} \\
\hline
\multirow{4}{*}{Trust} & 1. I trust the extracted comments the system generated. \\
 & 2. I trust the overall support of the system. \\
 & 3. The overall support of the system is accurate and helpful. \\
 & 4. I trust the mapping from review comments to the corresponding paragraph generated by the system. \\
\hline\hline
\end{tabular}
\caption{Post-task survey for participants in 7-point Likert scale(1: Strongly Disagree, 7: Strongly Agree).}
\label{tab:Post_task_questions}
\end{table}

\begin{table}[h]
\centering
\begin{tabular}{l p{0.6\linewidth}}
\hline\hline
\textbf{Category} & \textbf{Question} \tabularnewline
\hline
\multirow{9}{*}{Comment Extraction} 
& 1. I observed that you read the original review comments after they were automatically extracted, why? \tabularnewline
\cline{2-2}
& 2. I observed that you read through the reviews of different reviewers first before automatically generating the extracted review comments, why? \tabularnewline
\cline{2-2}
& 3. Why were some of the comment cards deleted after they were generated? \tabularnewline
\cline{2-2}
& 4. I observed that you read the original opinion before automatically generating the extracted opinion, why? \tabularnewline
\cline{2-2}
& 5. I observed that you manually added some comment cards after the fully automated generation of them, why? \tabularnewline
\hline
\multirow{11}{*}{Categorization and mapping} 
& 1. I noticed that you added many subcategories under the content category while organizing, what's the reason? \tabularnewline
\cline{2-2}
& 2. Why did you write the revision outline of a review immediately after you have finished mapping and categorizing of the same review? \tabularnewline
\cline{2-2}
& 3. Why did you write the outline after you've categorized and mapped all your comment cards? \tabularnewline
\cline{2-2}
& 4. I observed that you didn't use the feature of categorization, why? \tabularnewline
\cline{2-2}
& 5. I observed that you strictly categorized the reviews based on predefined categorization and didn't add anything new, why? \tabularnewline
\cline{2-2}
& 6. It seems that you only used the mapping function but not the annotation function, why was that? \tabularnewline
\hline
\multirow{4}{*}{Organization and writing} 
& 1. I noticed you didn't follow the process exactly, why? \tabularnewline
\cline{2-2}
& 2. The outlines seem to be written in such a way that one issue corresponds to one of the reviews, and there is no integration of reviews, so what is the reason for that? \tabularnewline
\cline{2-2}
& 3. I realized that you were doing the integration based on content, why? \tabularnewline
\hline\hline
\end{tabular}
\caption{Questions in the controlled user study.}
\label{tab:Controlled_study_questions}
\end{table}

\begin{table}[h]
\centering
\begin{tabular}{l p{0.6\linewidth}}
\hline\hline
\textbf{Category} & \textbf{Question} \tabularnewline
\hline
\multirow{4}{*}{Usability } 
& 1. What is your impression of the system, why? \tabularnewline
\cline{2-2}
& 2. Is the tool's interface and the way it operates easy to get started?  \tabularnewline
\cline{2-2}
& 3. Is there anything in particular that you are not satisfied with? \tabularnewline
\cline{2-2}
& 4. Have you encountered anything that has confused you when using the tools? \tabularnewline
\hline
\multirow{2}{*}{Effectiveness} 
& 1. Has this system helped you to better understand review comments and revise your paper? Can you give me an example? \tabularnewline
\cline{2-2}
\hline
\multirow{6}{*}{Acceptance and collaboration} 
& 1. What do you think of the idea of using large language models to assist humans in review comment extraction? Do you accept this type of collaboration? \tabularnewline
\cline{2-2}
& 2. Are there any security or privacy concerns about sharing review comments and paper content with LLM? \tabularnewline
\cline{2-2}
& 3. Do you think that this system can improve the efficiency or quality of collaboration between you and your team members? \tabularnewline
\hline\hline
\end{tabular}
\caption{Questions in field deployment study.}
\label{tab:Qualitative_study_questions}
\end{table}

\begin{table}[h]
\centering
\begin{tabular}{p{0.3\textwidth} p{0.3\textwidth} p{0.02\textwidth} p{0.3\textwidth}}
\hline\hline
\multicolumn{1}{c}{\textbf{Scenario}} & \multicolumn{1}{c}{\textbf{Design Concept}} & \multicolumn{1}{c}{} & \multicolumn{1}{c}{\textbf{Storyboard}} \\
\hline
Users are overwhelmed by seeing the original review and want to extract the review individually.
& Provide diverse methods for comment extraction, using comment cards as a vehicle.
& 1 & Provides a semi-automatic method of adding comment cards. \\ \cline{3-4}
& & 7 & The system summarizes reviews and provides a centralized overview, which can be efficiently processed and responded to. \\ \cline{3-4}
& & 10 & Provides an automatic method of adding comment cards. \\ \cline{3-4}
\hline
Users want to categorize and sort different reviews after extracting reviews.
& Provide methods for customized categorization, classification, and rating.
& 2 &  The system allows the user to design their own categorization criteria for reviews\\ \cline{3-4}
& & 3 &  The system allows the user to categorize reviews according to gradient criteria when designing.\\ \cline{3-4}
& & 9 &  The system allows users to sort by specific criteria. \\ \cline{3-4}
\hline
Users find that the reviews are too scattered, so it is hard to map them to the paper.
& Provide ways to link the comments to the original paper directly.
& 6 & The system can analyze location information in the review and provide locations in the original text that may correspond to the review.\\ 
\hline
Users want to add their own revision plans to the review and map them to the corresponding place in the paper.
&Engaging the user in forming actual connections between reviews and paper through interactions.
& 8 & User can drag and drop comment cards to establish a connection between the review and the original text.
\\ 
\hline
After processing the review, users want to make a revision plan and write a cover letter.
& Gantt chart to organize this information, allowing users to customize the chart to plan their revision time schedules.
& 4 & The system can integrate comment cards and notes to create a Gantt graph. \\ \cline{3-4}
& & 5 & The system can assist authors in writing cover letter drafts through ChatGPT.\\ \cline{3-4}
\hline\hline
\end{tabular}
\caption{The list of scenarios, design concepts, and storyboards.}
\label{tab:storyboards}
\end{table}

\clearpage
\subsection{Baseline System Interface}
\begin{figure}[h]
    \centering
    \includegraphics[width=\textwidth]{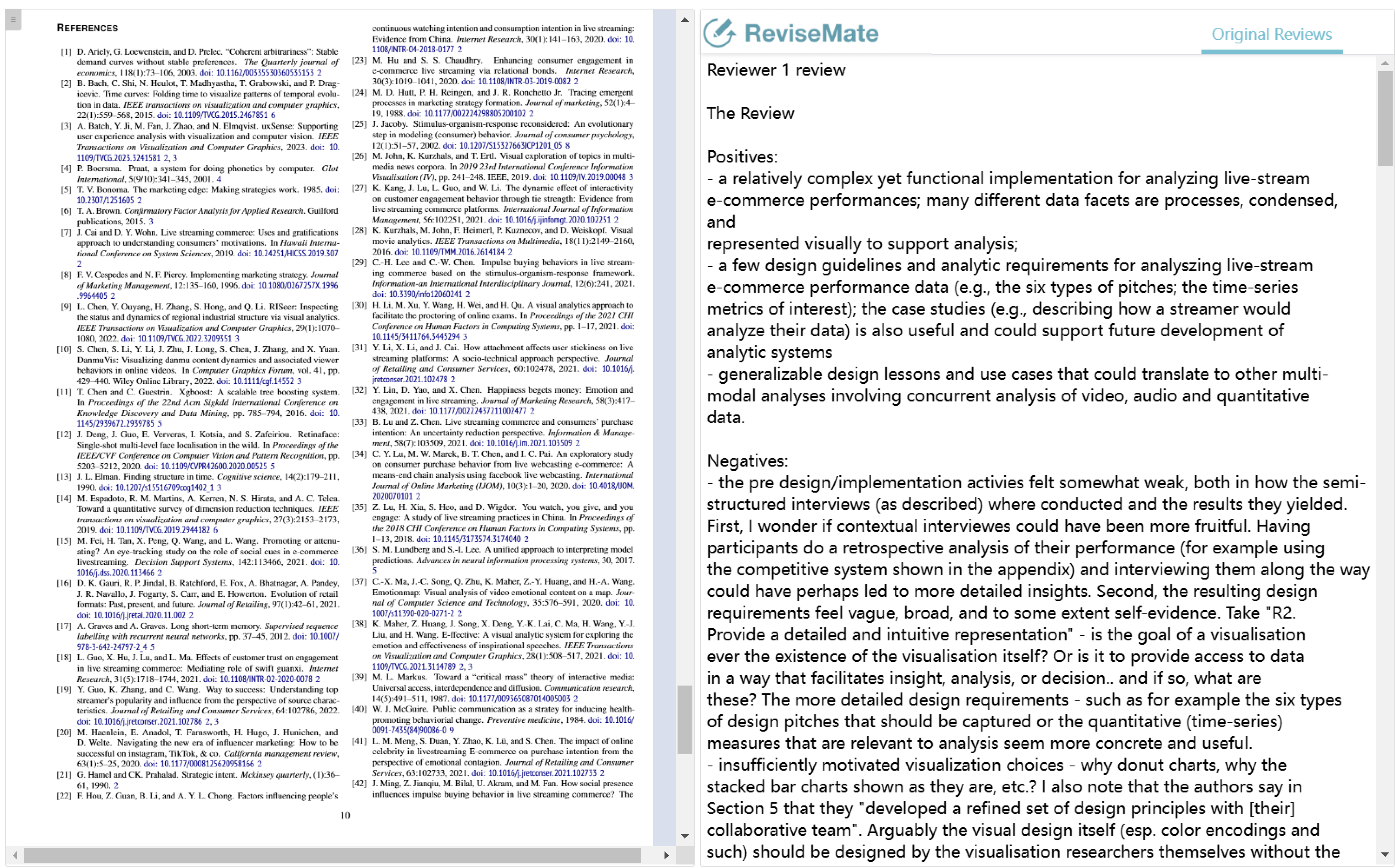}
    \caption{The interface of the baseline system. The interface displays the
original manuscript and the review text side by side, but without any
additional interactions or features.}
    \label{fig:baseline}
\end{figure}
\end{document}